\documentclass[
  journal=pasa,
  manuscript=research-paper, 
  year=2021,
  volume=37,
]{cup-journal}

\usepackage{graphicx}
\usepackage{adjustbox}
\usepackage{xcolor}
\usepackage{microtype,siunitx}
\usepackage{amsmath}
\usepackage{amssymb}
\usepackage{gensymb}
\usepackage[backref=page]{hyperref}
\definecolor{dodgerblue}{RGB}{30, 144, 255}
\definecolor{crimson}{RGB}{220, 20, 60}
\definecolor{darkerblue}{RGB}{0, 0, 139}
\hypersetup{colorlinks,citecolor=dodgerblue,linkcolor=blue,urlcolor=darkerblue}
\usepackage{enumitem, xcolor}
\let\svitem\item

\usepackage{booktabs} 
\usepackage{threeparttable,longtable}  
\usepackage{makecell}

\usepackage[skip=0.5ex]{subcaption}
\usepackage{cprotect}
\renewcommand\thesubfigure{(\roman{subfigure})}

\usepackage{multirow}

\title{The Rapid ASKAP Continuum Survey V: cataloguing the sky at 1367.5~MHz and the second data release of RACS-mid}

\author{S.~W.~Duchesne}
\affiliation{CSIRO Space and Astronomy, PO Box 1130, Bentley WA 6102, Australia}
\email[S.~W.~Duchesne]{stefan.duchesne.astro@gmail.com} 

\author{J.~A.~Grundy}
\affiliation{International Centre for Radio Astronomy Research, Curtin University, Bentley, WA 6102, Australia}
\alsoaffiliation{CSIRO Space and Astronomy, PO Box 1130, Bentley WA 6102, Australia}

\author{George~H.~Heald}
\affiliation{CSIRO Space and Astronomy, PO Box 1130, Bentley WA 6102, Australia}

\author{Emil~Lenc}
\affiliation{CSIRO Space and Astronomy, PO Box 76, Epping, NSW, 1710, Australia}

\author{James~K.~Leung}
\affiliation{Sydney Institute for Astronomy, School of Physics, University of Sydney, NSW 2006, Australia}
\alsoaffiliation{CSIRO Space and Astronomy, PO Box 76, Epping, NSW, 1710, Australia}

\author{David~McConnell}
\affiliation{CSIRO Space and Astronomy, PO Box 76, Epping, NSW, 1710, Australia}

\author{Tara~Murphy}
\affiliation{Sydney Institute for Astronomy, School of Physics, University of Sydney, NSW 2006, Australia}

\author{Joshua~Pritchard}
\affiliation{Sydney Institute for Astronomy, School of Physics, University of Sydney, NSW 2006, Australia}
\alsoaffiliation{CSIRO Space and Astronomy, PO Box 76, Epping, NSW, 1710, Australia}

\author{Kovi~Rose}
\affiliation{Sydney Institute for Astronomy, School of Physics, University of Sydney, NSW 2006, Australia}
\alsoaffiliation{CSIRO Space and Astronomy, PO Box 76, Epping, NSW, 1710, Australia}

\author{Alec~J.~M.~Thomson}
\affiliation{CSIRO Space and Astronomy, PO Box 1130, Bentley WA 6102, Australia}

\author{Yuanming~Wang}
\affiliation{Centre for Astrophysics and Supercomputing, Swinburne University of Technology, Hawthorn, Victoria 3122, Australia}
\alsoaffiliation{ARC Centre of Excellence for Gravitational Wave Discovery (OzGrav), Hawthorn, Victoria 3122, Australia}

\author{Ziteng~Wang}
\affiliation{Sydney Institute for Astronomy, School of Physics, University of Sydney, NSW 2006, Australia}
\alsoaffiliation{CSIRO Space and Astronomy, PO Box 76, Epping, NSW, 1710, Australia}

\author{Matthew~T.~Whiting}
\affiliation{CSIRO Space and Astronomy, PO Box 76, Epping, NSW, 1710, Australia}

\doi{10.1017/pasa.20XX.XX}

\received {dd Mmm YYYY}
\revised  {dd Mmm YYYY}
\accepted {dd Mmm YYYY}
\published{22 September 2020}

\keywords{catalogues; surveys; radio continuum: general; radio continuum: galaxies; radio continuum: stars} 

\definecolor{dodgerblue}{RGB}{30, 144, 255}

\begin{document}

\begin{abstract}

The Australian SKA Pathfinder (ASKAP) has surveyed the sky at multiple frequencies as part of the Rapid ASKAP Continuum Survey (RACS). The first two RACS observing epochs, at 887.5 (RACS-low) and 1367.5 (RACS-mid) MHz, have been released \citep{racs1,racs-mid}. A catalogue of radio sources from RACS-low has also been released, covering the sky south of declination $+30\degree$ \citep{racs2}. With this paper, we describe and release the first set of catalogues from RACS-mid, covering the sky below declination $+49\degree$. The catalogues are created in a similar manner to the RACS-low catalogue, and we discuss this process and highlight additional changes. The general purpose primary catalogue covering 36\,200\,deg$^2$ features a variable angular resolution to maximise sensitivity and sky coverage across the catalogued area, with a median angular resolution of $11.2^{\prime\prime} \times 9.3^{\prime\prime}$. The primary catalogue comprises 3\,105\,668  radio sources, including those in the Galactic Plane (2\,861\,923 excluding Galactic latitudes of $|b|<5\degree$) and we estimate the catalogue to be 95\% complete for sources above 1.6 mJy. With the primary catalogue, we also provide two auxiliary catalogues. The first is a fixed-resolution, 25-arcsec catalogue approximately matching the sky coverage of the RACS-low catalogue. This 25-arcsec catalogue is constructed identically to the primary catalogue, except images are convolved to a less-sensitive 25-arcsec angular resolution. The second auxiliary catalogue is designed for time-domain science, and is the concatenation of source-lists from the original RACS-mid images with no additional convolution, mosaicking, or de-duplication of source entries to avoid losing time-variable signals. All three RACS-mid catalogues, and all RACS data products, are available through the CSIRO ASKAP Science Data Archive \footnote{\url{https://research.csiro.au/casda/}.}.
\end{abstract}

\section{Introduction}
\label{sec:int}

\defcitealias{racs1}{Paper~I}
\defcitealias{racs2}{Paper~II}
\defcitealias{Thomson2023}{Paper~III}
\defcitealias{racs-mid}{Paper~IV}

The Australian SKA Pathfinder \citep[ASKAP;][]{Hotan2021}, located at Inyarrimanha Ilgari Bundara, the CSIRO \footnote{Commonwealth Scientific and Industrial Research Organisation.} Murchison Radio-astronomy Observatory in Western Australia, is a survey radio telescope that operates with a 288-MHz bandwidth between 700 and 1800\,MHz. ASKAP features a unique Phased Array Feed \citep[PAF;][]{Hotan2014,McConnell2016} which allows the formation of 36 primary beams covering up to $\sim 31$\,deg$^{2}$. This instantaneous wide field of view coupled with $36\times12$\,m antennas---with baselines ranging from 22\,m to 6\,km---provides the basis for an instrument optimised for survey speed, with the capability to rapidly survey $\sim 80$\% of the sky to heretofore unseen sensitivity.

The Rapid ASKAP Continuum Survey (RACS) is a multi-wavelength survey covering the sky up to $\delta_\text{J2000} \lesssim +50\degree$ \citep[][hereinafter, \citetalias{racs1}]{racs1}. The survey is being conducted within the three operating bands used by ASKAP. For RACS, these three bands are centered on effective frequencies of 887.5, 1367.5, and 1632.5\,MHz. As the name suggests, RACS is a shallow survey with 15-min total integration per pointing reaching median root-mean-square (rms) noise between $\sim 200$--260\,\textmu Jy\,PSF$^{-1}$ across the three bands. Due to its baseline distribution, ASKAP has excellent snapshot $(u,v)$ coverage, providing a well-constrained point-spread function at zenith ($\delta_\text{J2000} \approx -27\degree$) with elongation of the major axis for lower-elevation pointings. The resultant median angular resolution of the whole survey across the three bands is in the range $\sim 9$--$18$\,arcsec. The goal of RACS is to provide a global sky model for calibration and validation of other ASKAP observations, and once images and source catalogues are available work can be done matching and spectrally modelling sources across the three bands.  

The first pass of the sky in the lowest frequency band at 887.5 MHz (RACS-low) was released in 2020 and is described in \citetalias{racs1}. RACS-low features a median rms noise of $\sim 260$\,\textmu Jy\,PSF$^{-1}$ and median angular resolution of 18\,arcsec. Following the initial data release, a radio source catalogue was constructed and described by \citet[][hereinafter \citetalias{racs2}]{racs2}. This radio source catalogue contains $\sim 2.1$ million sources, excluding Galactic latitudes of $|b| < 5\degree$. The catalogue was constructed after combining neighbouring images from the original data release to provide a more uniform sensitivity across the region covering $-80\degree \leq \delta_\text{J2000} \leq +30\degree$. This catalogue is the first large-area catalogue produced as part of RACS. Alongside the Stokes I total intensity data releases, RACS-low data products are also being used to investigate the polarized sky. Using circular polarization data products from RACS-low, \citet{Pritchard2021} identified radio emission coincident with 33 stars, including 23 new radio star associations in this sample. In linear polarization, RACS data products are being used to generate Spectra and Polarisation In Cutouts of Extragalactic Sources \citep[SPICE-RACS;][\citetalias{Thomson2023}]{Thomson2023}. The first data release from SPICE-RACS covers a 1\,300~deg$^2$ pilot region and features 5\,818 sources with Faraday rotation measures.

The second observing epoch of the sky at 1367.5\,MHz (RACS-mid) was completed in 2021, with follow-up observations taken over 2021--2022, and the first data release is described by \citet[][hereinafter \citetalias{racs-mid}]{racs-mid}. RACS-mid data products, along with all available RACS data, are publicly available through the CSIRO ASKAP Science Data Archive \citep[CASDA;][]{casda,Huynh2020}. RACS-mid features a number of changes from RACS-low that helped to improve the overall quality of the survey. These changes include: fully-autonomous scheduling with a limit to the hour angle of $\pm 1^\text{h}$, peeling and subtraction of bright off-axis sources, bespoke primary beam modelling, and correction of both on-axis and off-axis leakage of Stokes I into V. While RACS-mid has an effective bandwidth of 144-MHz due to RFI, the survey still boasts a median rms noise of $\sim 200$\,\textmu Jy\,PSF$^{-1}$, which varies smoothly across the sky as a function of declination, except around bright sources. Similarly, because of the close-to-meridian scheduling the angular resolution varies smoothly as a function of declination ranging from 8.1\,arcsec near zenith and 47.5\,arcsec in the lowest-elevation pointings at high declination. The first RACS-mid data release also included leakage-corrected Stokes V continuum images alongside the Stokes I images for each individual observation.

RACS-mid is both competitive with and complementary to existing completed large-area 1.4-GHz radio surveys in the Northern sky such as the NRAO \footnote{National Radio Astronomy Observatory.} VLA \footnote{Very Large Array.} Sky Survey \citep[NVSS;][]{ccg+98} and the Faint Images of the Radio Sky at Twenty Centimeters \citep[FIRST;][]{Becker95,White1997,hwb15}. In the highest declination strip covered by RACS-mid ($\delta_\text{J2000} = +46\degree$), the resolution and sensitivity approaches that of the NVSS. For the remainder of the survey the resolution and sensitivity can be expressed as an average of the survey properties of the NVSS and FIRST, with a median angular resolution of $10.1$\,arcsec and moderate sensitivity to extended sources up to a few arcmin. RACS-mid data products have so far been used to identify circularly polarized emission from 52 stars that had not previously been detected at radio frequencies (Driessen et al. in prep), including the detection of circularly polarized emission from a T8 brown dwarf \citep{Rose2023}, and have been used in conjunction with FIRST and the VLA Sky Survey \citep[VLASS;][]{Gordon2021} in a search for radio stars via proper motion \citep{Driessen2023}. RACS-low and RACS-mid images are also used as an extra epoch for the Variability and Slow Transients (VAST) survey being performed by ASKAP \citep{Murphy2013,Murphy2021}. 

At 1367.5\,MHz, ASKAP is also undertaking the Widefield ASKAP L-band Legacy All-sky Blind surveY (WALLABY), which is a neutral hydrogen survey being performed over the next five years with using the same frequency band as RACS-mid \citep{wallaby1}. While WALLABY is primarily a spectral line survey, the 10-hour observations and subsequent imaging also result in deep 1367.5-MHz continuum images. Currently pre-pilot observations of the Eridanus supergroup \citep{For2021} have been utilised to create an initial WALLABY continuum catalogue \citep[hereinafter the WALLABY pre-pilot catalogue;][]{Grundy2023} containing 9\,416 sources covering $\sim 42$\,deg$^{2}$. 

In this fifth paper in the RACS series, we describe the cataloguing of the RACS-mid Stokes I data. Although Stokes V continuum images are available, cataloguing circularly polarized sources with RACS-mid will be described in a future work.

\section{Catalogue creation}\label{sec:catalogue}

\begin{figure*}[p]
    \centering
    \begin{subfigure}[b]{1\linewidth}
    \includegraphics[width=1\linewidth]{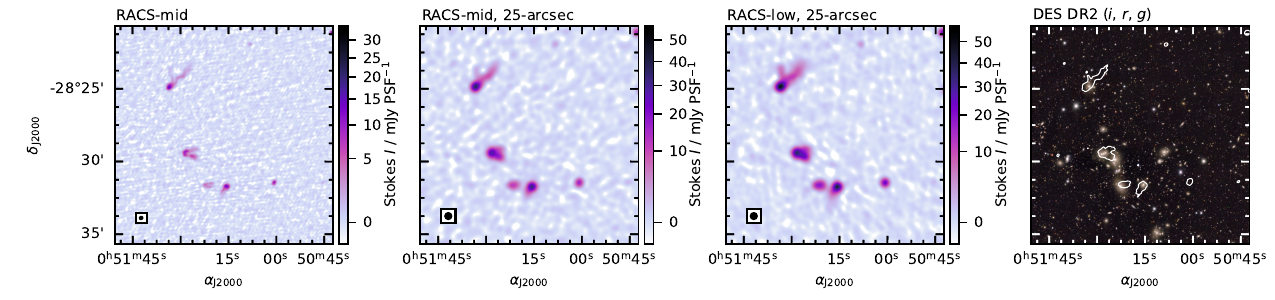}
    \caption{RACS images of radio galaxies near the galaxy cluster Abell 2829. \label{fig:images:1}}
    \end{subfigure}\\[0.5em]%
    \begin{subfigure}[b]{1\linewidth}
    \includegraphics[width=1\linewidth]{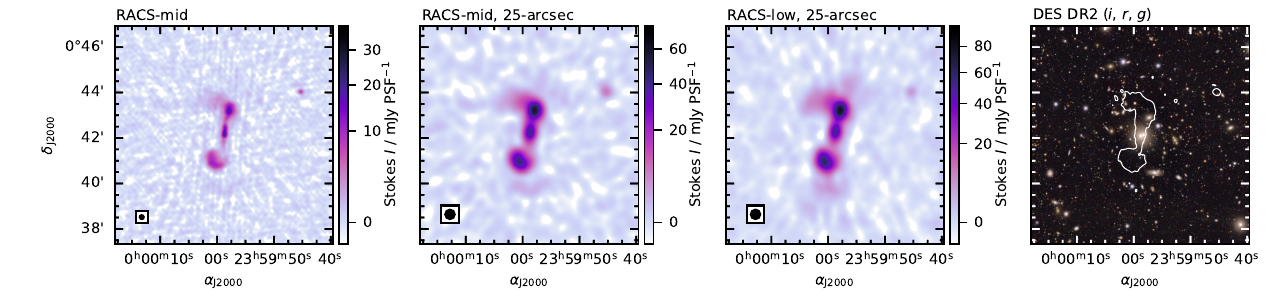}
    \caption{RACS images of PKS~2357$+$00. \label{fig:images:2}}
    \end{subfigure}\\[0.5em]%
    \begin{subfigure}[b]{1\linewidth}
    \includegraphics[width=1\linewidth]{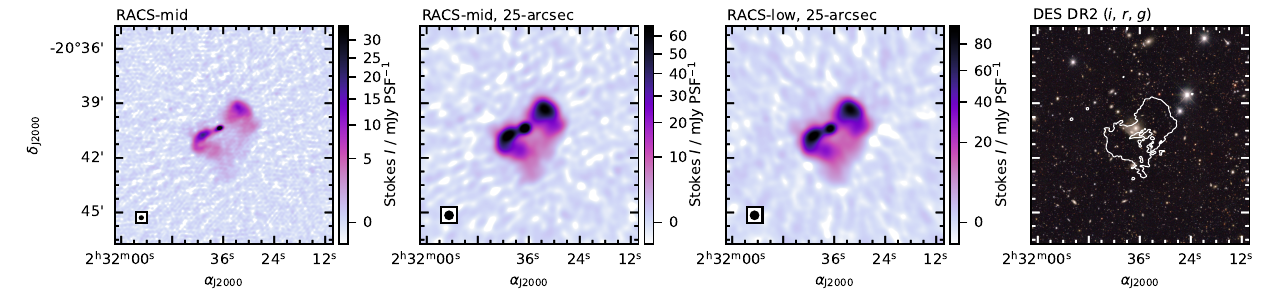}
    \caption{RACS images of PKS~0229$-$208. \label{fig:images:3}}
    \end{subfigure}\\[0.5em]%
    \begin{subfigure}[b]{1\linewidth}
    \includegraphics[width=1\linewidth]{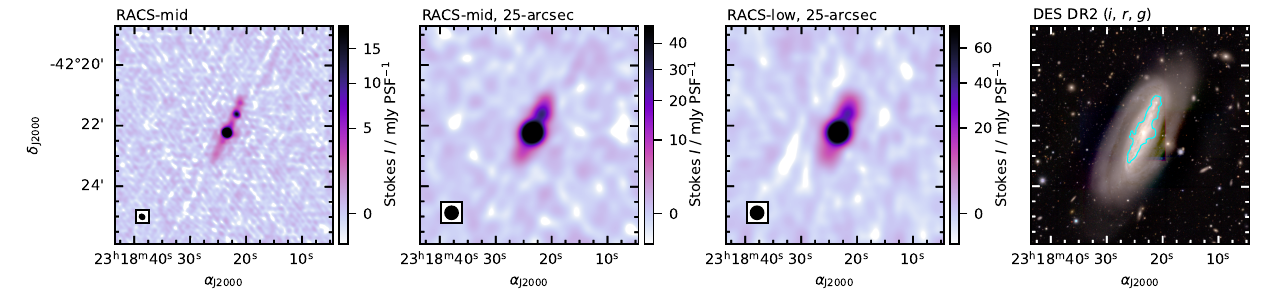}
    \caption{RACS images of NGC~7582. \label{fig:images:4}}
    \end{subfigure}\\[0.5em]%
    \caption{\label{fig:images} Example cutouts of sources in RACS images. The images are the RACS-mid full-sensitivity image (\emph{left}), the RACS-mid 25-arcsec image (\emph{centre left}), the RACS-low 25-arcsec image (\emph{centre right}, \citetalias{racs2}), and a three-color DES DR2 image ($i$, $r$, $g$ assigned to RGB, \emph{right}). The radio colour scales follow a square-root stretch between $[-2, 200]\sigma_{\text{rms}}$. The ellipses in the lower left of the RACS images show the size of the PSF. $5\sigma_{\text{rms}}$ contours from the left panel are drawn on the DES images for reference.}
\end{figure*}

\begin{figure*}[p]
    \centering
    \begin{subfigure}[b]{1\linewidth}
    \includegraphics[width=1\linewidth]{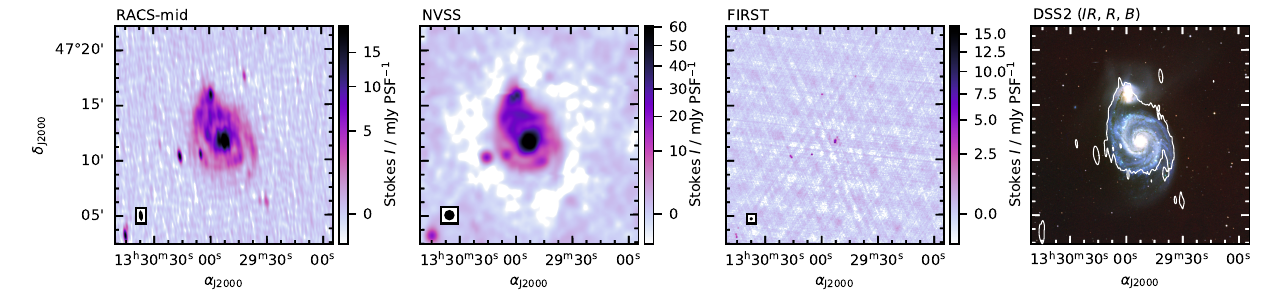}
    \caption{Images of M 51. \label{fig:highdec:1}}
    \end{subfigure}\\[0.5em]%
    \begin{subfigure}[b]{1\linewidth}
    \includegraphics[width=1\linewidth]{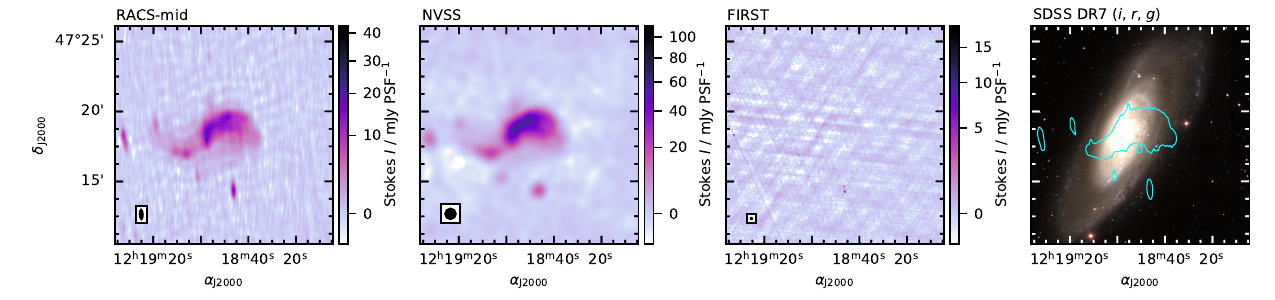}
    \caption{Images of M 106. \label{fig:highdec:2}}
    \end{subfigure}\\[0.5em]%
    \begin{subfigure}[b]{1\linewidth}
    \includegraphics[width=1\linewidth]{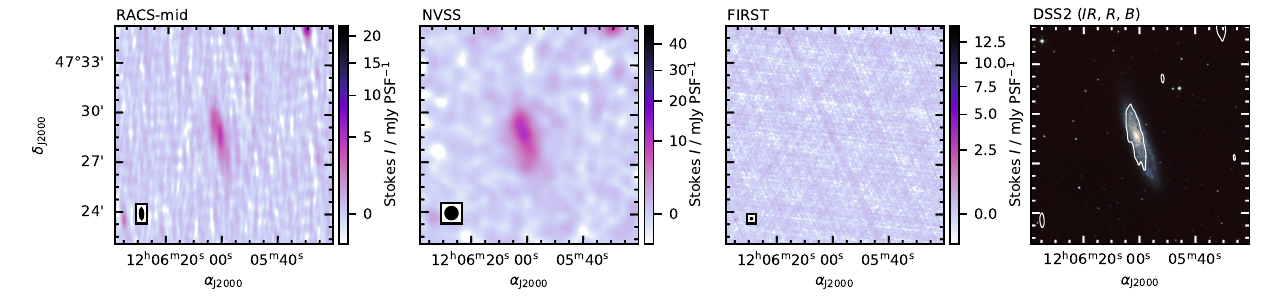}
    \caption{Images of NGC~4096. \label{fig:highdec:3}}
    \end{subfigure}\\[0.5em]%
    \begin{subfigure}[b]{1\linewidth}
    \includegraphics[width=1\linewidth]{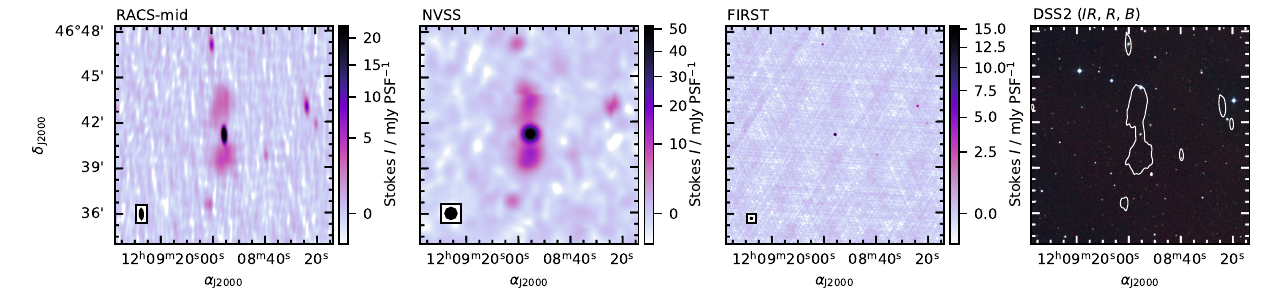}
    \caption{Images of B3~1206$+$469. \label{fig:highdec:4}}
    \end{subfigure}\\[0.5em]%
\caption{\label{fig:highdec} Examples cutouts of high-declination sources. The images are the RACS-mid full-sensitivity image (\emph{left}), the NVSS (\emph{centre left}), FIRST (\emph{centre right}), and a three-color DSS2 or SDSS DR7 image \emph{right}). The radio colour scales follow a square-root stretch between $[-2, 100]\sigma_{\text{rms}}$. The ellipses in the lower left of the RACS images show the size of the PSF. $5\sigma_{\text{rms}}$ contours from the RACS-mid image are drawn on the DSS2 and SDSS DR7 images for reference.}
\end{figure*}

In cataloguing the RACS-mid Stokes I data, we follow the steps outlined for the construction of the RACS-low catalogue described by \citetalias{racs2}. We make some changes to this process to better suit the current data products. As a departure from \citetalias{racs2}, we opted to create a primary catalogue for general purpose use, as well as two auxiliary catalogues for more specific uses. The three catalogues cover the normal uses for Stokes I source catalogues, each with their own limitations and strengths. In the following sections, we describe the general steps taken to construct the catalogues, including selection of which observations to include (Section~\ref{sec:tileselection}), making full-sensitivity images (Section~\ref{sec:images}), source-finding (Section~\ref{sec:sourcefinding}), and merging source-lists and duplicate removal (Section~\ref{sec:duplicates}). The primary catalogue and the two auxiliary catalogues are described in Section~\ref{sec:catalogues} and are created following the relevant subset of the following steps.

\subsection{Image selection}\label{sec:tileselection}

As part of the RACS-mid observing process, some fields were observed and imaged several times. Fields have the naming convention \texttt{RACS\_HHMM$\pm$DD}, and most duplicate field observations and their resultant Stokes I and V images are made available through CASDA. As described in section 2.1 in \citetalias{racs-mid}, some of these re-observations were intended to provide a higher-resolution or to account for observations of $<12$\,min (see section 2.1 in \citetalias{racs-mid}). Fields that were re-observed due to fatal errors with the instrument are not considered. We define a `good' subset of observed fields to use in the cataloguing process. Each RACS-mid observation has a dedicated scheduling block ID (SBID) which is used as a unique identifier. The selection of which SBID to use when multiple observations exist for a given field is based on the following criteria:
\begin{itemize}
    \item[1] If either duplicate SBID had bright sources peeled (as described in section~2.3.3 of \citetalias{racs-mid}), then use the SBID that has undergone peeling,
    \item[2] Else if either duplicate SBIDs have $< 12$\,m total integration time, then Use the SBID with the longest integration time,
    \item[3] Else if the difference in PSF major axes is $>2^{\prime\prime}$ between duplicate SBIDs, then select the SBID with the smaller PSF major axes,
    \item[4] Else select the SBID with the lowest median rms noise.
\end{itemize}

The ordering is important, as peeling in a small number of cases was required to remove CLEAN divergence artefacts but was not necessarily done on the highest-resolution image. The SBIDs with $<12$\,m total integration time are significantly less sensitive than their re-observations so naturally should not be used when more sensitive observations exists. No short observations had better PSF or noise properties than their respective re-observations, and none of these short observations underwent peeling. Finally, resolution and noise properties are often closely related due to their direct relationship to $(u,v)$ coverage and we find that generally the selection is based on the PSF size rather than the median rms noise if criteria (1) and (2) are not met. A single field was observed three times (\texttt{RACS\_1844+46}), though only in the second observation was 3C~380 peeled and is chosen as the good SBID. The selection of good SBIDs is noted in the RACS database file \footnote{\url{https://bitbucket.csiro.au/projects/ASKAP_SURVEYS/repos/racs/browse}.} under the column \texttt{SELECT}, where \texttt{SELECT=1} indicates the SBID is selected as part of the good SBID subset. We note that SBIDs not part of this subset can still be used for other science cases and may be particularly useful for variability or transient source studies. The selection is based on the Stokes I images only, and future work with Stokes V may wish to select different SBIDs if noise properties vary between them. 

\subsection{Full-sensitivity images}\label{sec:images}

A 12\% cut-off in the primary beam attenuation was used for primary beam correction and weighting the PAF beam mosaics for each observation. One of the unfortunate features of the linear mosaicking of individual PAF beam images is a tile mosaic with non-rectangular image boundaries and the usual primary beam roll-off. This results in an irregular tile image border that follows the 12\% contour of the outer beams. From a science standpoint, the primary beam roll-off limits detection of sources towards image boundaries due to a drop in sensitivity. From a user-perspective, rectangular images with large, irregular areas of blanked pixels can make identifying which sources lie in which image difficult. 

As RACS is observed with an overlap between adjacent observations that is at least the PAF beam separation (0.9~degrees for RACS-mid), we follow \citetalias{racs2} and use \texttt{SWarp} \citep{swarp} to form a linear mosaic of adjacent tiles to create rectangular images with full sensitivity towards the image edges. We use the same weight maps used during PAF beam linear mosaicking to ensure the sensitivity profile is consistent across the full image. All adjacent tiles within a group are convolved to a common resolution \footnote{Using \texttt{beamcon\_2D} from \url{https://github.com/AlecThomson/RACS-tools}, which is a generic radio image 2-D convolution tool.} prior to mosaicking. We create one set of images convolved to the lowest common resolution of neighbouring images, and a second set convolved to a fixed 25-arcsec resolution. These images are provided as part of this data release. Convolving to the lowest common resolution results in mosaics that are close to the full resolution of the survey. These full-sensitivity images are created for Stokes I and V, though only Stokes I images are part of this data release. For the 25-arcsec images, we also increase the pixel size by a factor two in each celestial coordinate (i.e. from $2^{\prime\prime} \times 2^{\prime\prime}$ to $4^{\prime\prime} \times 4^{\prime\prime}$) to reduce the image file size.  {The 25-arcsec full-sensitivity images (and catalogue) are restricted to $\delta_\text{J2000} \lesssim +30^\circ$, similar to RACS-low, as the PSF major axis for some of the original images becomes larger than 25\,arcsec beyond this declination.} 

Figure~\ref{fig:images} shows example sources as they appear in the full-sensitivity images compared to the 25-arcsec image from RACS-low \citepalias{racs2}, and comparison optical data from from the Dark Energy Survey data release 2 \citep[DES DR2;][]{decam,des1,des:dr2}. Figure~\ref{fig:images:1} shows an example of radio sources projected onto the galaxy cluster Abell 2829, highlighting a range of angular scales and features. Figures~\ref{fig:images:2} and \ref{fig:images:3} shows two radio galaxies, PKS~2357$+$00 and PKS~0229$-$208, and \ref{fig:images:4} shows the spiral galaxy NGC~7582. Note these examples highlight situations where the angular resolution of the full-sensitivity image ($9.8^{\prime\prime} \times 8.5^{\prime\prime}$ and $11.4^{\prime\prime} \times 9.2^{\prime\prime}$) is significantly reduced to form the 25-arcsec images. 

Similarly, Figure~\ref{fig:highdec} shows example sources in the highest-declination strip of the survey ($\delta_\text{J2000} = +46\degree$) with comparison to the NVSS, FIRST, and either the Digitized Sky Survey (DSS2) or the Sloan Digitized Sky Survey data release 7 \citep[SDSS DR7;][]{sdssdr7}. Figures~\ref{fig:highdec:1}--\ref{fig:highdec:3} show the spiral galaxies M~51, M~106, and NGC~4096, and Figure~\ref{fig:highdec:4} shows the giant radio galaxy B3~1206$+$469 \citep{Dabhade2020}. Due to the declination, the 25-arcsec images from RACS-mid and RACS-low are not available, and the angular resolution of the RACS-mid data approaches that of the NVSS.

\subsection{Source finding and measuring}\label{sec:sourcefinding}

Source-finding is performed on both the full-sensitivity  images as well as the original  images using \texttt{PyBDSF} \citep{pybdsf}. To help limit the number of artefacts recovered during source-finding, we modify the box and grid size for the position-dependent rms calculations to account for the varying PSF (as a function of the minor axis, $\theta_\text{minor}$). We find ($15\theta_\text{minor}, 3\theta_\text{minor}$) provides a good balance in removing artefacts around bright sources though removes a similar number of faint ($\sim 5\sigma_\text{rms}$) sources from the resulting source-lists. This box and grid size is functionally the same (150, 30) pixel box used by \citetalias{racs2} for their resolution and pixel size. The choice to scale with the minor axis of the PSF is to account for the smallest angular scale in the image. 

For each of the individual source-lists, we identify sources with \begin{equation}
\frac{ab}{\theta_\text{M}\theta_\text{m}} < \text{median}\left(\frac{ab}{\theta_\text{M}\theta_\text{m}}\right)
\end{equation}
for sources with major and minor axes $a$ and $b$ in images with PSF major and minor axes $\theta_\text{M}$ and $\theta_\text{m}$ that are comprised of more than one Gaussian component. Such sources are either point sources (or close to) and should not be decomposed into multiple components. Such decomposition is an artefact created during \texttt{PyBDSF} fitting where noise fluctuations lead to irregular boundaries of pixel islands. We find this to affect sources at all signal-to-noise ratios (SNRs). We group these components into a single source with the sum of integrated flux densities and a 2-D elliptical Gaussian to represent the size and shape.

For each source list the PSF of the image (major and minor axes FWHM and the position angle) is recorded for each source. For the individual image source lists these are single-valued over the individual source lists, but are retained when merging the individual source lists to ensure that each catalogued source has the correct PSF associated with it from the image within which it is found.

\subsection{Merging source-lists and duplicate removal}\label{sec:duplicates}

As there is significant overlap between the full-sensitivity images (by construction of the survey) these overlapping regions contain similar information. For the primary catalogue, the overlap regions may differ as the PSF for each full-sensitivity image may be different. This difference is small, but can result in two effects: (1) a difference to detected sources due to a change in SNR, and (2) a difference in components fit to detected sources. For (1), this results in overlap regions containing different sources at the low-SNR end of the source counts. For (2), this becomes particularly noticeable for, e.g., double and triple radio sources where the components in one image are connected and catalogued as a single source in one image but decompose into multiple sources in an adjacent image. Additionally, a combination of (1) and (2) can occur where components from multi-component sources get pushed below SNR limits.

We construct the catalogues by merging individual image source-lists one-by-one, and cross-match sources in the incoming source-list to the partially constructed catalogue. Sources in the partially constructed catalogue that match to a source in the incoming source lists with an angular separation, $s$, satisfying \begin{equation}
    s < \frac{1}{2} \left(\theta_\text{M,1} + \theta_\text{M,2}\right) \,
\end{equation}
are considered duplicates, where $\theta_\text{M,1}$ is the PSF major axis of the source in the partially constructed catalogue and $ \theta_\text{M,2}$ is the PSF major axis of the source in the incoming source list. When a duplicate is found, the source that has the smallest separation from an image centre is kept. While constructing the catalogue, we also avoid sources within 2~arcmin of an image boundary to avoid cross-matching of extended sources that might be cut off at the edge of an image. The Gaussian components associated with the final set of sources are then collected into a separate component list.

\section{The catalogues}\label{sec:catalogues}

\begin{figure}[t]
    \centering
    \begin{subfigure}[b]{1\linewidth}
    \includegraphics[width=1\linewidth]{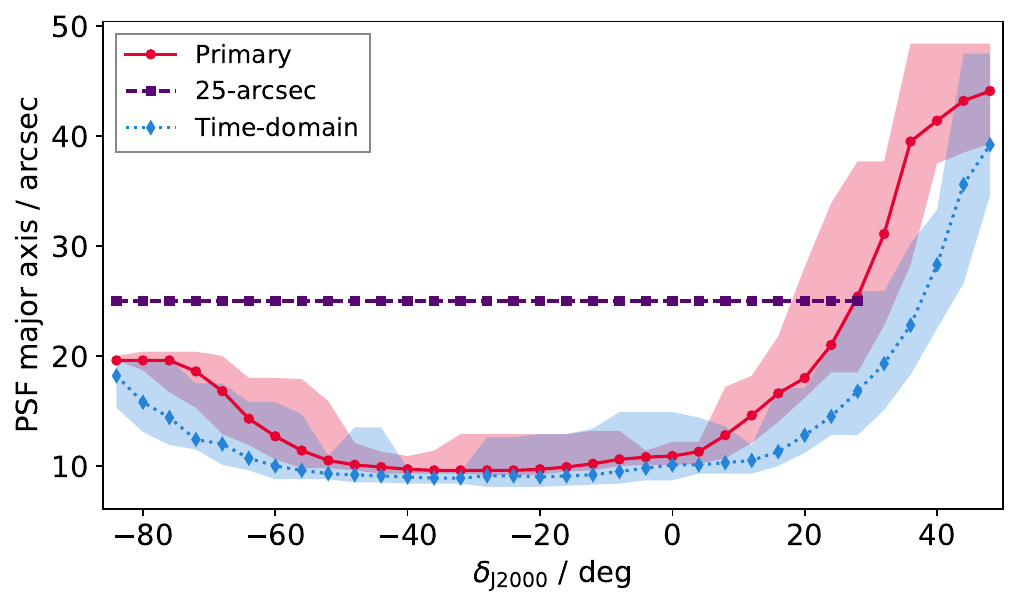}
    \caption{\label{fig:psf:major}}
    \end{subfigure}\\%
    \begin{subfigure}[b]{1\linewidth}
    \includegraphics[width=1\linewidth]{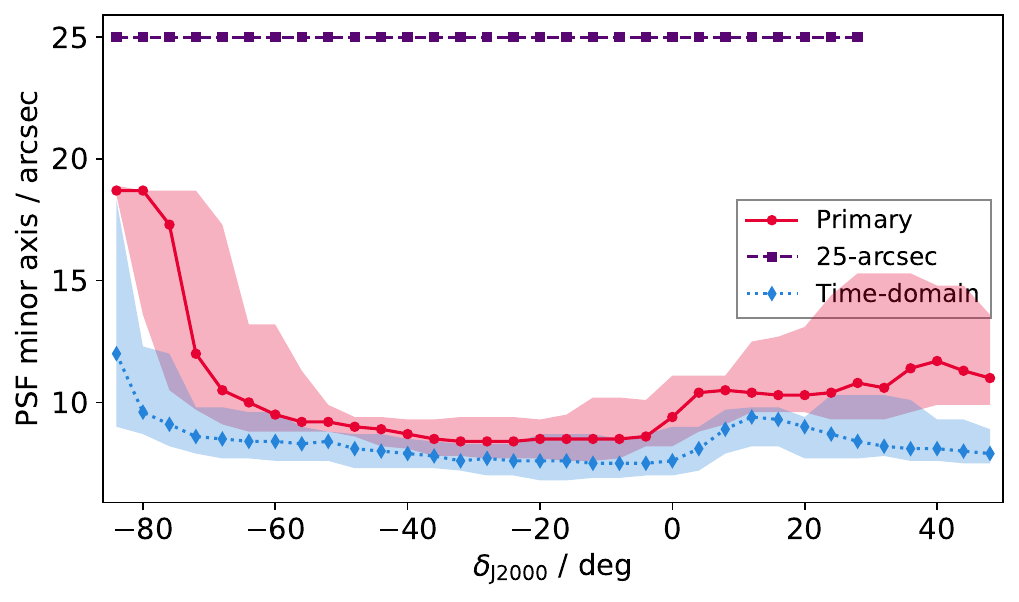}
    \caption{\label{fig:psf:minor}}
    \end{subfigure}\\%
    \caption{\label{fig:psf} Variation of the {major \subref{fig:psf:major} and minor \subref{fig:psf:minor} axes of the PSF over the RACS-mid primary catalogue (red), the 25-arcsec catalogue (purple), and the time-domain catalogue (blue). The filled markers are medians within 4-deg bins, and the shaded regions show the range of PSF axes within the given bins.}}
\end{figure}

\begin{table*}
    \centering
    \begin{threeparttable}
    \caption{\label{tab:surveyregions} RACS-mid all-sky catalogues and image properties.}
    \begin{tabular}{c c c c c c c c}\toprule
         Label & $\delta_\text{J2000}$ limit & Median resolution & Pixel size & Area \tnote{b} & Median $\sigma_\text{rms}$ \tnote{a} & $N_\text{sources}$ \tnote{b} & $N_\text{components}$ \tnote{b} \\
         & ($^\circ$) & ($^{\prime\prime} \times ^{\prime\prime}$) & ($^{\prime\prime} \times ^{\prime\prime}$)&  (deg$^2$) & (\textmu Jy\,PSF$^{-1}$) &  & \\\midrule
        Primary & $ \leq +49$ & $11.2 \times 9.3$ & $2 \times 2$ & 36\,200 (33\,242) & $182_{-23}^{+41}$  & 3\,105\,668 (2\,861\,923) & 4\,199\,578 (3\,869\,149) \\
        25 arcsec & $\leq +30$ & $25 \times 25$ & $4\times 4$& 30\,900 (28\,467)  & $278_{-47}^{+68}$ & 2\,154\,585 (1\,990\,598) & 2\,521\,038 (2\,324\,196) \\
        Time-domain & $\lesssim +49$ & $10.0 \times 8.1$ &$2 \times 2$ &$\sim$ 36\,200 (33\,242) & $203_{-33}^{+140}$ & 4\,087\,417 (3\,766\,945) & 5\,530\,478 (5\,094\,689) \\[0.5em]
        \bottomrule
    \end{tabular}
    \begin{tablenotes}[flushleft]
    {\footnotesize \item[a] Uncertainties are reported from the $16^\text{th}$ and $84^\text{th}$ percentiles. \item[b] In parenthesis excluding the Galactic Plane ($b\pm5^\circ$).
    }
    \end{tablenotes}
    \end{threeparttable}
\end{table*}

The catalogues produced as part of this data release are summarised in Table~\ref{tab:surveyregions}. {These catalogues include} (1) the primary catalogue comprising our best characterisation of each source at the observed and position-dependent resolution; (2) an auxiliary fixed 25-arcsec resolution catalogue, prepared similarly to the primary catalogue but with images  convolved to a resolution of 25-arcsec; (3) an auxiliary time-domain catalogue, which includes all sources including those detected multiple times in adjacent images or duplicate observations of a particular field. The catalogues have identical columns, which are described in \ref{app:columns}. Note that in \ref{app:columns} we also detail some changes to the columns and names used between the RACS-mid catalogues and the RACS-low catalogue. Tables~\ref{tab:app:exsource} and \ref{tab:app:excomp} in \ref{app:columns} also show example rows from the primary source and component catalogues, respectively. The catalogues are detailed further in the following sections.

\subsection{The primary catalogue}

The first catalogue is considered the primary catalogue and will be suitable for most general users. This catalogue covers the full region observed as part of RACS-mid and remains close to the angular resolution of the original RACS-mid images described in \citetalias{racs-mid}. Construction of this catalogue begins with identifying adjacent fields, convolving to the lowest common resolution of that subset of fields, then forming full-sensitivity images as described in Section~\ref{sec:images}. Examples of the image quality is shown in Figures~\ref{fig:images} and \ref{fig:highdec} for a range of declinations, and Figure~\ref{fig:psf} shows the variation of the PSF major {[\ref{fig:psf:major}] and minor [\ref{fig:psf:minor}] axes as a function of declination for sources in the catalogue. The PSF for the primary catalogue is elliptical, and the major axis reaches 48.4\,arcsec (at high declination) while the minor axis has a maximum of $ 18.9$\,arcsec (at low declination).}

\subsection{Auxiliary fixed-resolution, 25-arcsec catalogue}

The second catalogue is an auxiliary `fixed-resolution' catalogue, with a position-independent resolution of $25 \times 25$~arcsec$^2$. Figure~\ref{fig:psf} shows the 25-arcsec PSF compared to other catalogues' PSF variation for reference. This catalogue is designed to match the RACS-low catalogue described by \citetalias{racs2}, including the same resolution and sky coverage up to $\delta_\text{J2000} \leq +30^\circ$ \footnote{Note that the RACS-low catalogue described in \citetalias{racs2} features a hole in its coverage below $\delta_\text{J2000} < -80\degree$ as some images did not meet the 25-arcsec resolution requirement. For RACS-mid we provide full coverage below $\delta_\text{J2000} \leq +30\degree$ for the 25-arcsec catalogue as all images below this declination could be convolved to $25^{\prime\prime}\times 25^{\prime\prime}$.}.

\subsection{Auxiliary time-domain catalogue}

As the previously mentioned catalogues are created after linearly mosaicking neighbouring observations, variable/transient radio sources may have their emission averaged between what may be two disjoint epochs. For many time-domain applications, it may be preferable to retain source detections and measurements at specific epochs. Searches for and characterisation of variable and transient sources have already been performed using individual images and sources lists from RACS-low \citep[e.g.][]{Leung2021} and RACS-mid \citep[e.g.][]{Driessen2023,Gulati2023}. To help enable further time-domain science, we provide another auxiliary catalogue for Stokes I that is simply the concatenation of source-lists from the individual RACS-mid images. The individual images are not convolved prior to source-finding so they retain the original angular resolution. Figure~\ref{fig:psf} shows the variation of the PSF at the locations of sources in the catalogue compared to the primary and 25-arcsec catalogues. {As with the primary catalogue the PSF is elliptical and has a similar range of values.} Sources that are detected in multiple images in overlap regions (i.e. duplicates) are not removed.

\section{Analysis of the catalogues and images}\label{sec:analysis}

\begin{figure}[tp!]
    \centering
    \begin{subfigure}[b]{1\linewidth}
    \includegraphics[width=1\linewidth]{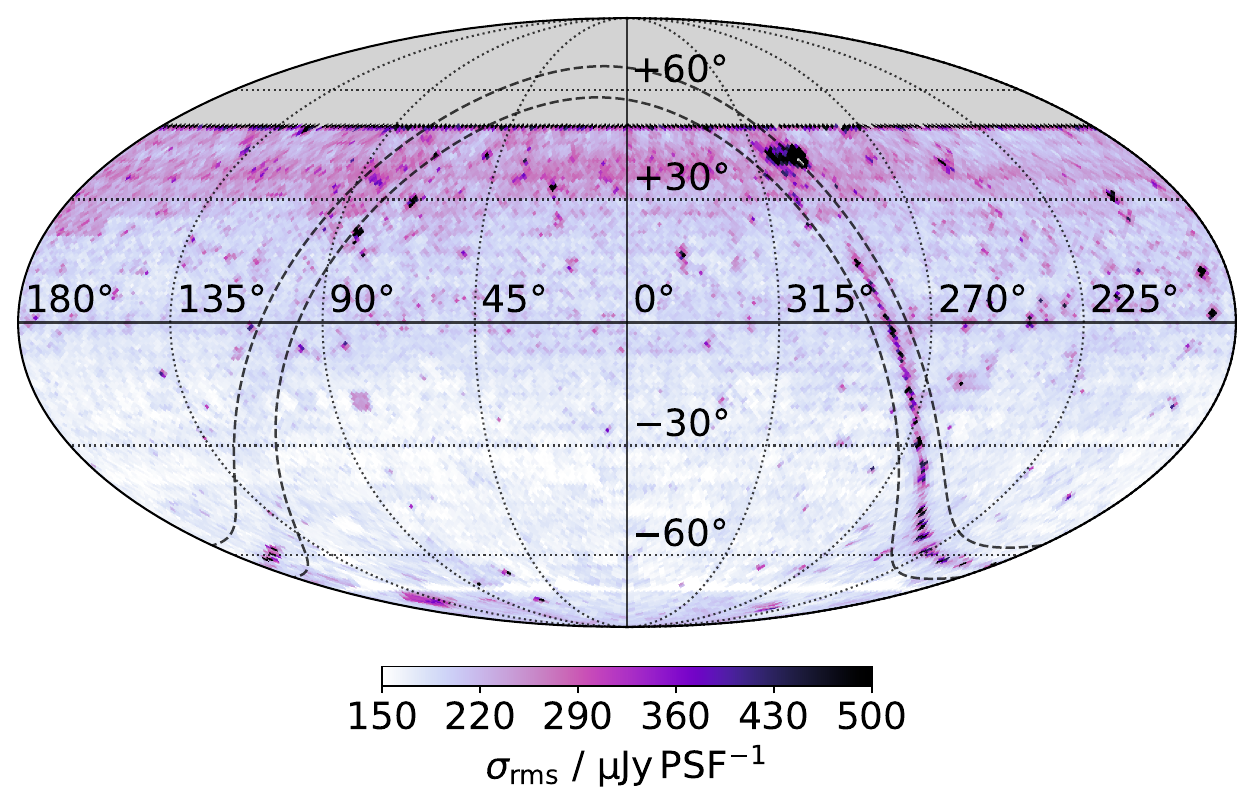}
    \caption{\label{fig:noise:fr} Primary catalogue.}
    \end{subfigure}\\[2em]%
    \begin{subfigure}[b]{1\linewidth}
    \includegraphics[width=1\linewidth]{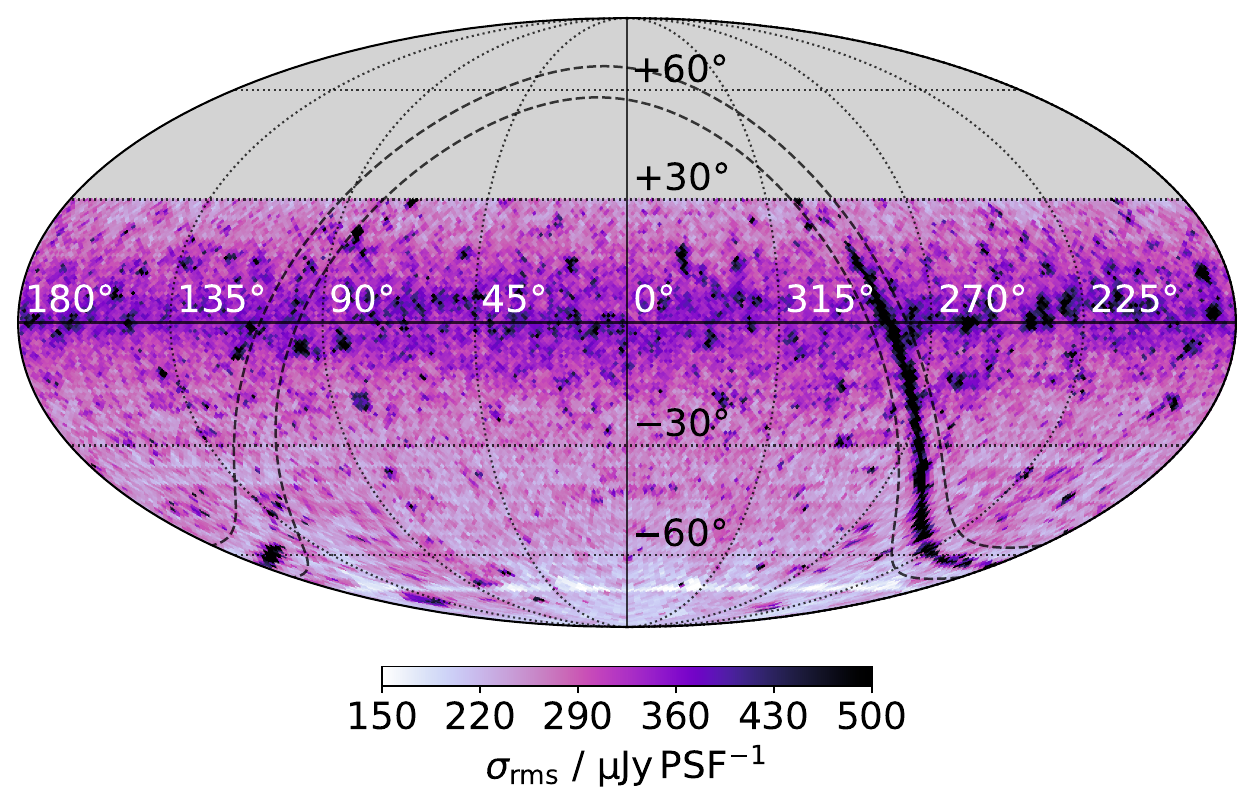}
    \caption{\label{fig:noise:25} 25-arcsec catalogue.}
    \end{subfigure}\\[2em]%
    \begin{subfigure}[b]{1\linewidth}
    \includegraphics[width=1\linewidth]{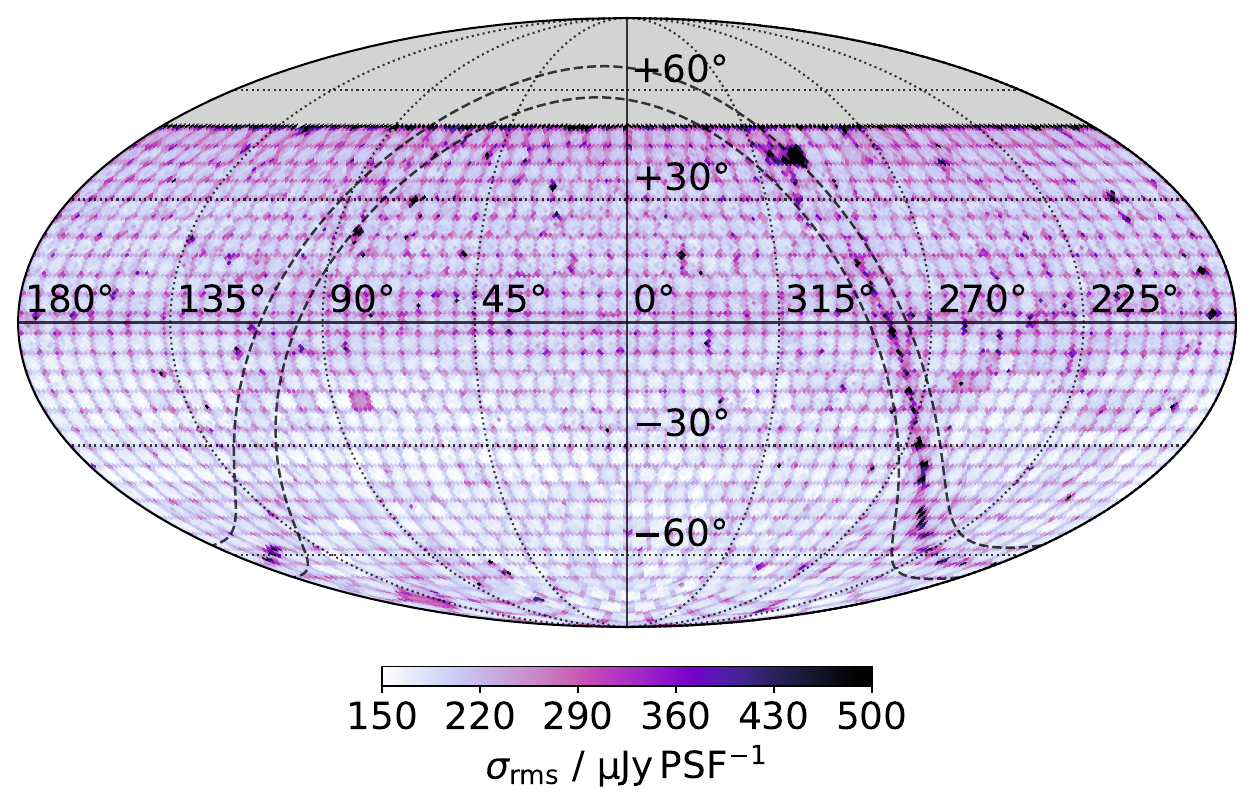}
    \caption{\label{fig:noise:td} Time-domain catalogue.}
    \end{subfigure}\\[2em]%
    \caption{\label{fig:noise} Median-binned HEALPix representation of the root-mean-square noise distributions of the primary catalogue \subref{fig:noise:fr}, the 25-arcsec catalogue \subref{fig:noise:25}, and the time-domain catalogue \subref{fig:noise:td}. The black, dashed lines are drawn at Galactic latitudes $b \pm 5^\circ$.}
\end{figure}

\begin{figure}
    \centering
    \includegraphics[width=1\linewidth]{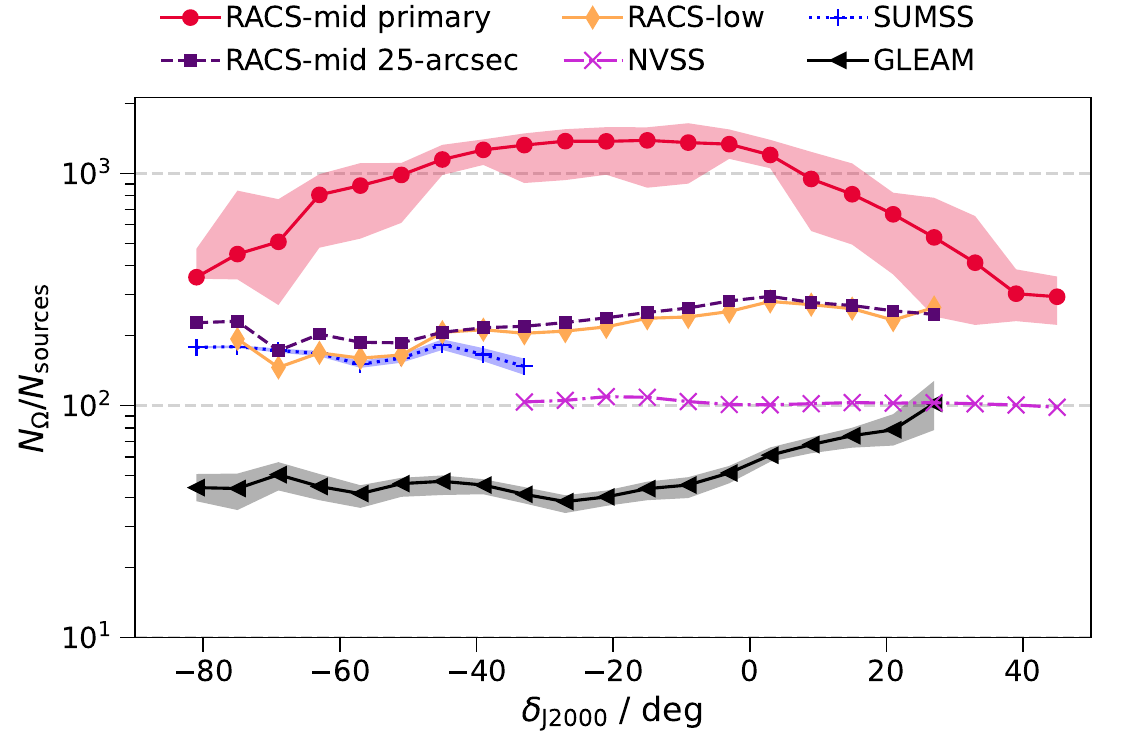}
    \caption{\label{fig:beta} {The number of PSF solid angles per source above $5\sigma_\text{rms}$, binned as a function of declination for the RACS-mid primary and 25-arcsec catalogues, alongside equivalent products from RACS-low, the NVSS, SUMSS, and GLEAM. The shaded regions indicate maximum and minimum values corresponding to a varying PSF solid angle within declination bins.}}
\end{figure}

\begin{figure*}[p]
\centering
\begin{subfigure}[b]{0.49\linewidth}
\includegraphics[width=1\linewidth]{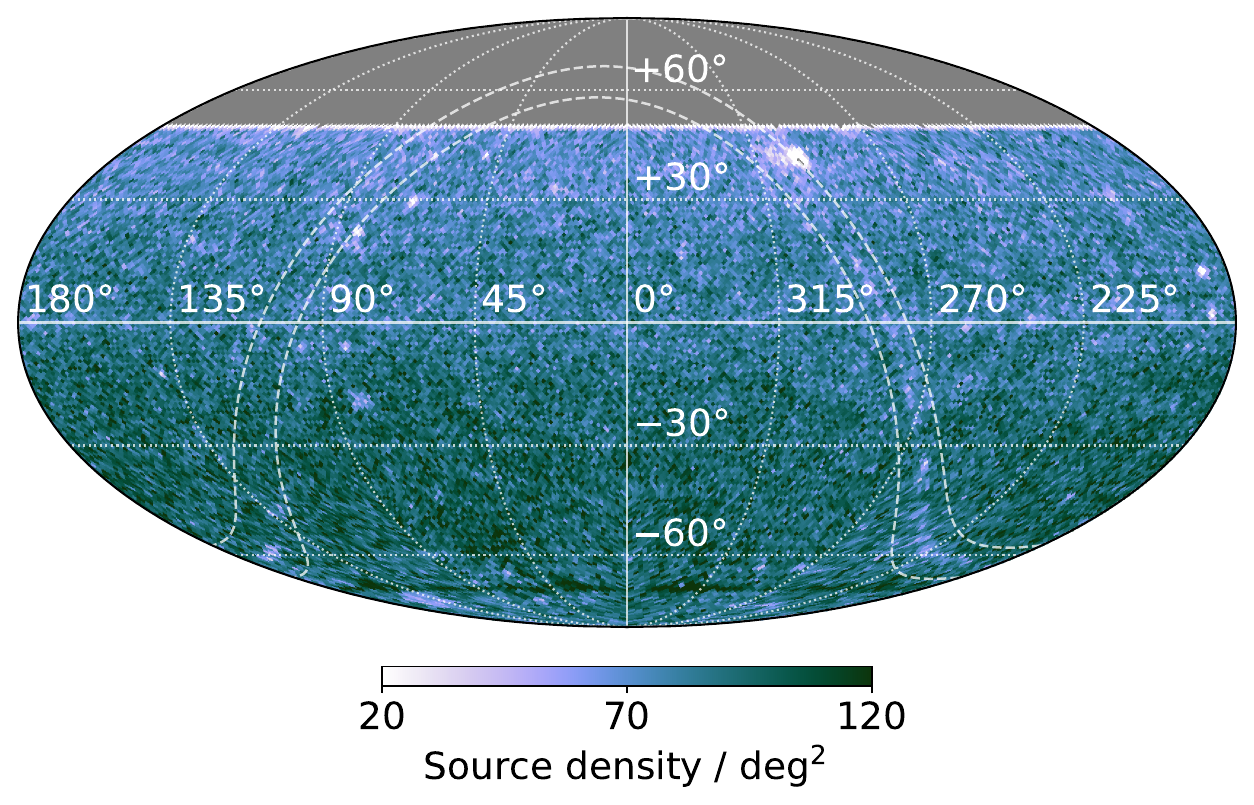}%
\caption{\label{fig:sourcedensity:1} Primary catalogue.}
\end{subfigure}\hfill%
\begin{subfigure}[b]{0.49\linewidth}
\includegraphics[width=1\linewidth]{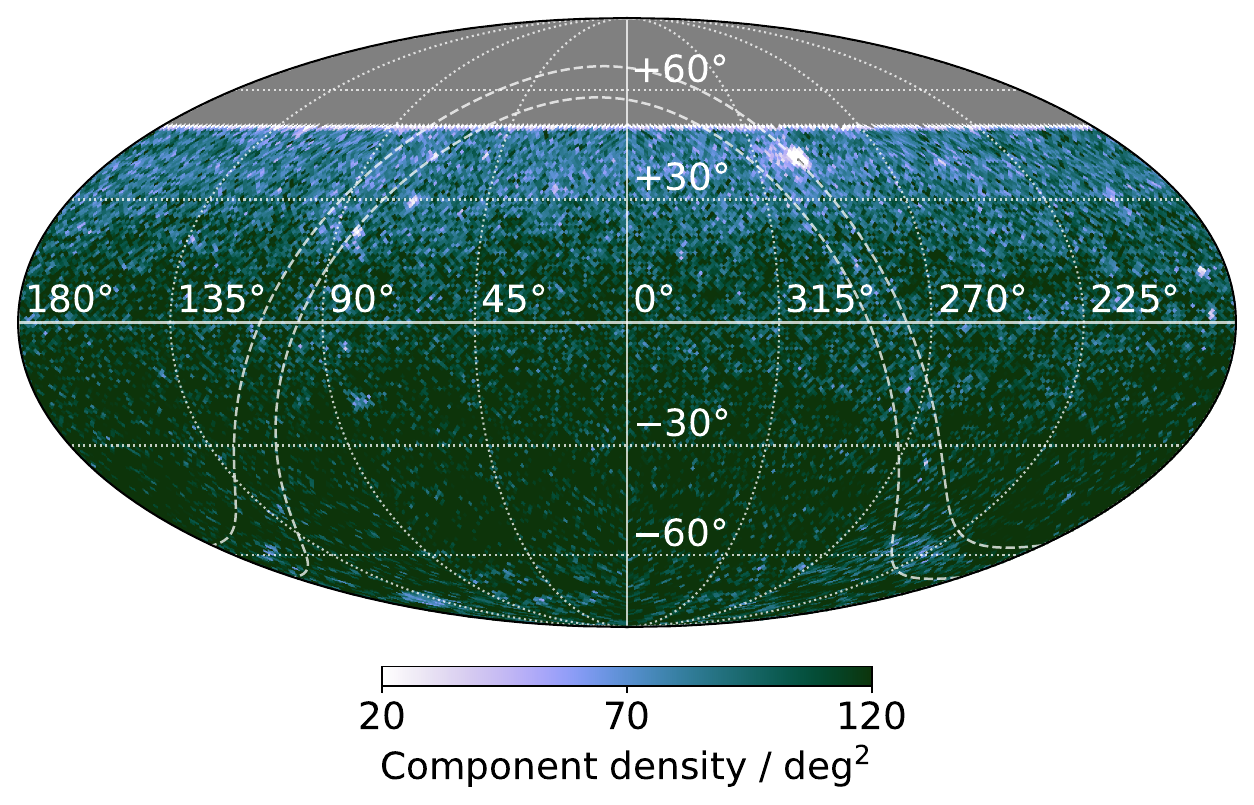}
\caption{\label{fig:sourcedensity:2} Primary catalogue.}
\end{subfigure}\\[1em]%
\begin{subfigure}[b]{0.49\linewidth}
\includegraphics[width=1\linewidth]{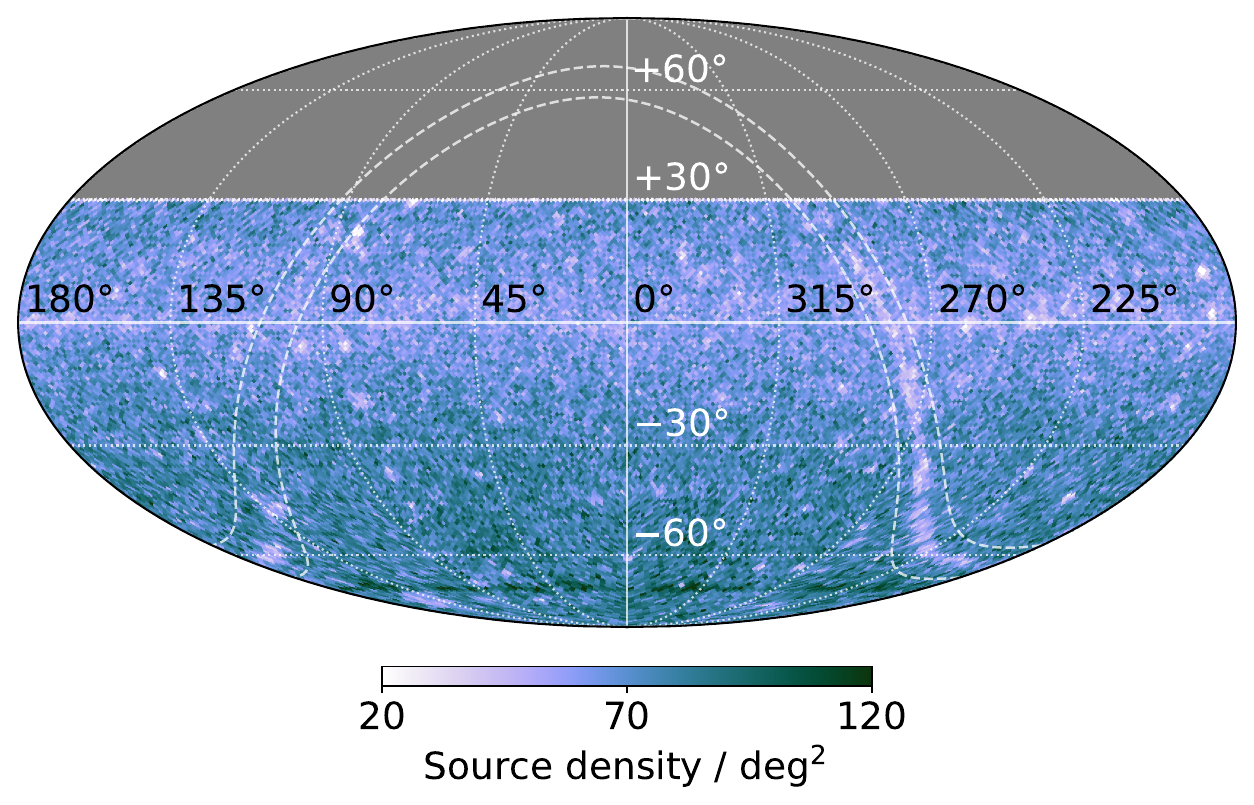}%
\caption{\label{fig:sourcedensity:3}25~arcsec catalogue.}
\end{subfigure}\hfill%
\begin{subfigure}[b]{0.49\linewidth}
\includegraphics[width=1\linewidth]{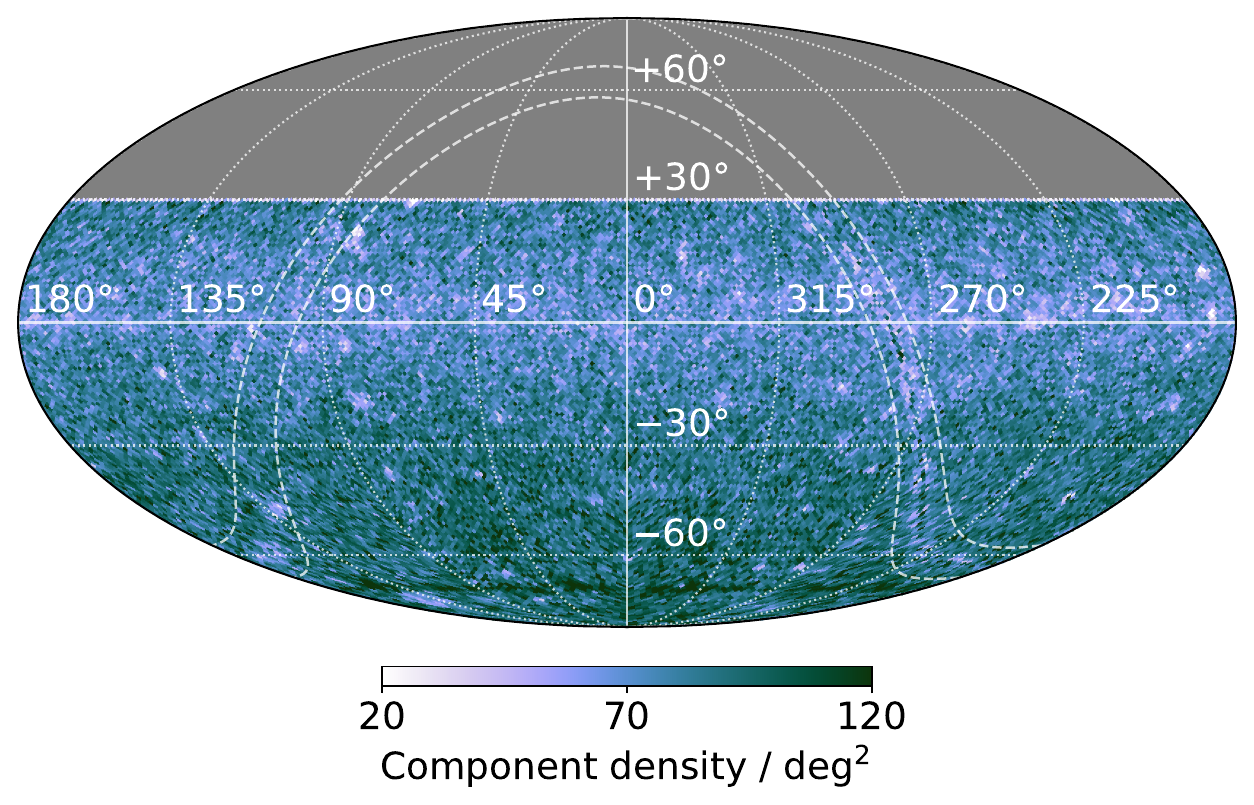}
\caption{\label{fig:sourcedensity:4}25~arcsec catalogue}
\end{subfigure}\\[1em]%
\begin{subfigure}[b]{0.49\linewidth}
\includegraphics[width=1\linewidth]{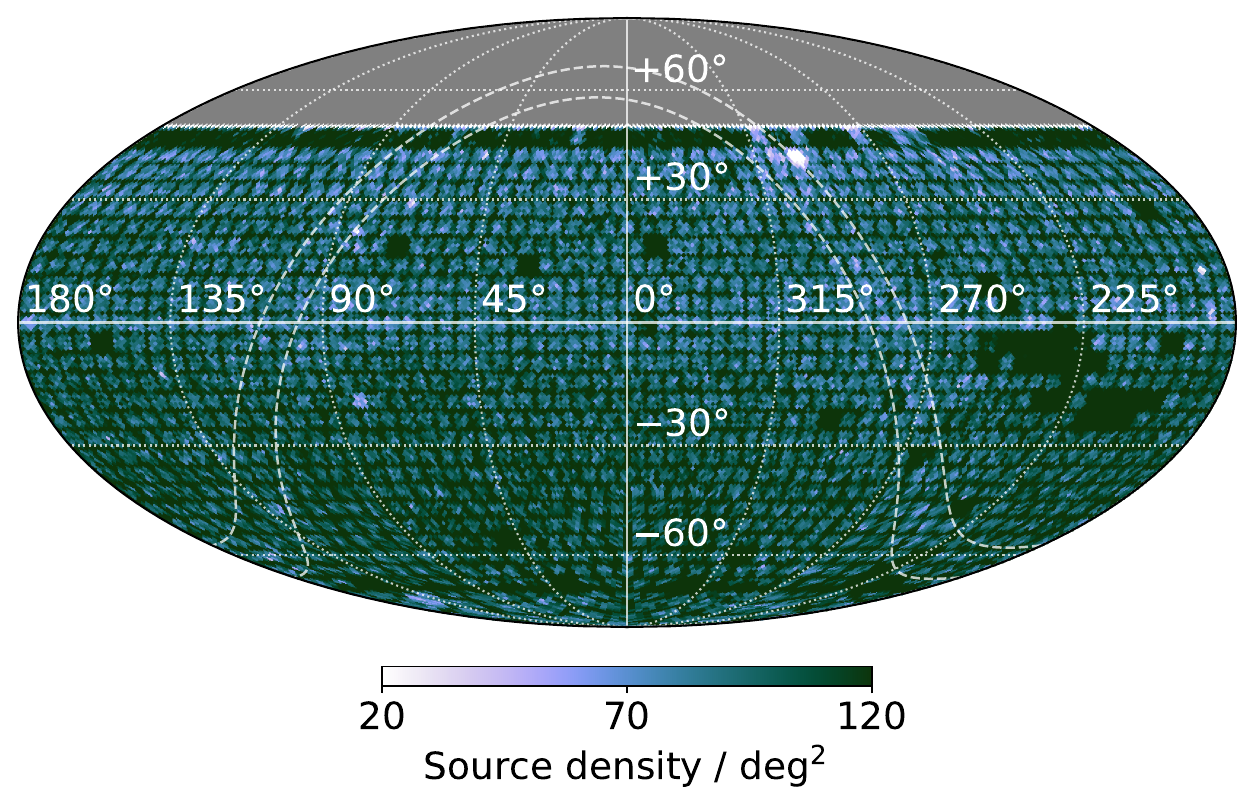}%
\caption{\label{fig:sourcedensity:5}Time-domain catalogue.}
\end{subfigure}\hfill%
\begin{subfigure}[b]{0.49\linewidth}
\includegraphics[width=1\linewidth]{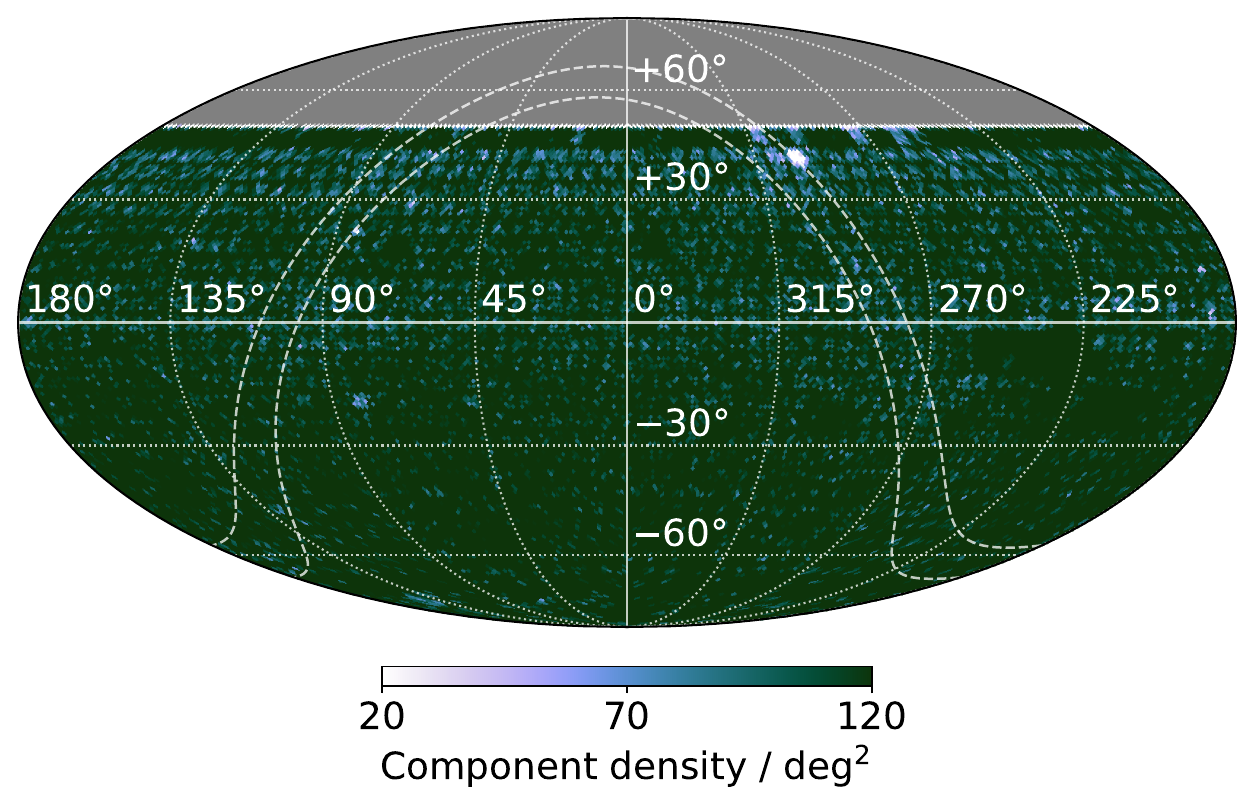}
\caption{\label{fig:sourcedensity:6}Time-domain catalogue.}
\end{subfigure}\\[1em]%
\caption{\label{fig:sourcedensity} HEALPix representation of the source and component density of the primary catalogue [\subref{fig:sourcedensity:1} and \subref{fig:sourcedensity:2}], the 25-arcsec catalogue [\subref{fig:sourcedensity:3} and \subref{fig:sourcedensity:4}], and the time-domain catalogue [\subref{fig:sourcedensity:5} and \subref{fig:sourcedensity:6}]. The black, dashed lines are drawn at Galactic latitudes $b \pm 5^\circ$. Regions with no sources are coloured grey. The colourscales are kept the same to highlight the differences in source/component densities.}
\end{figure*}

\begin{figure*}[!t]
    \centering
    \begin{subfigure}[b]{0.32\linewidth}
    \includegraphics[width=1\linewidth]{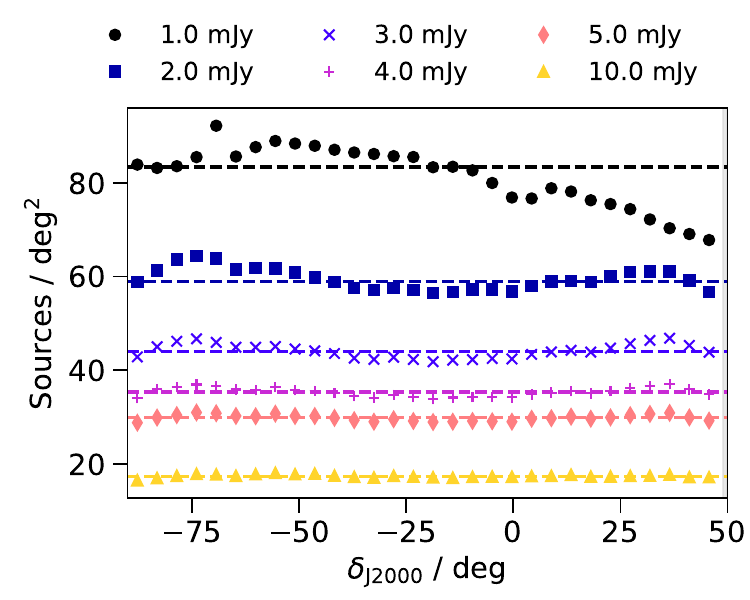}
    \caption{\label{fig:dec_density:full} Primary catalogue.}
    \end{subfigure}%
    \begin{subfigure}[b]{0.32\linewidth}
    \includegraphics[width=1\linewidth]{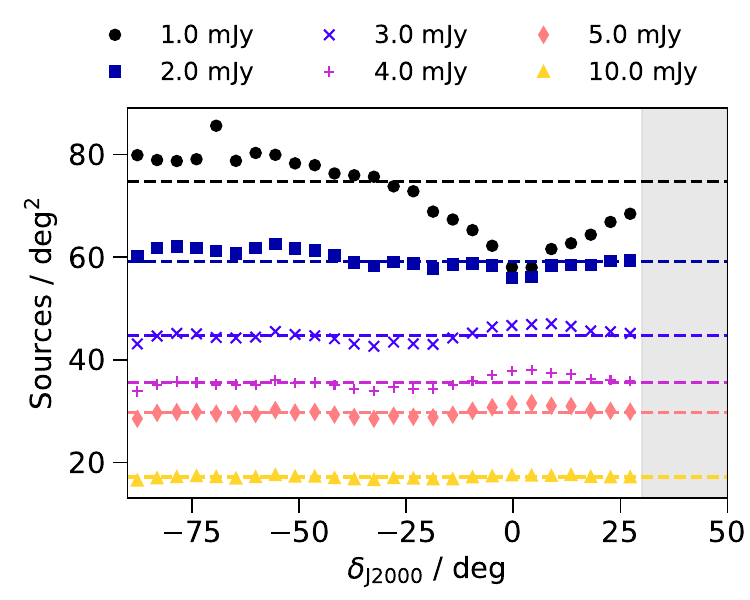}
    \caption{\label{fig:dec_density:25} 25-arcsec catalogue.}
    \end{subfigure}%
    \begin{subfigure}[b]{0.32\linewidth}
    \includegraphics[width=1\linewidth]{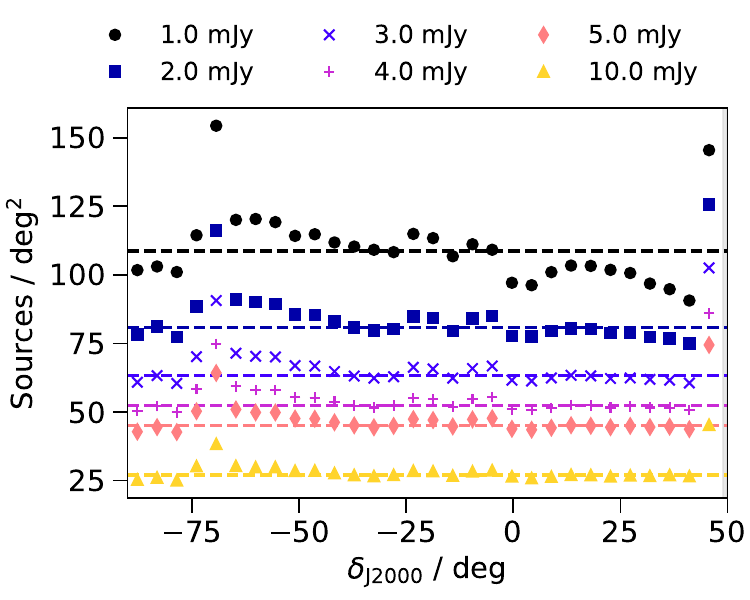}
    \caption{\label{fig:dec_density:td} Time-domain catalogue.}
    \end{subfigure}%
    \caption{\label{fig:dec_density} Source density as a function of declination for {six} flux density limits ($[1, 2, 3, 4, 5, 10]$\,mJy) for the primary catalogue \subref{fig:dec_density:full}, the 25-arcsec catalogue \subref{fig:dec_density:25}, and the time-domain catalogue \subref{fig:dec_density:td}. Dashed lines correspond to median source densities for the associated flux density limit. The grey, shaded region in \subref{fig:dec_density:25} is not covered in the 25-arcsec catalogue.}
\end{figure*}

\begin{figure}[tp!]
\centering
\begin{subfigure}[b]{1\linewidth}
\includegraphics[width=1\linewidth]{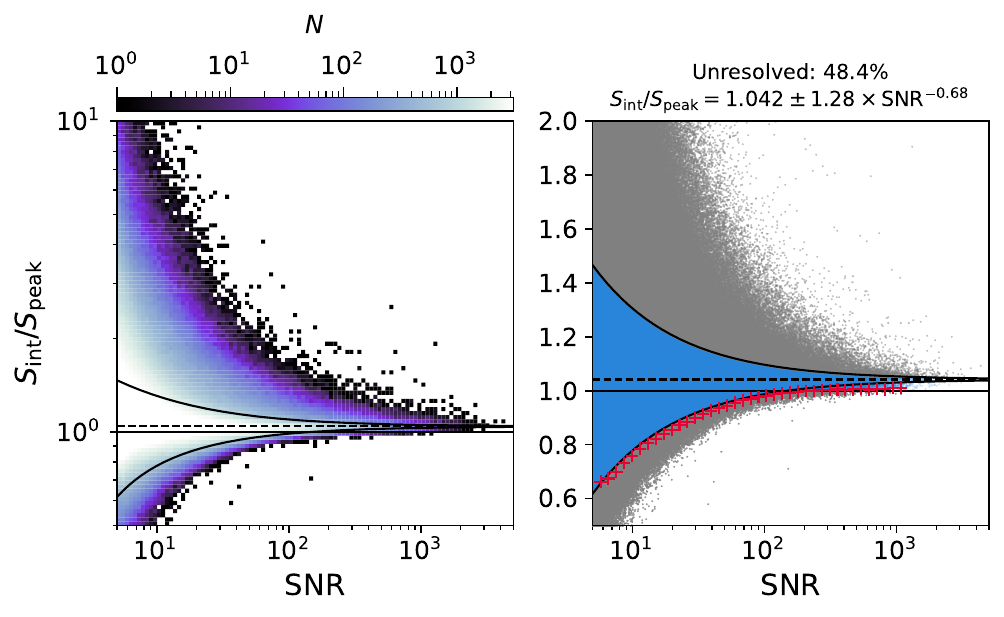}
\caption{\label{fig:resolved:fr} Primary catalogue.}
\end{subfigure}\\%
\begin{subfigure}[b]{1\linewidth}
\includegraphics[width=1\linewidth]{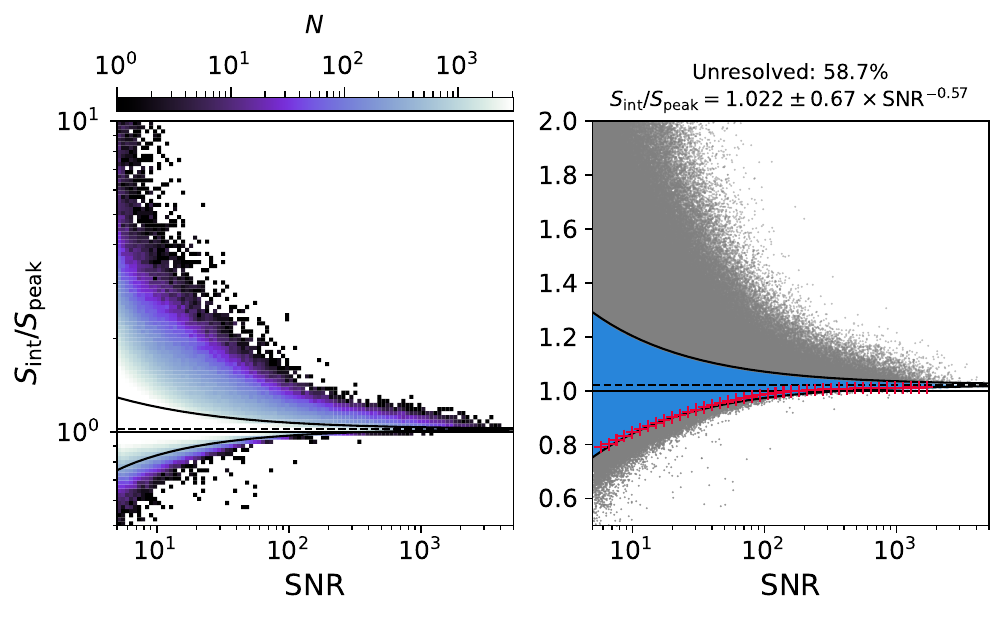}
\caption{\label{fig:resolved:25arcsec}25-arcsec catalogue.}
\end{subfigure}\\%
\begin{subfigure}[b]{1\linewidth}
\includegraphics[width=1\linewidth]{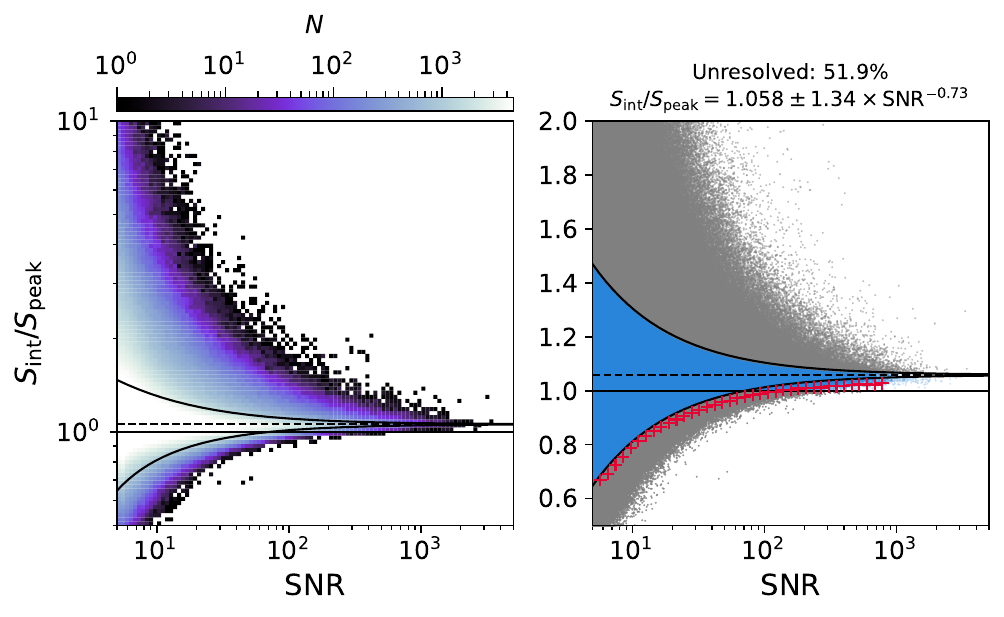}
\caption{\label{fig:resolved:td}Time-domain catalogue.}
\end{subfigure}\\%
\caption{\label{fig:resolved} The ratio of integrated flux density ($S_\text{int}$) to peak flux density ($S_\text{peak}$) as a function of signal-to-noise ratio ($\text{SNR} = S_\text{peak} / \sigma_\text{rms}$). The left panels show binned 2-D histograms of the sources while the right panel shows a zoomed-in region with unresolved sources coloured blue. Red crosses in the right panel indicate the binned values used to fit the lower envelope. The lower and upper envelope are drawn as solid black lines, and the median $S_\text{int}/S_\text{peak}$ value is shown as a black, dashed line. The single solid, black horizontal line indicates $S_\text{int}/S_\text{peak} = 1$.}
\end{figure}

\begin{figure*}[t]
    \begin{subfigure}[b]{0.33\linewidth}
        \includegraphics[width=1\linewidth]{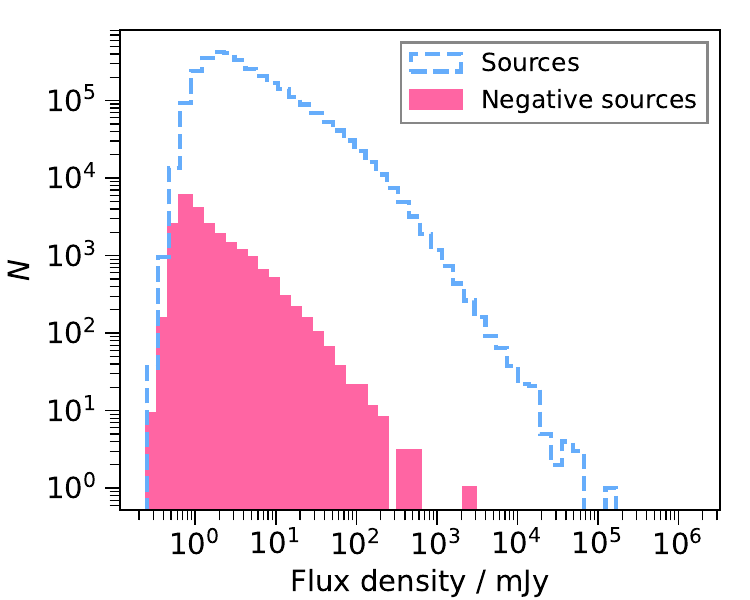}
        \caption{Primary catalogue.\label{fig:negative_hist:fr}}
    \end{subfigure}
    \begin{subfigure}[b]{0.33\linewidth}
        \includegraphics[width=1\linewidth]{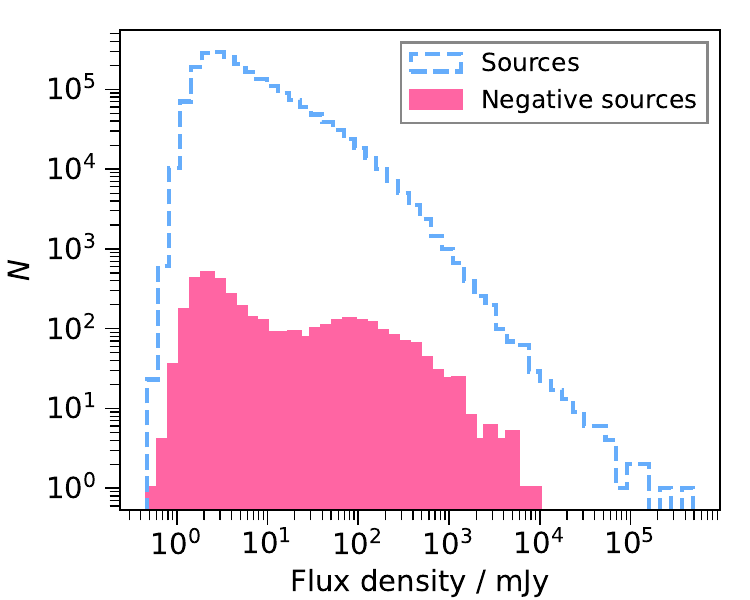}
        \caption{25-arcsec catalogue.\label{fig:negative_hist:25arcsec}}
    \end{subfigure}
    \begin{subfigure}[b]{0.33\linewidth}
        \includegraphics[width=1\linewidth]{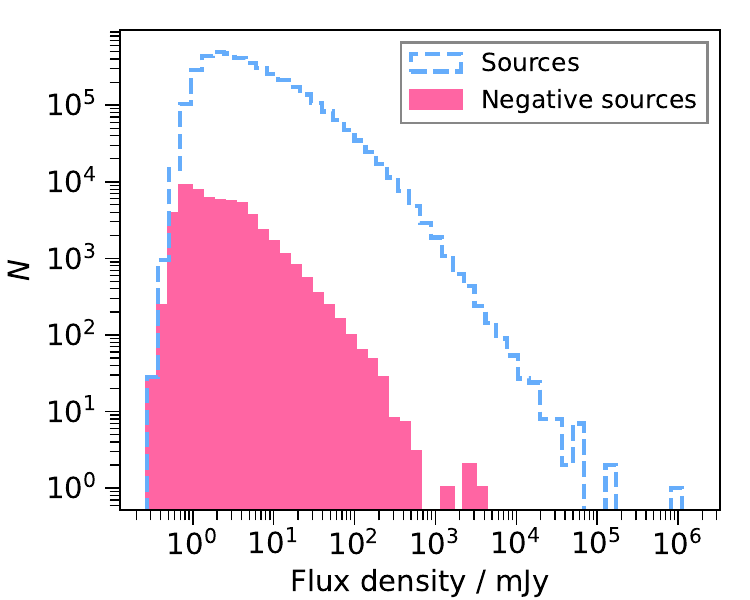}
        \caption{Time-domain catalogue.\label{fig:negative_hist:td}}
    \end{subfigure}
    \caption{\label{fig:negative_hist} {Histograms of the negative (pink) and positive (blue) components found over the survey images used in the primary \subref{fig:negative_hist:fr}, 25-arcsec \subref{fig:negative_hist:25arcsec}, and time-domain \subref{fig:negative_hist:td} catalogues.}}
\end{figure*}

\begin{figure*}[t]
    \centering
    \begin{subfigure}[b]{0.33\linewidth}
    \includegraphics[width=1\linewidth]{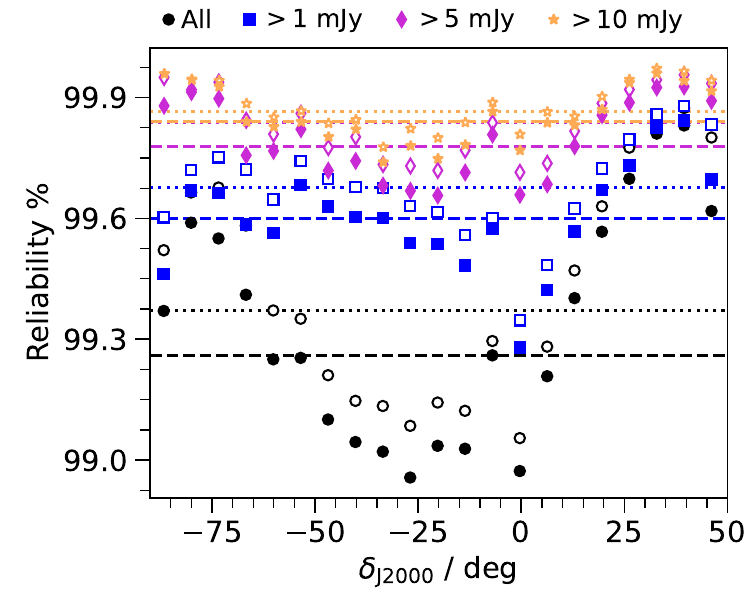}
    \caption{Primary catalogue, $|b| >5^\circ$.\label{fig:negative:fr}}
    \end{subfigure}%
    \begin{subfigure}[b]{0.33\linewidth}
    \includegraphics[width=1\linewidth]{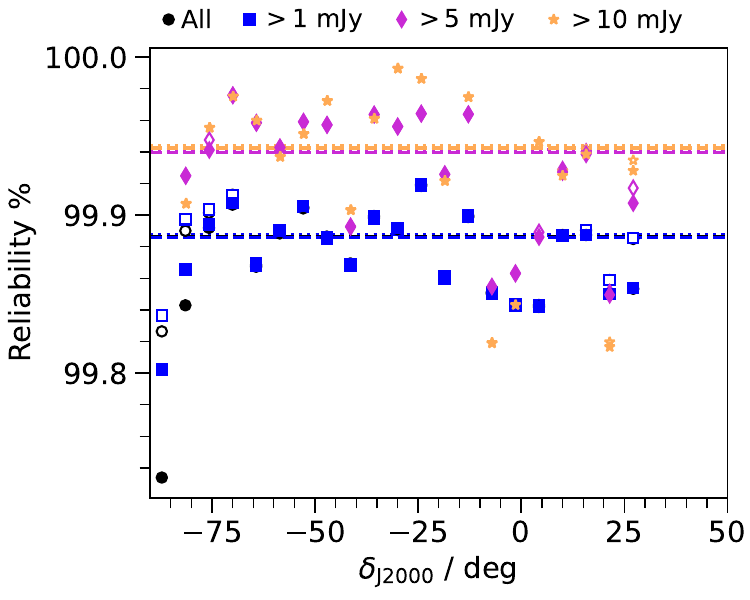}
    \caption{25-arcsec catalogue, $|b| >5^\circ$.\label{fig:negative:25}}
    \end{subfigure}%
    \begin{subfigure}[b]{0.33\linewidth}
    \includegraphics[width=1\linewidth]{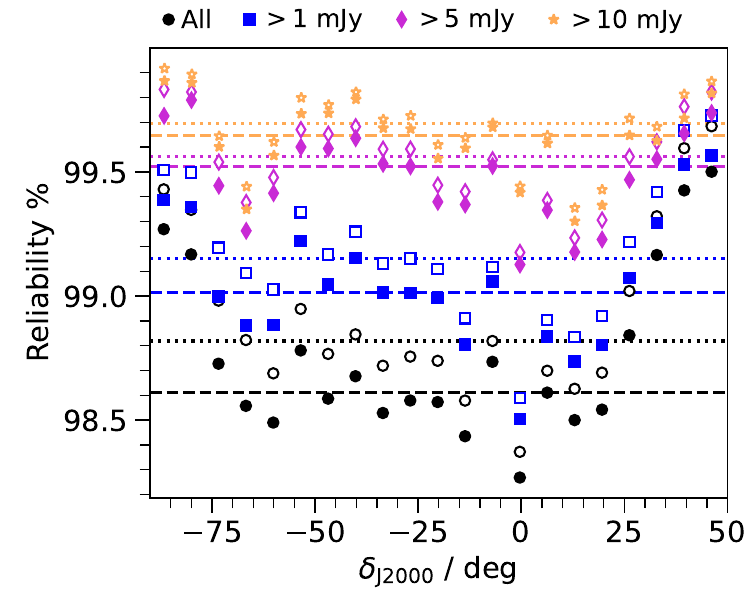}
    \caption{Time-domain catalogue, $|b| >5^\circ$.\label{fig:negative:td}}
    \end{subfigure}\\[1em]%
    \begin{subfigure}[b]{0.33\linewidth}
    \includegraphics[width=1\linewidth]{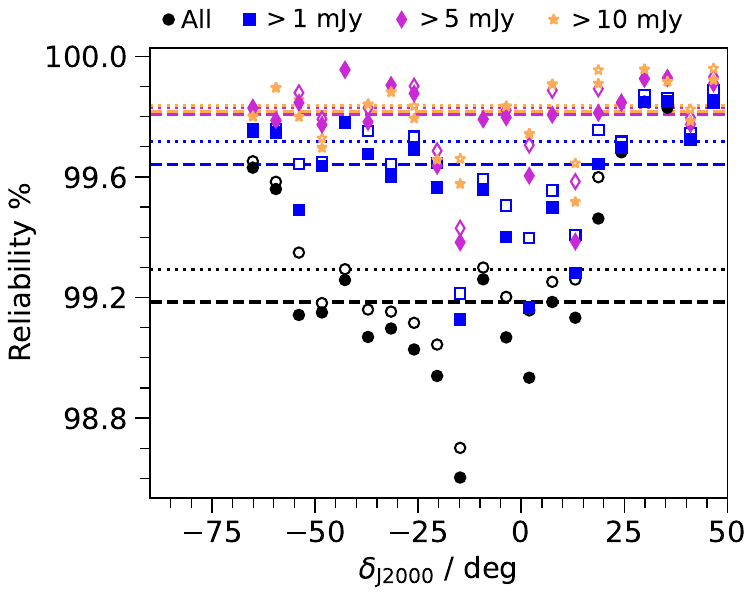}
    \caption{Primary catalogue, $|b| <5^\circ$ .\label{fig:negative:gp:fr}}
    \end{subfigure}%
    \begin{subfigure}[b]{0.33\linewidth}
    \includegraphics[width=1\linewidth]{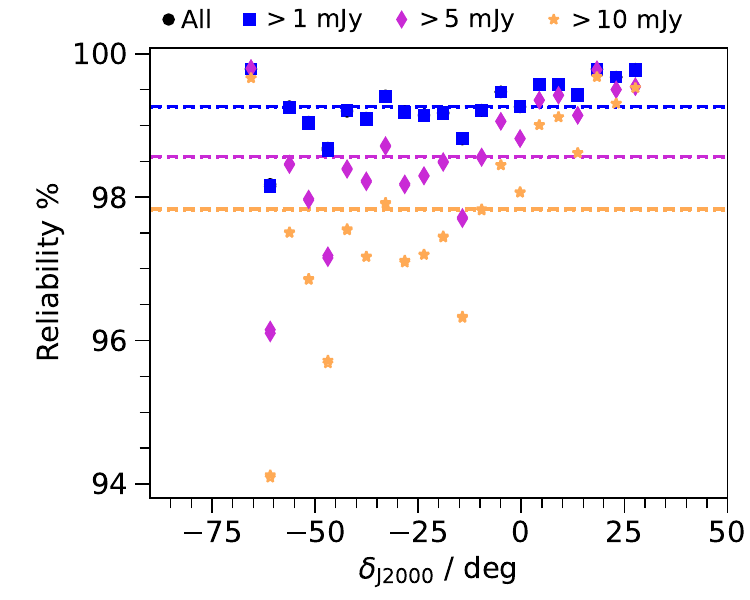}
    \caption{25-arcsec catalogue, $|b| <5^\circ$.\label{fig:negative:gp:25}}
    \end{subfigure}%
    \begin{subfigure}[b]{0.33\linewidth}
    \includegraphics[width=1\linewidth]{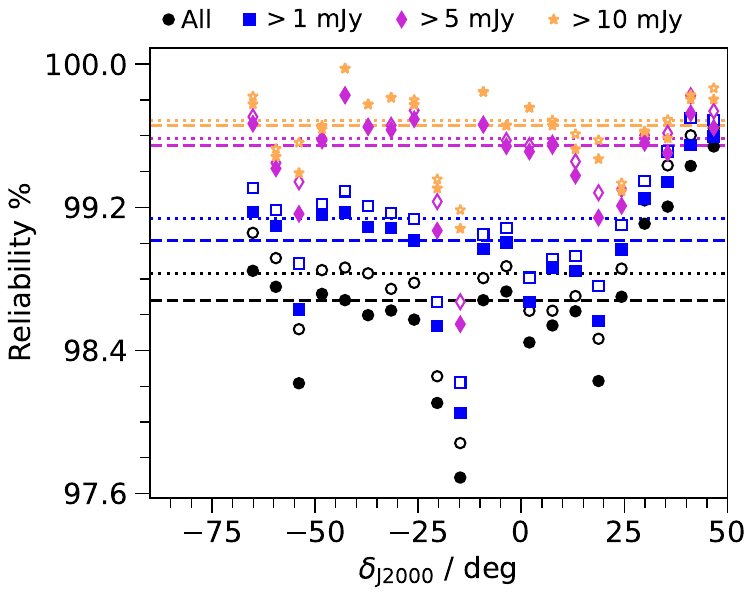}
    \caption{Time-domain catalogue, $|b| <5^\circ$.\label{fig:negative:gp:td}}
    \end{subfigure}
    \caption{\label{fig:negative} Reliability for the three catalogues as a function of declination for different flux density cuts. Empty markers correspond to the subsample that excludes the \texttt{Flag = 2} `spurious' sources. Horizontal lines correspond to medians for each flux density limit (dashed lines for the full samples, and dotted lines for the subsample). \emph{Top row.} [\subref{fig:negative:fr}--\subref{fig:negative:td}] Sources outside of the Galactic Plane ($|b|>5^\circ$). \emph{Bottom row.} [\subref{fig:negative:gp:fr}--\subref{fig:negative:gp:td}] Sources within the Galactic Plane ($|b|<5^\circ$). Note all plots show a different y-axis scale to highlight different declination-dependent features.}
\end{figure*}

In the following sections, we provide an analysis of the overall quality of the available catalogues and full-sensitivity image mosaics made. As the catalogue are created from the same original images, there is some redundancy to the following analysis and validation work. For consistency we show results for all catalogues  where appropriate. 

\subsection{{Overall noise properties}}\label{sec:noise}

The position-dependent rms noise properties of the images that are used for the three catalogues varies due to the difference in angular resolution and their overall construction. A local rms noise estimate is included for each source in the catalogue. Figure~\ref{fig:noise} shows the rms noise median-binned using Hierarchical Equal Area isoLatitude Pixelation \citep[HEALPix;][]{Gorski2005} \footnote{\url{https://healpix.sourceforge.io/}.} with $N_\text{side}=64$ corresponding to $\sim 55 \times 55$~arcmin$^{2}$ bins. The rms values are reported at the location of the source, so will on average be marginally elevated compared to off-source regions.

The rms noise distribution of the primary catalogue [Figure~\ref{fig:noise:fr}] follows closely the median Stokes I noise per tile shown in figure~24 of \citetalias{racs-mid}, which was constructed by mosaicking position-dependent rms maps. The primary catalogue has a median rms noise of $\sigma_{\text{rms}} = 182_{-23}^{+41}$\,\textmu Jy\,PSF$^{-1}$ (with uncertainties derived from the $16^\text{th}$ and $84^\text{th}$ percentiles of the rms noise distribution). This median rms noise is a decrease from the median rms noise reported over the original survey images reported in \citetalias{racs-mid} due to the removal of the primary beam roll-off at the edge of convolved images. The larger spread in the distribution is due to the bias in local rms estimates taken from source locations, including the Galactic Plane and other areas of bright, extended sources. {While not part of this data release, we note that the Stokes V images mosaicked and convolved in the same way achieve a median rms noise of $\sigma_{\text{rms},V} = 144_{-14}^{+20}$\,\textmu Jy\,PSF$^{-1}$ at the catalogued source positions.}

The 25-arcsec catalogue [Figure~\ref{fig:noise:25}] has a median rms noise $\sigma_{\text{rms}} = 278_{-47}^{+68}$\,\textmu Jy\,PSF$^{-1}$, generally showing an increase in rms noise over most of the sky due to the decrease in angular resolution. Of particular note is an additional increase in noise near the celestial equator, corresponding to residuals from the poorer equatorial PSF becoming significant after convolution to the lower 25-arcsec resolution. {Convolving the images to the lower $25\times25$\,arcsec$^{2}$ PSF also results in a general loss of information in the images, translating to an overall increase in the image rms noise. The difference in the median Stokes I and Stokes V noise ($\sigma_{\text{rms},V} = 167_{-17}^{+23}$\,\textmu Jy\,PSF$^{-1}$) is comparatively larger than for the primary catalogue, suggesting another contribution to the increase in noise is the increase in classical source confusion at this lower resolution \citep[][]{Condon1974} though the 25-arcsec catalogue is not confusion-limited (see Section~\ref{sec:confusion}).}

Finally, the time-domain catalogue [Figure~\ref{fig:noise:td}] features increased noise at the boundaries between fields due to overlapping uncorrected primary beam roll-off. The remainder of the survey area has similar position-dependent noise characteristics to the primary catalogue due to the similarities towards the image centres. The median rms noise, $\sigma_{\text{rms}} = 203_{-33}^{+140}$\,\textmu Jy\,PSF$^{-1}$ is closer to the tile median reported in \citetalias{racs-mid} with an increase in spread due to bias in the sampled noise locations (i.e. at the locations of sources).

\subsection{{Classical source confusion}}\label{sec:confusion}
{While we do not expect even the deeper 10-hr images from ASKAP to become limited by classical source confusion noise at 1.4\,GHz \citep[assuming a $10\times10$\,arcsec$^{2}$ PSF;][]{Condon2012}, as this noise is a function of PSF solid angle [$\Omega_\text{PSF} = \pi\theta_\text{M}\theta_\text{m}/\left(4\ln{2}\right)$] it is worth considering a possible limit given the variable PSFs of the RACS-mid catalogues. The number of PSF solid angles per source above some flux density threshold is often used as an estimate of the severity of confusion \citep[e.g.][]{Cohen2004,Condon2012,Heywood2016}, with values of order $\lesssim 10$ indicating that the images are limited in sensitivity by confusion.}

{Figure~\ref{fig:beta} shows the number of PSF solid angles per source above $5\sigma_\text{rms}$ ($N_\Omega / N_\text{sources}$) binned as a function of declination, excluding the area covering Galactic latitudes of $|b|<10^\circ$. We also show the equivalent product with the same threshold for some other surveys with largely contiguous sky coverage: the NVSS \citep{ccg+98} with a $45\times45$\,arcsec$^2$ PSF, the Sydney University Molonglo Sky Survey \citep[SUMSS;][]{bls99,mmb+03} with a position-dependent $45\times45/\text{sin}\lvert\delta_\text{J2000}\rvert$\,arcsec$^2$ PSF, and the Galactic and Extragalactic All-sky MWA \footnote{Murchison Widefield Array.} survey \citep[GLEAM;][]{wlb+15} extragalactic catalogue \citep{gleamegc} with a position-dependent PSF major axis that varies between $\sim 128$--240\,arcsec and minor axis that varies between $\sim 122$--173\,arcsec at 200\,MHz. Note we do not include the RACS-mid time-domain catalogue as the overlapping observations artificially increase the source density and are not reflective of the actual images used. The shaded regions in Figure~\ref{fig:beta} show the range of values within a declination bin for surveys with variable PSFs.}

{Despite the elongated PSF of the primary catalogue at high declination, the solid angle remains small due to the high ellipticity of the PSF ($\theta_\text{M}/\theta_\text{m} \lesssim 4.7$). The number of PSF solid angles per source is thus sufficiently high that we do not consider classical source confusion to be the predominant noise in the RACS-mid images and subsequent catalogues. We note that the 25-arcsec catalogue generally has a larger PSF solid angle than the primary catalogue (except in seven of the full-sensitivity images used for the primary catalogue) and so is overall more affected, though still not limited by classical source confusion.}

\subsection{Source density}\label{sec:density}

Figure~\ref{fig:sourcedensity} shows the HEALPix-binned ($N_\text{side}=64$, $\sim 55 \times 55$\,arcmin$^{2}$) source and component density across the sky for the primary catalogue [\ref{fig:sourcedensity:1} and \ref{fig:sourcedensity:2}], the 25-arcsec catalogue [\ref{fig:sourcedensity:3} and \ref{fig:sourcedensity:4}], and the time-domain catalogue [\ref{fig:sourcedensity:5} and \ref{fig:sourcedensity:6}]. 
The average source density across the primary catalogue is $\sim 86$\,deg$^{-2}$, over the catalogued area of $\sim 36\,200$\,deg$^{2}$. The 25-arcsec catalogue has a source density of $\sim 70$\,deg$^{-2}$ (coverage $\sim 30\,900$\,deg$^{2}$) and the time-domain catalogue has a source density of $\sim 113$\,deg$^{-2}$ (coverage $\sim 36\,200$\,deg$^{2}$, though note that by construction there are many duplicate sources in the time-domain catalogue). 

In the primary and 25-arcsec catalogues the Galactic Plane features a lower density of sources, largely due to higher noise on multiple scales. This is a combination of both the real extended sources and the artefacts on multiple angular scales from the poor modelling of the real extended sources. A similar decrease in source density is not as obvious in the time-domain catalogue due to the double-counting of sources in overlap regions. We also see a decrease in source density at the celestial equator corresponding to a reduction in image sensitivity. This is most prominent in the 25-arcsec catalogue. The component density of all three catalogues is higher than the source density, corresponding to the resolved sources being modelled with multiple Gaussian components by \texttt{PyBDSF}. Smaller-scale low-density regions can be seen around other bright sources, though generally restricted to the 1.2~degree radius left from peeling described in \citetalias{racs-mid}. The regions in all three catalogues around $\delta_\text{J2000} \approx -75$ show significantly higher average source density caused by the large overlap in adjacent observations where the Celestial tiling changes to cover the South Celestial Pole. The time-domain catalogue features increased source density regions with duplicate observations as we do not remove the duplicated sources.

In Figure~\ref{fig:dec_density} we also show the source density, binned as a function of declination for six flux density limits (1, 2, 3, 4, 5, and 10\,mJy) with the Galactic Plane excluded. The features described for the HEALPix maps are clear in the 1- and 2-mJy flux density populations but largely begin to disappear for the brighter source populations with the 10-mJy source density showing no variation as a function of declination. The exception to this is the time-domain catalogue, for which the 10-mJy population still shows variation in source density as a function of declination---a consequence of including duplicated source entries.

\subsection{The fraction of unresolved sources}\label{sec:resolved}

For certain comparisons to external surveys, it is generally preferable to consider isolated and unresolved sources to avoid biases introduced by differences in $(u,v)$ coverage, and by extension angular resolution. While some comparison could be done for resolved/extended radio sources, comparison in that case becomes more heavily affected by differences in $(u,v)$ coverage and frequency. 

To create an unresolved subset of the primary catalogue and two auxiliary catalogues, we opt to follow the procedure outlined in section 5.2.1 of \citetalias[][]{racs2} following similar methods employed by e.g., \citet[][]{Bondi2008,Smolcic2017a,lotss-dr1}. Firstly, this involves assuming the ratio of integrated flux density, $S_\text{int}$, and peak flux density, $S_\text{peak}$, is close to unity for an unresolved source. In practice, $S_\text{int}/S_\text{peak} > 1$ for a variety of reasons, including: source positions may move during the observation or due to mosaicking adjacent images with small astrometric offsets resulting in a blurred source in final images; self-calibration processes may pull sources in different directions in adjacent beam images (see section 3.4.3 in \citetalias{racs1} or section 3.7.1 in \citetalias{racs-mid}), resulting in blurring of the source. In these situations integrated flux density is generally preserved and peak flux density decreases. Other factors such as as bandwidth or time smearing cause similar problems, which for RACS-mid can be up to $\sim 2$--4~\% reduction towards the edges of primary beam mainlobe, depending on pointing. The blurring, or smearing, of sources more heavily affects those in the low-SNR regime and low-SNR sources may also be pushed to $S_\text{int}/S_\text{peak} < 1$ due to image noise and uncertainty while fitting Gaussian components to sources. 

As in \citetalias{racs2}, we begin by assuming that unresolved sources sit within an envelope defined by \begin{equation}
    S_\text{int}/S_\text{peak} = C + A\times \text{SNR}^{B} \, , \label{eq:snr}
\end{equation}
where $C$ is an estimate of the median $S_\text{int}/S_\text{peak}$ in the high-SNR regime unique to each catalogue as it will depend on the average PSF. $A$ and $B$ are found by fitting Equation~\ref{eq:snr} as in \citetalias{racs2}, after binning values below $C$ and obtaining the $100-95^{\text{th}}$ percentile in each bin. 

Figure~\ref{fig:resolved} shows the ratio $S_\text{int}/S_\text{peak}$ as a function of SNR for single-component sources, along with the fitted envelopes highlighting the resolved sources for each of the three catalogues. The percentage of unresolved sources naturally increases as the catalogue resolution decreases, ranging from $48.4\%$ in the primary catalogue to $58.7\%$ in the 25-arcsec catalogue. Despite the lower rms noise in the RACS-mid data compared to RACS-low, we find a higher fraction of unresolved sources (cf. $\sim 40\%$ for RACS-low; \citetalias{racs2}). We suggest this is largely due to the loss in sensitivity to extended sources in the RACS-mid images, and the lower number of resolved sources in the 25-arcsec catalogue is further evidence of this.

Identifying unresolved/resolved sources in this way also highlights a subset of the catalogued sources that fall both below $ S_\text{int}/S_\text{peak} < 1$ and $S_\text{int}/S_\text{peak} <  C + A\times \text{SNR}^{B}$ (i.e. below the envelope). Such sources are unlikely to be real and may be spurious source-finder detections or artefacts. We therefore add a column in each catalogue called \texttt{Flag} which can be one of the following:
\paragraph{\texttt{Flag = 0}} An unresolved source, satisfying: \begin{equation}
    S_\text{int} / S_\text{peak} > \left(C + A\times \text{SNR}^{B}\right) \, , 
\end{equation}
or \begin{equation}
    1 \leq S_\text{int} / S_\text{peak} < \left(C - A\times \text{SNR}^{B}\right) \, .
\end{equation}
{The second condition covers high-SNR sources that have $S_\text{int}/S_\text{peak}$ below the median $S_\text{int}/S_\text{peak}$ but are not spurious detections. The requirement to include such sources suggests the fitted envelope is not a completely accurate in determining unresolved sources.} 
\paragraph{\texttt{Flag = 1}} A resolved source, satisfying: \begin{equation}
   S_\text{int} / S_\text{peak} > \left(C - A\times \text{SNR}^{B}\right) \, .
\end{equation}

\paragraph{\texttt{Flag = 2}} A spurious source, not satisfying the above (i.e. $S_\text{int}/S_\text{peak}$ below 1 and below the envelope), likely to be an artefact.

\subsection{Reliability}\label{sec:reliability}

\def\relfr{99.28}
\def\relfrgp{99.37}
\def\reltf{99.88}
\def\reltfgp{99.88}
\def\reltd{98.68}
\def\reltdgp{98.86}

\begin{table}[t!]
    \centering
    \caption{\label{tab:reliability} Reliability estimates for the RACS-mid catalogue, for different flux density limits, $S_\text{limit}$.}
    \begin{tabular}{c c c c c}\toprule
    $S_\text{limit}$ & \multicolumn{2}{c}{Reliability \%} & \multicolumn{2}{c}{Reliability \%, $|b| < 5^\circ$} \\ 
 & All & \texttt{Flag != 2} & All & \texttt{Flag != 2} \\\midrule
    \multicolumn{5}{c}{Primary catalogue} \\\midrule
    All & \relfr & \relfrgp & 99.30 & 99.36 \\
$>1$ mJy & 99.61 & 99.68 & 99.66 & 99.71 \\
$>5$ mJy & 99.80 & 99.84 & 99.80 & 99.81 \\
$>10$ mJy & 99.84 & 99.86 & 99.79 & 99.79 \\\midrule
    \multicolumn{5}{c}{25-arcsec catalogue} \\\midrule
    All & \reltf & \reltfgp & 99.30 & 99.29 \\
$>1$ mJy & 99.88 & 99.88 & 99.30 & 99.29 \\
$>5$ mJy & 99.95 & 99.95 & 98.67 & 98.66 \\
$>10$ mJy & 99.95 & 99.95 & 97.88 & 97.87 \\\midrule
    \multicolumn{5}{c}{Time-domain catalogue} \\\midrule
    All & \reltd & \reltdgp & 98.75 & 98.87 \\
    $>1$ mJy & 99.05 & 99.18 & 99.02 & 99.14 \\
    $>5$ mJy & 99.53 & 99.59 & 99.55 & 99.59 \\
    $>10$ mJy & 99.66 & 99.71 &  99.68 & 99.72 \\\bottomrule
    \end{tabular}
\end{table}

Following \citetalias{racs2}, we investigate the reliability of our images and source-finding by quantifying the number of sources detected by \texttt{PyBDSF} below $-5\sigma_\text{rms}$. We assume the noise is symmetric in the image and define the reliability as $1-N_\text{negative}/N$, where $N$ is the number of sources found above $5\sigma_\text{rms}$ and $N_\text{negative}$ are sources detected below $-5\sigma_\text{rms}$. We repeat source-finding on all images (used for the primary catalogue and the two auxiliary catalogues) after multiplying the images by $-1$ and setting the source-finding threshold to $5\sigma_\text{rms}$ as in the normal source-finding procedure. Due to the choice of box size for rms calculations, for many images no sources are found in these inverted images. The source-lists generated from the inverted images are then merged following the same process used for each catalogue. The source flags outlined in Section~\ref{sec:resolved} are also added assuming the same criteria for each catalogue. Figures~\ref{fig:negative_hist:fr}--\ref{fig:negative_hist:td} show flux density histograms of sources detected in the RACS-mid catalogues in both the normal catalogues and negative catalogues.

Figures~\ref{fig:negative:fr}--\ref{fig:negative:td} show an estimate of the catalogues' reliability for sources outside of the Galactic Plane ($|b| > 5^\circ$), binned as a function of declination for a range of flux density cuts (including 1, 5, and 10\,mJy). We find that reliability is not constant in declination. In Table~\ref{tab:reliability} we report median reliability as a percentage of sources without a negative counterpart for the given flux density limits and the full catalogues. For sources outside of the Galactic Plane, we find overall reliability of \relfr\ (\relfrgp) \%, \reltf\ (\reltfgp) \%, and \reltd\ (\reltdgp) \%, for the primary, 25-arcsec, and time-domain catalogues, respectively, with (and without) the `spurious' sources included. The time-domain catalogue shows the least reliability, which is a consequence of including the high-noise regions at the edges of images. Conversely, the 25-arcsec catalogue has least number of negative sources and closely reflects the reliability of the RACS-low catalogue, also at 25-arcsec (cf. $\sim 99.7$\% reliability reported in \citetalias{racs2}). We find that the primary catalogue suffers from its increased angular resolution, where smaller-scale, low-significance artefacts (not present in the 25-arcsec images) are harder to avoid during rms-thresholding and source-finding. As the reliability is lowest at declinations where the PSF is smallest, it is clear the angular resolution plays a significant role in affecting the reliability of the source-finding and catalogues. 

We repeat this comparison for sources within the Galactic Plane ($|b| < 5^\circ$), with medians reported in Table~\ref{tab:reliability} and in Figures~\ref{fig:negative:gp:fr}--\ref{fig:negative:gp:td}. Generally there is an increase in negative sources detected in the Galactic Plane. This is most notable in the 25-arcsec catalogue, where reliability gets worse for brighter sources. This is due to artefacts caused by unmodelled (and poorly modelled) extended emission. When convolved to 25-arcsec, these artefacts become significant with respect to the background noise and the source-finder treats them as real sources. We suggest users of the 25-arcsec catalogue (and 25-arcsec images) be cautious when selecting sources in the Galactic Plane. 

\subsection{Completeness}\label{sec:completeness}

\begin{figure*}[t!]
    \centering
    \begin{subfigure}[b]{0.33\linewidth}
    \includegraphics[width=1\linewidth]{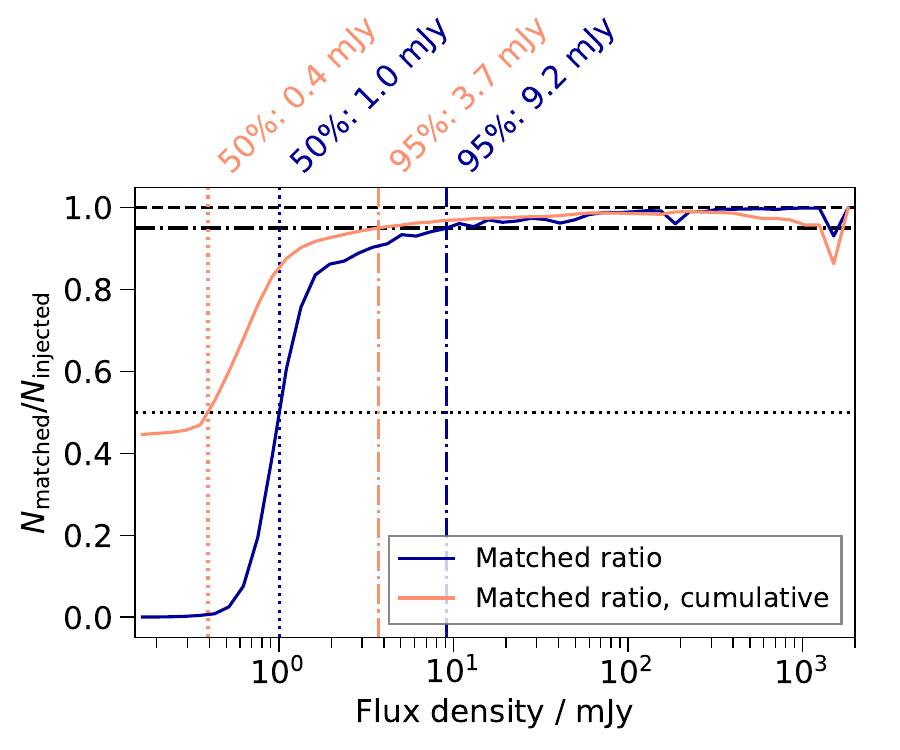}
    \caption{\label{fig:detections:fr} Primary catalogue, detection fraction.}
    \end{subfigure}%
    \begin{subfigure}[b]{0.33\linewidth}
    \includegraphics[width=1\linewidth]{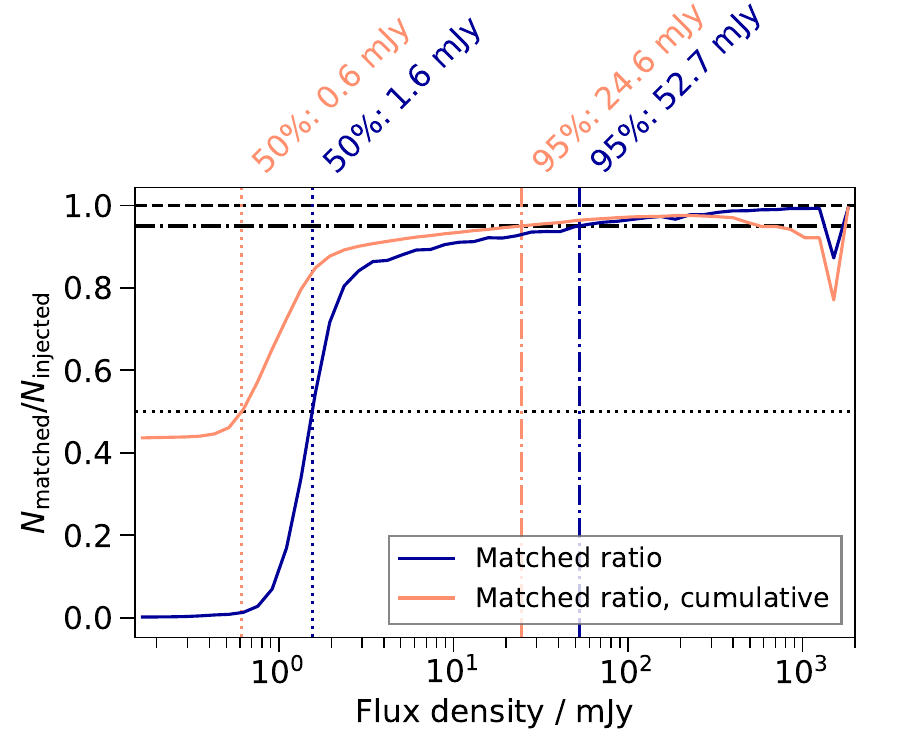}
    \caption{\label{fig:detections:25arcsec}25-arcsec catalogue, detection fraction.}
    \end{subfigure}%
    \begin{subfigure}[b]{0.33\linewidth}
    \includegraphics[width=1\linewidth]{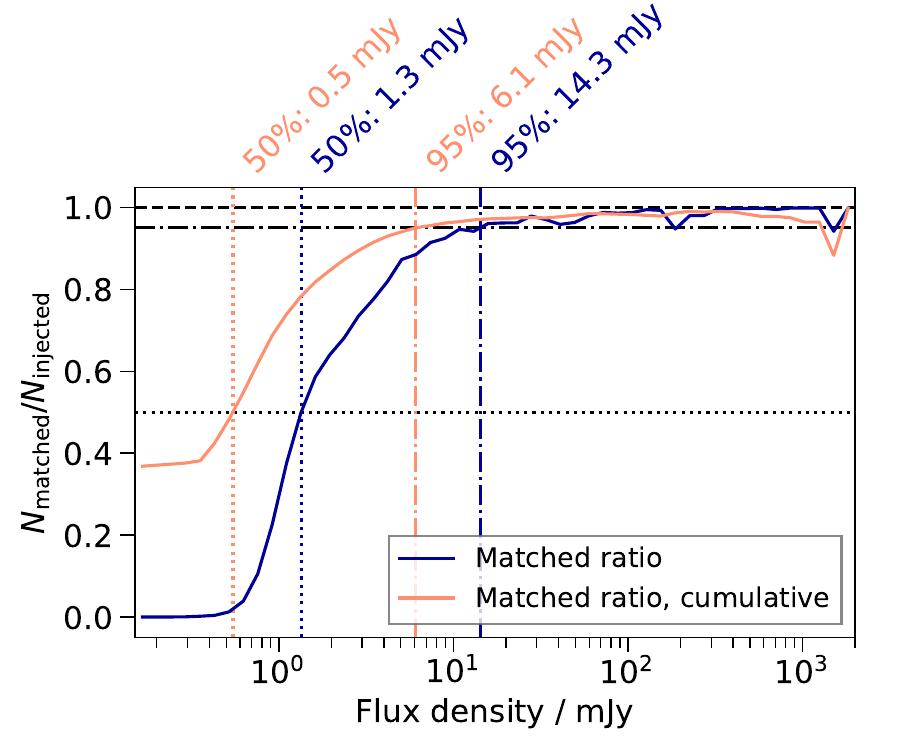}
    \caption{\label{fig:detections:td}Time-domain catalogue, detection fraction.}
    \end{subfigure}\\[1em]%
    \begin{subfigure}[b]{0.33\linewidth}
    \includegraphics[width=1\linewidth]{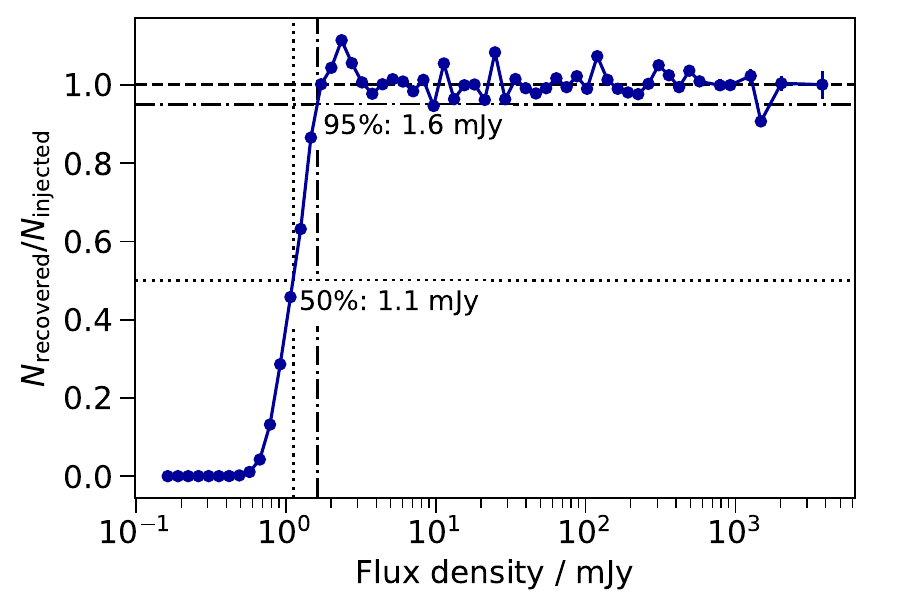}
    \caption{\label{fig:completeness:fr}Primary catalogue, completeness.}
    \end{subfigure}%
    \begin{subfigure}[b]{0.33\linewidth}
    \includegraphics[width=1\linewidth]{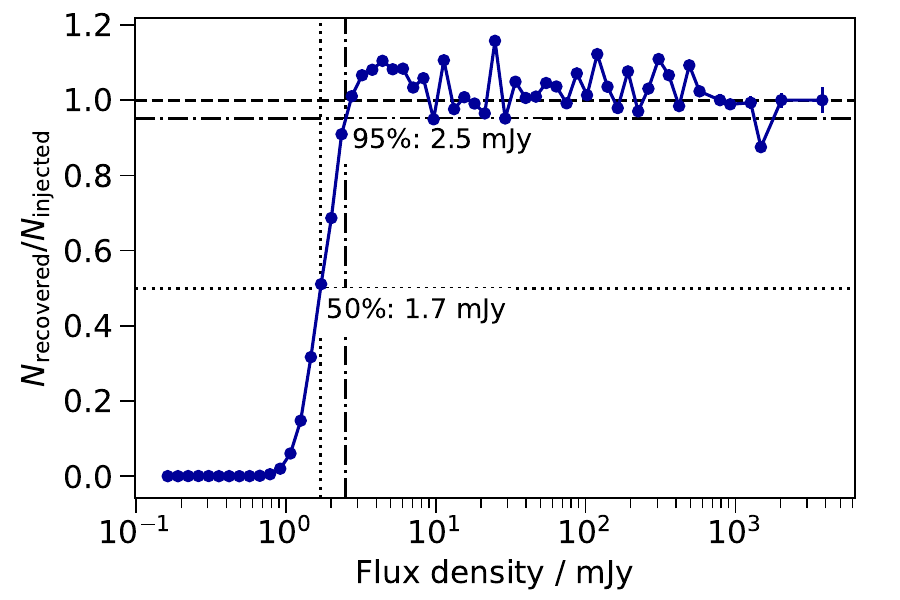}
    \caption{\label{fig:completeness:25arcsec}25-arcsec catalogue, completeness.}
    \end{subfigure}%
    \begin{subfigure}[b]{0.33\linewidth}
    \includegraphics[width=1\linewidth]{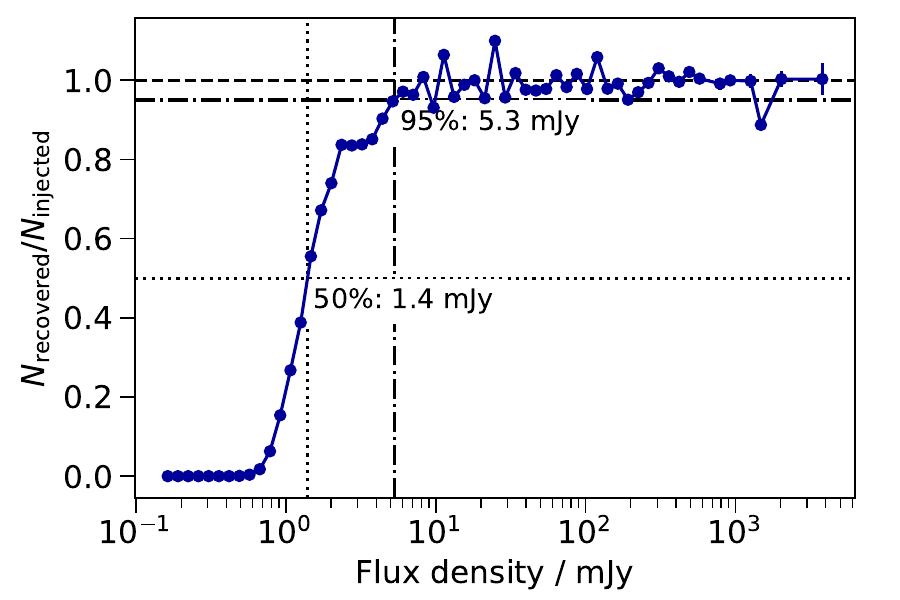}
    \caption{\label{fig:completeness:td}Time-domain catalogue, completeness.}
    \end{subfigure}\\[0.5em]%
    \begin{subfigure}[b]{0.33\linewidth}
    \includegraphics[width=1\linewidth]{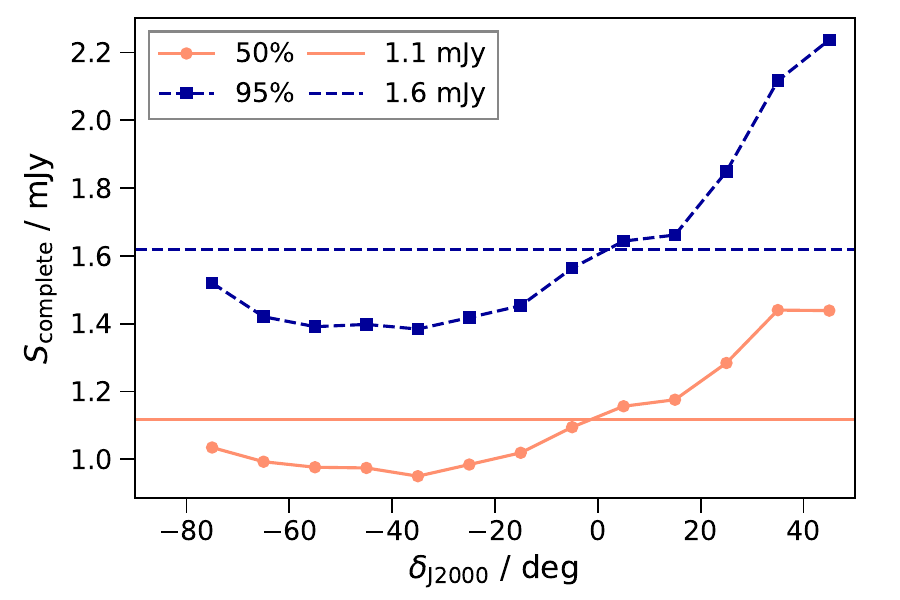}
    \caption{\label{fig:completeness:dec:fr}Primary catalogue, completeness.}
    \end{subfigure}%
    \begin{subfigure}[b]{0.33\linewidth}
    \includegraphics[width=1\linewidth]{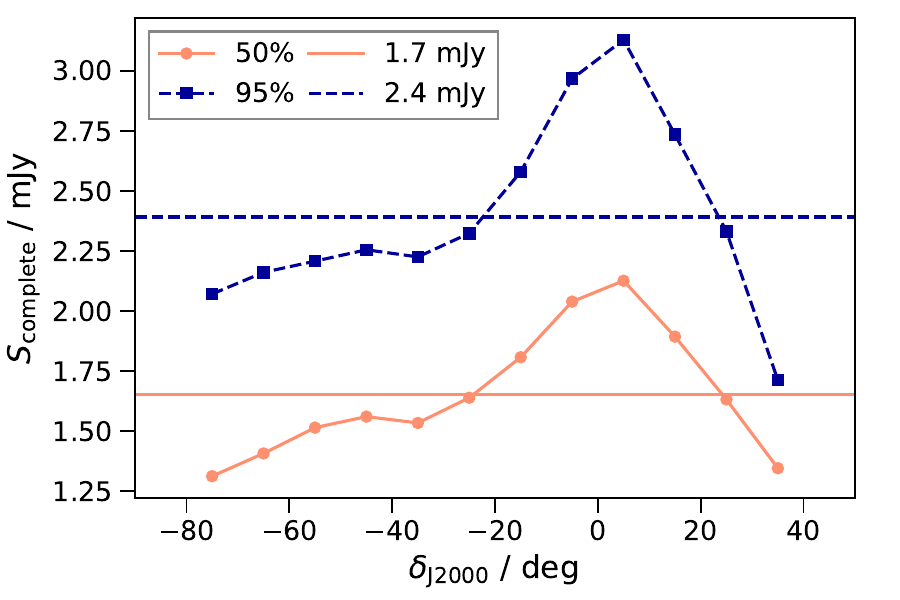}
    \caption{\label{fig:completeness:dec:25arcsec}25-arcsec, completeness.}
    \end{subfigure}%
    \begin{subfigure}[b]{0.33\linewidth}
    \includegraphics[width=1\linewidth]{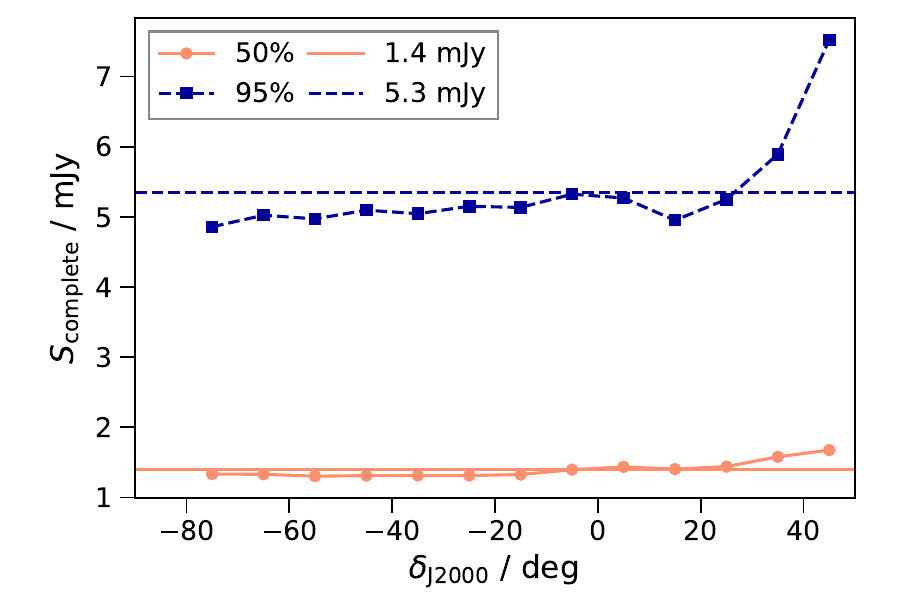}
    \caption{\label{fig:completeness:dec:td}Time-domain catalogue, completeness.}
    \end{subfigure}\\%
    \caption{\label{fig:completeness} Assessment of the completeness of the three RACS-mid catalogues. \subref{fig:detections:fr}--\subref{fig:detections:td}: The fraction of detected model sources matched to the input SKA sky model. The red lines correspond to matched fraction in flux density bins, while the blue line corresponds to the integrated matched fraction. 50 and 95 \% matched fractions are indicated on the plots. \subref{fig:completeness:fr}--\subref{fig:completeness:td}: An estimate of completeness as the fraction of flux density recovered in the model images as a function of flux density. 50 and 95 \% completeness fractions are shown on the plots. {\subref{fig:completeness:dec:fr}--\subref{fig:completeness:dec:td}: An estimate of the 50\% and 95\% completeness ($S_\text{complete}$) as a function of declination, with the mean completeness shown as horizontal lines.}}
\end{figure*}

To estimate the completeness of the RACS-mid catalogues, we follow \citetalias{racs2} and inject a realistic sky model into the individual images after subtracting the Gaussian components found and modelled by \texttt{PyBDSF}. For this purpose, we use the semi-empirical sky model from the SKA Simulated Skies project \citep{Wilman2008,Levrier2009}. We use a base sky model that covers a 100\,deg$^{2}$ region including sources considered both resolved and unresolved at the angular resolution of the RACS-mid data. For each image, we generate five separate sub-sky models by randomising positions of sources from the 100-deg$^{2}$ sky model. We clip the sub-sky models at $3\sigma_\text{rms}$ and at $2 \times S_\text{max}$ where $S_\text{max}$ is the maximum flux density of real sources in the image. We do not expect full recovery of sources at the brightest flux densities, as the brightest sources are typically extended at the RACS-mid resolution \citep[e.g.][]{Perley2017} and RACS-mid has poorer sensitivity to extended sources \citepalias[see e.g. section 3.5 and figure 26 in][]{racs-mid}. The semi-empirical SKA sky model may not be accurate below $\sim 0.1$\,mJy \citep[e.g.][]{Smolcic2017a,Hale2023}, which is generally below the rms noise in the RACS-mid images. The low flux density limit still allows sources below our detection thresholds to be modelled, but are typically above $0.1$\,mJy. This provides a total sub-sky model flux density of $\sim 150$\,Jy for each image. We do not distinguish between unresolved and resolved sources for these tests.

The sub-sky models are injected onto model images and then restored in a similar way to normal imaging: the model is first convolved with the image PSF then the convolved model is added to the residual map. We use \texttt{PyBDSF} to source-find these model-injected images using the same settings as we used for the real images. The individual output source-lists are then cross-matched to the original sub-sky model catalogues for each image independently before concatenation. For this purpose we use the source list output by \texttt{PyBDSF} rather than the component list. This process is done separately for all image set used to generate the catalogues. 

In Figures~\ref{fig:detections:fr}--\ref{fig:detections:td} we show the fraction of sources that are cross-matched to the input sky model after source-finding the three RACS-mid catalogues. As in \citetalias{racs2}, the detection fraction is calculated in logarithmically-sampled flux density bins, and we show both cumulative (orange line) and independent (blue line) detection fractions with 50\% and 95\% detection fractions highlighted. The detection fractions at 50\% and 95\% are higher than in RACS-low for the primary catalogue and conversely lower for for the 25-arcsec catalogue. The drop in detection fraction for the 25-arcsec catalogue is on account of increased blending of sources at the lower resolution of the 25-arcsec catalogue. 

Figures~\ref{fig:completeness:fr}--\ref{fig:completeness:td} show the completeness of each catalogue as a function of flux density. This is defined as the binned ratio of recovered flux density as a function of flux density. The primary catalogue is found to be 50\% complete at 1.1\,mJy and 95\% complete at 1.6\,mJy while we find 1.7 and 2.5\,mJy completeness (at 50\% and 95\%, respectively) for the 25-arcsec catalogue and 1.4 and 5.3\,mJy for the time-domain catalogue. Comparing to the `resolved' source case from \citetalias{racs2} \footnote{Where `resolved' in this case includes all sources, resolved or otherwise.} this suggests the primary RACS-mid catalogue is complete to lower flux densities, accounting for frequency differences (assuming a spectral index \footnote{We define the spectral index as $S_\nu \propto \nu^\alpha$.} of $\alpha=-0.8$) whereas the 25-arcsec and time-domain catalogues are less complete due to lower resolution in the 25-arcsec catalogue and the high-noise image edges of the time-domain catalogue. 

{In addition to the overall completeness, we also repeat the estimate of completeness in 10-degree declination bins for each catalogue. Figures~\ref{fig:completeness:dec:fr}--\ref{fig:completeness:dec:td} show the 50\% and 95\% completeness flux density limit ($S_\text{complete}$) as a function of declination. In all cases there are changes in the completeness as a function of declination, with the primary and time-domain catalogues being less complete beyond $\delta_\text{J2000} \gtrsim +20^\circ$. The 25-arcsec catalogue shows a decrease in completeness around the equator, but with an increase beyond $\delta_\text{J2000} \gtrsim +20^\circ$. These features correspond well to the variation in the noise properties seen in Figure~\ref{fig:noise}.}  

\subsection{Source counts}\label{sec:counts}

\begin{table*}[t!p]
    \centering
    \caption{Tabulated source counts for the primary and 25-arcsec RACS-mid catalogues.\label{tab:counts}}
    \begin{adjustbox}{max width=\textwidth}
    \begin{tabular}{l c c c c c c c}\toprule
    $S_\text{1.4\,GHz}$ & $S_\text{1.4\,GHz,centre}$ & \multicolumn{2}{c}{$N$} & \multicolumn{2}{c}{Normalised $N$} & \multicolumn{2}{c}{Normalised, corrected $N$} \\
    (mJy) & (mJy) & & & \multicolumn{2}{c}{(Jy$^{1.5}$\,sr$^{-1}$)} & \multicolumn{2}{c}{(Jy$^{1.5}$\,sr$^{-1}$)} \\\midrule
    \multicolumn{2}{c}{} & Primary & 25-arcsec & Primary & 25-arcsec & Primary & 25arcsec \\\midrule 
$0.47$--$0.65$ & $0.56$ & $14\,991 \pm 122$ & $45 \pm 6$ & $0.06173 \pm 0.00050$ & $0.000216 \pm 0.000029$ & $4.27 \pm 0.41$ & $0.27 \pm 0.27$ \\
$0.65$--$0.89$ & $0.77$ & $97\,699 \pm 312$ & $1\,647 \pm 40$ & $0.6485 \pm 0.0021$ & $0.01277 \pm 0.00031$ & $4.258 \pm 0.020$ & $0.769 \pm 0.062$ \\
$0.89$--$1.2$ & $1.1$ & $237\,539 \pm 487$ & $28\,462 \pm 168$ & $2.5413 \pm 0.0052$ & $0.3556 \pm 0.0021$ & $6.643 \pm 0.016$ & $4.308 \pm 0.041$ \\
$1.2$--$1.7$ & $1.4$ & $343\,781 \pm 586$ & $144\,807 \pm 380$ & $5.928 \pm 0.010$ & $2.9160 \pm 0.0077$ & $8.147 \pm 0.015$ & $9.885 \pm 0.033$ \\
$1.7$--$2.3$ & $2.0$ & $404\,630 \pm 636$ & $282\,259 \pm 531$ & $11.247 \pm 0.018$ & $9.162 \pm 0.017$ & $11.895 \pm 0.020$ & $16.173 \pm 0.035$ \\
$2.3$--$3.2$ & $2.7$ & $375\,891 \pm 613$ & $318\,281 \pm 564$ & $16.841 \pm 0.027$ & $16.652 \pm 0.030$ & $15.995 \pm 0.027$ & $17.624 \pm 0.033$ \\
$3.2$--$4.4$ & $3.8$ & $297\,424 \pm 545$ & $272\,410 \pm 521$ & $21.479 \pm 0.039$ & $22.972 \pm 0.044$ & $20.937 \pm 0.041$ & $22.423 \pm 0.045$ \\
$4.4$--$6.0$ & $5.2$ & $232\,006 \pm 481$ & $211\,747 \pm 460$ & $27.006 \pm 0.056$ & $28.783 \pm 0.063$ & $27.071 \pm 0.060$ & $26.619 \pm 0.061$ \\
$6.0$--$8.2$ & $7.1$ & $185\,967 \pm 431$ & $164\,555 \pm 405$ & $34.892 \pm 0.081$ & $36.054 \pm 0.089$ & $34.994 \pm 0.085$ & $34.220 \pm 0.088$ \\
$8.2$--$11$ & $9.8$ & $151\,841 \pm 389$ & $129\,486 \pm 359$ & $45.92 \pm 0.12$ & $45.73 \pm 0.13$ & $46.22 \pm 0.13$ & $44.38 \pm 0.13$ \\
$11$--$16$ & $13$ & $123\,870 \pm 351$ & $102\,833 \pm 320$ & $60.38 \pm 0.17$ & $58.54 \pm 0.18$ & $61.17 \pm 0.18$ & $58.13 \pm 0.19$ \\
$16$--$21$ & $18$ & $98\,286 \pm 313$ & $81\,701 \pm 285$ & $77.23 \pm 0.25$ & $74.96 \pm 0.26$ & $77.97 \pm 0.26$ & $74.66 \pm 0.27$ \\
$21$--$29$ & $25$ & $77\,345 \pm 278$ & $64\,723 \pm 254$ & $97.96 \pm 0.35$ & $95.72 \pm 0.38$ & $98.54 \pm 0.37$ & $95.21 \pm 0.39$ \\
$29$--$40.$ & $35$ & $59\,679 \pm 244$ & $50\,131 \pm 223$ & $121.83 \pm 0.50$ & $119.51 \pm 0.53$ & $122.09 \pm 0.52$ & $117.29 \pm 0.54$ \\
$40.$--$56$ & $48$ & $45\,395 \pm 213$ & $38\,799 \pm 196$ & $149.37 \pm 0.70$ & $149.08 \pm 0.75$ & $149.33 \pm 0.74$ & $145.64 \pm 0.77$ \\
$56$--$76$ & $66$ & $34\,751 \pm 186$ & $29\,705 \pm 172$ & $184.32 \pm 0.99$ & $184.0 \pm 1.1$ & $183.8 \pm 1.1$ & $179.3 \pm 1.1$ \\
$76$--$110$ & $91$ & $25\,448 \pm 159$ & $21\,929 \pm 148$ & $217.6 \pm 1.4$ & $218.9 \pm 1.5$ & $215.4 \pm 1.4$ & $210.6 \pm 1.5$ \\
$110$--$140$ & $120$ & $18\,095 \pm 134$ & $15\,761 \pm 125$ & $249.4 \pm 1.8$ & $253.6 \pm 2.0$ & $246.7 \pm 1.9$ & $242.8 \pm 2.0$ \\
$140$--$200$ & $170$ & $12\,765 \pm 112$ & $11\,004 \pm 104$ & $283.5 \pm 2.5$ & $285.4 \pm 2.7$ & $283.9 \pm 2.6$ & $276.9 \pm 2.7$ \\
$200$--$270$ & $240$ & $8\,682 \pm 93$ & $7\,459 \pm 86$ & $310.8 \pm 3.3$ & $311.8 \pm 3.6$ & $309.6 \pm 3.5$ & $301.9 \pm 3.7$ \\
$270$--$380$ & $320$ & $5\,612 \pm 74$ & $4\,908 \pm 70$ & $323.9 \pm 4.3$ & $330.7 \pm 4.7$ & $319.2 \pm 4.5$ & $318.4 \pm 4.8$ \\
$380$--$520$ & $450$ & $3\,741 \pm 61$ & $3\,270 \pm 57$ & $348.0 \pm 5.7$ & $355.2 \pm 6.2$ & $344.4 \pm 5.9$ & $346.0 \pm 6.3$ \\
$520$--$710$ & $610$ & $2\,185 \pm 46$ & $1\,884 \pm 43$ & $327.6 \pm 6.9$ & $329.9 \pm 7.5$ & $325.2 \pm 7.5$ & $324.3 \pm 8.0$ \\
$710$--$970$ & $840$ & $1\,324 \pm 36$ & $1\,105 \pm 33$ & $320.0 \pm 8.7$ & $311.8 \pm 9.3$ & $317.8 \pm 9.1$ & $307.9 \pm 9.7$ \\
$970$--$1\,300$ & $1\,200$ & $841 \pm 29$ & $740 \pm 27$ & $328 \pm 11$ & $337 \pm 12$ & $333 \pm 12$ & $348 \pm 13$ \\
$1\,300$--$1\,800$ & $1\,600$ & $436 \pm 20$ & $391 \pm 19$ & $274 \pm 13$ & $287 \pm 14$ & $285 \pm 14$ & $311 \pm 17$ \\
$1\,800$--$2\,500$ & $2\,200$ & $266 \pm 16$ & $233 \pm 15$ & $269 \pm 16$ & $275 \pm 18$ & $269 \pm 17$ & $275 \pm 19$ \\
$2\,500$--$3\,500$ & $3\,000$ & $173 \pm 13$ & $145 \pm 12$ & $282 \pm 21$ & $276 \pm 23$ & $282 \pm 22$ & $276 \pm 24$ \\
$3\,500$--$4\,800$ & $4\,100$ & $78 \pm 8$ & $71 \pm 8$ & $205 \pm 21$ & $218 \pm 25$ & $205 \pm 21$ & $218 \pm 25$ \\
$4\,800$--$6\,600$ & $5\,700$ & $49 \pm 7$ & $47 \pm 6$ & $208 \pm 30$ & $233 \pm 30$ & $208 \pm 30$ & $233 \pm 30$ \\
$6\,600$--$9\,000$ & $7\,800$ & $25 \pm 5$ & $25 \pm 5$ & $171 \pm 34$ & $199 \pm 40$ & $171 \pm 34$ & $199 \pm 40$ \\
$9\,000$--$12\,000$ & $11\,000$ & $9 \pm 3$ & $8 \pm 2$ & $99 \pm 33$ & $103 \pm 26$ & $99 \pm 33$ & $103 \pm 26$ \\
$12\,000$--$17\,000$ & $15\,000$ & $16 \pm 4$ & $8 \pm 2$ & $284 \pm 71$ & $166 \pm 41$ & $284 \pm 71$ & $166 \pm 42$ \\
$17\,000$--$23\,000$ & $20\,000$ & $4 \pm 2$ & $4 \pm 2$ & $114 \pm 57$ & $134 \pm 67$ & $114 \pm 57$ & $134 \pm 67$ \\
$23\,000$--$32\,000$ & $28\,000$ & $2 \pm 1$ & $3 \pm 1$ & $92 \pm 46$ & $162 \pm 54$ & $92 \pm 46$ & $162 \pm 54$ \\
$32\,000$--$44\,000$ & $38\,000$ & $3 \pm 1$ & $1 \pm 1$ & $223 \pm 74$ & $87 \pm 87$ & $223 \pm 74$ & $87 \pm 87$ \\
$44\,000$--$61\,000$ & $53\,000$ & $3 \pm 1$ & $5 \pm 2$ & $359 \pm 120$ & $700 \pm 280$ & $359 \pm 120$ & $700 \pm 280$ \\
$61\,000$--$84\,000$ & $72\,000$ &  - & $1 \pm 1$ &  - & $226 \pm 226$ &  - & $226 \pm 226$ \\
$84\,000$--$120\,000$ & $100\,000$ &  - &  - &  - &  - &  - &  - \\
$120\,000$--$160\,000$ & $140\,000$ & $1 \pm 1$ & $2 \pm 1$ & $502 \pm 502$ & $1\,172 \pm 586$ & $502 \pm 502$ & $1\,172 \pm 586$ \\
$160\,000$--$220\,000$ & $190\,000$ &  - &  - &  - &  - &  - &  - \\
$220\,000$--$300\,000$ & $260\,000$ &  - & $1 \pm 1$ &  - & $1\,522 \pm 1\,522$ &  - & $1\,522 \pm 1\,522$ \\
$300\,000$--$410\,000$ & $360\,000$ &  - &  - &  - &  - &  - &  - \\
$410\,000$--$570\,000$ & $490\,000$ &  - & $1 \pm 1$ &  - & $3\,955 \pm 3\,955$ &  - & $3\,955 \pm 3\,955$ \\\bottomrule
    
    \end{tabular}
    \end{adjustbox}
\end{table*}

\begin{figure*}[t!]
    \centering
    \includegraphics[width=0.8\linewidth]{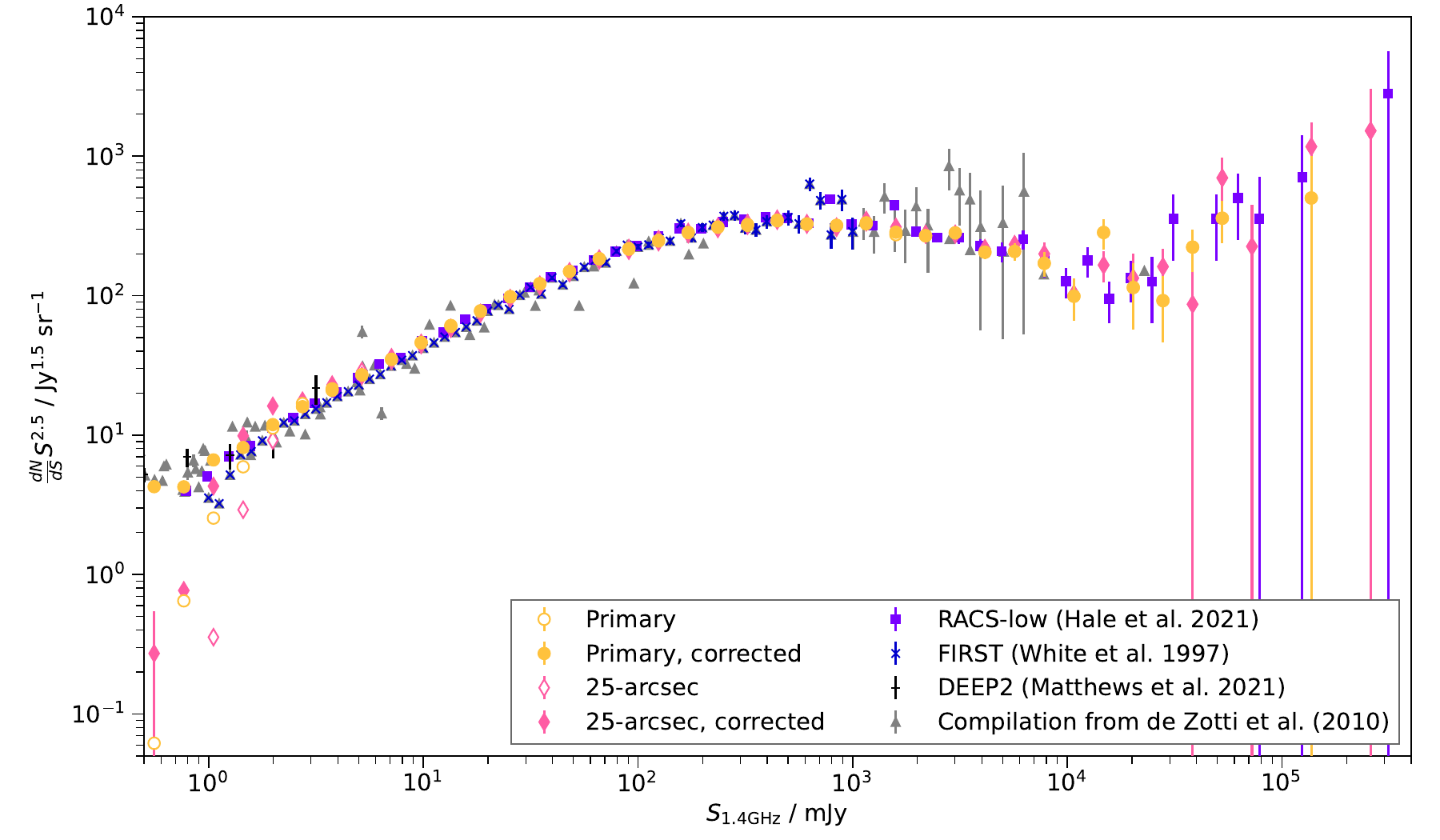}
    \caption{\label{fig:counts} The corrected and uncorrected source counts from the primary and 25-arcsec RACS-mid catalogues (excluding $|b|<5^\circ$). Source counts from a selection of other surveys are shown for reference, including the 1.4-GHz compilation from \citet{deZotti2010}, the 1997 FIRST catalogue \citet{White1997}, \citet{Matthews2021}, and RACS-low \citepalias{racs2}.} 
\end{figure*}

RACS provides catalogues at sensitivities, angular resolutions, and frequencies not previously available over such wide areas and flux density ranges. This makes the RACS releases valuable in exploring the radio source counts, which are typically restricted to high sensitivity and small areas \citep[e.g.][]{Vernstrom2016b,Smolcic2017a,Smolcic2017b,Mandal2021,Matthews2021,Hale2023} or lower sensitivity and wide areas \citep[e.g.][]{White1997,Franzen2021,Matthews2021,racs2}. \citetalias{racs2} reported the normalised differential source counts for RACS-low, and here we do the same for the RACS-mid for both the primary and 25-arcsec catalogues for all sources outside of the Galactic Plane ($|b| > 5^\circ$). 

We summarise the source counts (and normalised source counts with and without completeness corrections) in Table~\ref{tab:counts}. The normalised source counts are also shown in Figure~\ref{fig:counts} for the RACS-mid catalogues, scaled to 1.4\,GHz assuming $\alpha=-0.8$ alongside similar source counts from a range of other surveys, including RACS-low, FIRST (reported by \citet{White1997}), the compilation of 1.4-GHz source counts by \citet[][see references therein]{deZotti2010}, and source counts from the DEEP2 catalogue \citep{Matthews2021}. We also show the `corrected' RACS-mid source counts, after accounting for the flux density recovery fraction from Figures~\ref{fig:completeness:fr} and \ref{fig:completeness:25arcsec}. After applying corrections for incompleteness, we find generally good agreement with existing source counts, with some deviation at low and high flux densities. 

At low flux densities ($<2$\,mJy), the source counts from the primary catalogue follow the low-flux density bins from the \citet{deZotti2010} compilation and the DEEP2 source counts from \citet{Matthews2021} until flux density bins reach $\sim 0.6$\,mJy. This mirrors what is seen with the RACS-low catalogue. Conversely, the 25-arcsec RACS-mid catalogue shows a decrease in the corrected source counts at $\sim 1$\,mJy, though this is at the detection limit of the survey.  In the high-flux density bins, the primary catalogue fails to report some of the brighter sources in the sky. Some of the brightest extended sources such as Fornax~A are not recovered well in the data as described in section 3.5 in \citetalias{racs-mid} and may be split into smaller sources during source-finding. Such bright sources are therefore under-represented in the RACS-mid catalogues. We do not correct these high flux density bins for incompleteness as this requires a complex visibility-based process rather than our image-based sky model injection described in Section~\ref{sec:completeness} and is beyond the scope of this brief source count assessment. 

\subsection{Matching to external catalogues}\label{sec:matching}

\begin{table*}[t!]
    \centering
    \begin{threeparttable}
    \caption{\label{tab:photo} Median reported properties of sources cross-matched between the RACS-mid catalogues and other surveys, with and without flux density limits applied.}
    \begin{tabular}{c c c  c c c c c c c c}\toprule
    RACS-mid & Survey & $\nu$\tnote{a} & $S_\text{RACS-mid}^\text{limit}$ & $S_\text{survey}^\text{limit}$ & \multicolumn{2}{c}{$N_\text{sources}$} & \multicolumn{2}{c}{$\alpha$} & \multicolumn{2}{c}{$S^\text{RACS-mid}_\text{survey}$} \\
            &        & (MHz) & (mJy & (mJy) &  All & Limit & All & Limit & All & Limit \\\midrule
        & RACS-low & 888 & 3.7 & 4.2 & 464\,286 & 274\,373 & $-0.81_{-0.42}^{+0.61}$ & $-0.82_{-0.34}^{+0.51}$ & - & - \\
        & SUMSS & 843 & 12.1 & 4.4 & 57\,054 & 26\,505 & $-0.86_{-0.51}^{+0.51}$ & $-0.74_{-0.33}^{+0.58}$ & -& -  \\
        & TGSS & 150 & 82.9 & 164.4 & 135\,776 & 8\,265 & $-0.70_{-0.23}^{+0.25}$ & $-0.61_{-0.22}^{+0.29}$ & - & -\\
        & LoTSS-DR2 & 144 & 1.8 & 193.9 & 16\,840 & 1\,539 & $-0.53_{-0.28}^{+0.57}$ & $-0.78_{-0.19}^{+0.21}$ & -& - \\
        & WALLABY pre-pilot & 1367.5 & 10 & - & 1\,760 & 81 & - & - & $1.10_{-0.20}^{+0.32}$ &  $1.06_{-0.06}^{+0.11}$\\
        & FIRST & 1400\tnote{b} & 10 & - & 163\,149 & 38\,382 & - & - & $1.04_{-0.15}^{+0.27}$ & $0.99_{-0.07}^{+0.11}$ \\
        \multirow{-7}{*}{\makecell[c]{Primary}} & NVSS & 1400 & 10 & - &472\,136 & 120\,831 & - & - & $0.98_{-0.24}^{+0.21}$ & $1.00_{-0.08}^{+0.12}$ \\[0.5em]\midrule
        
        & RACS-low & 888 & 3.7 & 6.5 & 451\,954 & 305\,627 & $-0.76_{-0.34}^{+0.56}$ & $-0.83_{-0.30}^{+0.41}$ & - & - \\
        & SUMSS & 843 & 12.1 & 6.5 & 94\,498 & 48\,333 & $-0.84_{-0.43}^{+0.45}$ & $-0.77_{-0.30}^{+0.47}$ & -& -  \\
        & TGSS & 150 & 82.9 & 251.4 & 186\,522 & 14\,092 & $-0.70_{-0.21}^{+0.23}$ & $-0.67_{-0.19}^{+0.22}$ & - & -\\
        & LoTSS-DR2 & 144 & 1.8 & 235.4  & 4\,973 & 413 & $-0.53_{-0.28}^{+0.59}$ & $-0.77_{-0.17}^{+0.20}$ & -& - \\
        & WALLABY pre-pilot & 1367.5 & 10 & - & 1\,280 & 93 & - & - & $1.15_{-0.18}^{+0.35}$ & $1.07_{-0.06}^{+0.08}$ \\
        & FIRST & 1400\tnote{b} & 10 & - & 163\,149 & 38\,382 & - & - & $1.09_{-0.16}^{+0.35}$ & $1.01_{-0.08}^{+0.14}$ \\
        \multirow{-7}{*}{\makecell[c]{25-arcsec}} & NVSS & 1400 & 10 & - & 498\,956 & 158\,591 & - & - & $1.01_{-0.17}^{+0.22}$ & $1.01_{-0.07}^{+0.12}$ \\[0.5em]\midrule

        & RACS-low & 888 & 3.7 & 4.2 & 268\,310 & 135\,637 & $-0.80_{-0.43}^{+0.62}$ & $-0.82_{-0.34}^{+0.51}$ & - & - \\
        & SUMSS & 843 & 12.1 & 4.5 & 26\,535 & 11\,722 & $-0.88_{-0.56}^{+0.52}$ & $-0.73_{-0.33}^{+0.59}$ & -& -  \\
        & TGSS & 150 & 82.9 & 160.2 & 60\,401 & 3\,051 & $-0.71_{-0.25}^{+0.26}$ & $-0.59_{-0.22}^{+0.28}$ & - & -\\
        & LoTSS-DR2 & 144 & 1.8 & 171.6  & 10\,334 & 654 & $-0.53_{-0.28}^{+0.53}$ & $-0.78_{-0.19}^{+0.21}$ & -& - \\
        & WALLABY pre-pilot & 1367.5 & 10 & - & 2\,003 & 120 & - & - & $1.12_{-0.18}^{+0.30}$ & $1.09_{-0.10}^{+0.09}$ \\
        & FIRST & 1400\tnote{b} & 10 & - & 105\,742 & 16\,790 & - & - & $1.05_{-0.17}^{+0.28}$ & $0.99_{-0.07}^{+0.11}$ \\
        \multirow{-7}{*}{\makecell[c]{Time-domain}} & NVSS & 1400 & 10 & - & 273\,990 & 51\,180 & - & - & $0.95_{-0.29}^{+0.22}$ & $0.99_{-0.08}^{+0.12}$ \\[0.5em]
        
    \bottomrule
    \end{tabular}
    \begin{tablenotes}[flushleft]
    {\footnotesize
    \item[a] Frequency of the comparison survey. \item[b] Sources detected in data from 2011 are scaled from 1\,335\,MHz assuming a power law with spectral index $-0.7$.
    }
    \end{tablenotes}
    \end{threeparttable}
\end{table*}

For the following sections, we investigate the brightness scale and astrometry of the primary catalogue and the 25-arcsec catalogue. Due to similarities with the source-lists used for validation in \citetalias{racs-mid}, some comparisons with the time-domain catalogue are not repeated. We expect similarities with the primary catalogue and the 25-arcsec catalogue, though due to the additional mosaicking and image-based convolution it is worth confirming we have not introduced any significant systematic effects into the data. 

 For direct comparisons of flux density measurements and astrometry we look to the  NVSS \citep{ccg+98} at 1.4\,GHz, FIRST \citep{Becker95,White1997,hwb15}, and the 42-deg$^{2}$ region of the WALLABY pre-pilot catalogue \citep{Grundy2023}. FIRST is catalogued at both the same frequency as NVSS (for data prior to 2011) and at 1\,335\,MHz (for data from 2011). FIRST sources detected in data from 2011 onward comprise $\sim 6\%$ of the full FIRST catalogue but are exclusively in the region that overlaps with RACS-mid. To account for this in the FIRST data, we scale the flux density measurements, $S$, of the 2011-onward data to match the 1\,400-MHz data assuming a power law of the form $S \propto \nu^\alpha$ with {$\alpha = -0.8$}. For comparison to catalogues at other frequencies, we use the RACS-low catalogue \citepalias{racs2}, SUMSS \citep{bls99,mmb+03}, the TIFR \footnote{Tata Insititute for Fundamental Research.} GMRT \footnote{Giant Metrewave Radio Telescope.} Sky Survey alternate data release 1 \citep[TGSS;][]{ijmf16} catalogue, and the LOFAR \footnote{LOw Frequency ARray.} Two-metre Sky Survey data release 2 \citep[LoTSS-DR2;][]{lotss:dr2} catalogue. These catalogues provide a range of comparisons in the frequency range 144--1400~MHz. {For additional astrometry assessment, we also compare to the VLA Sky Survey \citep[VLASS;][]{vlass} first epoch `quick-look' catalogue \citep[hereinafter VLASS-QL;][]{Gordon2021}.}

For the remainder of the validation work, unless otherwise specified, we are only considering unresolved (\texttt{Flag = 0}) sources in the RACS-mid catalogues. For the comparison catalogues, we generally only look at unresolved sources if the survey angular resolution is on average higher than RACS-mid. In the case of RACS-low and the WALLABY pre-pilot catalogue, we use the same unresolved criteria described in \citetalias{racs2} and \citet{Grundy2023} (39.6\% and 76.5\% unresolved, respectively), but for the remaining comparison catalogues we employ the simpler criteria. For the FIRST, VLASS-QL, and LoTSS-DR2 catalogues we take sources that satisfy $|\log_{10}(S_\text{int}/S_\text{peak})| < \log_{10}(1.2)$ as unresolved. In the case of LoTSS-DR2, because of significantly blurring point sources show ratios closer to $\sim 1.25$, and for that catalogue we shift the ratio criterion by an additional $\log_{10}(0.25)$. With this criterion we find unresolved fractions for the FIRST, VLASS-QL, and LoTSS-DR2 of 44.9, 26.1, and 47.6\%. We note for LoTSS-DR2 \citet{lotss:dr2} find a resolved fraction of 8.0\%. In this case we are more concerned with confidence in determining unresolved sources so we remove potentially resolved sources, whereas \citet{lotss:dr2} were focused on resolved sources and conversely remove potentially unresolved sources with their method. 

Finally, when cross-matching \footnote{Using \texttt{match\_catalogues} packaged as part of \texttt{flux\_warp} (\url{https://gitlab.com/Sunmish/flux_warp}).} the catalogues we consider only isolated sources in the respective catalogues. Isolated sources in this case are defined as those with no neighbours within $2\theta_\text{major}$ for the given catalogue. Cross-matching is then done with a match radius equal to half the PSF major FWHM for the survey with the lowest angular resolution. After cross-matching with these strict criteria we find that the 25-arcsec catalogue results in the largest number of matches in catalogues with angular resolutions $\gtrsim 25 \times 25$\,arcsec$^{2}$. The time-domain catalogue results in the least matches simply because the duplication of sources makes the isolation criterion difficult to satisfy, and generally there will be few cross-matches in the overlap regions between tiles. For assessment of the overlap regions, we refer the reader to \citetalias{racs-mid} or Section~\ref{sec:tile:brightness}.

The numbers of sources cross-matched for each RACS-mid catalogue and external catalogue are reported in Table~\ref{tab:photo} .

\subsection{Photometry}\label{sec:photometry}

As part of the RACS-mid processing, we used cross-matches to NVSS and SUMSS to correct for time-dependent brightness scale variations (see section~2.6 in \citetalias{racs-mid} and further discussion of this issue in section~3.3.3 in \citetalias{racs1}). By construction this ensures the bulk brightness scale between RACS-mid and the NVSS and SUMSS is consistent within the assumed spectral index of $-0.82$ \footnote{We note that users of the original images can `unapply' the brightness scale correction factors as described in section~2.6 in \citetalias{racs-mid}, with scaling factors retained in the FITS headers of the images available on CASDA. The time-domain catalogue retains the scaling factors which users interested in variability may wish to revert for e.g. comparison with NVSS and/or SUMSS.}.  Despite this it is still worth comparing the catalogues produced here due to a difference in the images and source-finding.

\subsubsection{External brightness scale}\label{sec:brightness}

\begin{figure*}[t!]
    \centering
    \begin{subfigure}[b]{0.32\linewidth}
    \includegraphics[width=1\linewidth]{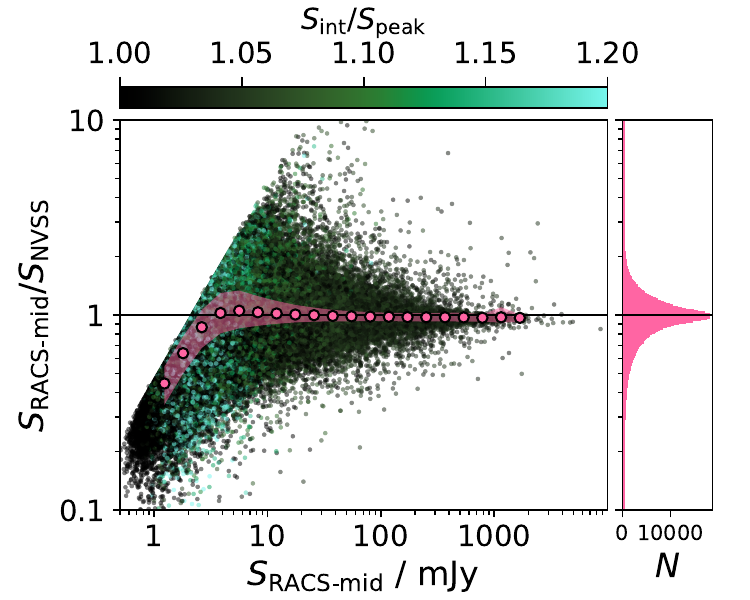}
    \caption{\label{fig:snr:fr:nvss} Primary catalogue and NVSS.}
    \end{subfigure}%
    \begin{subfigure}[b]{0.32\linewidth}
    \includegraphics[width=1\linewidth]{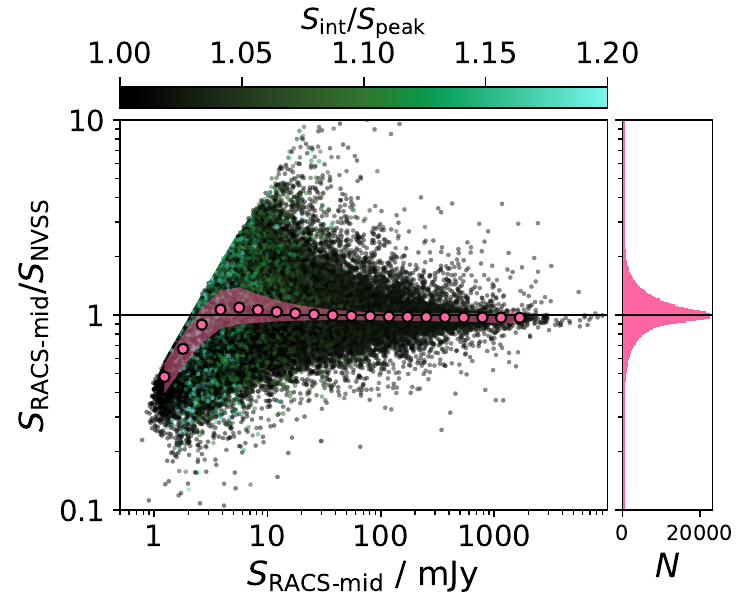}
    \caption{\label{fig:snr:25:nvss} 25-arcsec catalogue and NVSS.}
    \end{subfigure}%
    \begin{subfigure}[b]{0.32\linewidth}
    \includegraphics[width=1\linewidth]{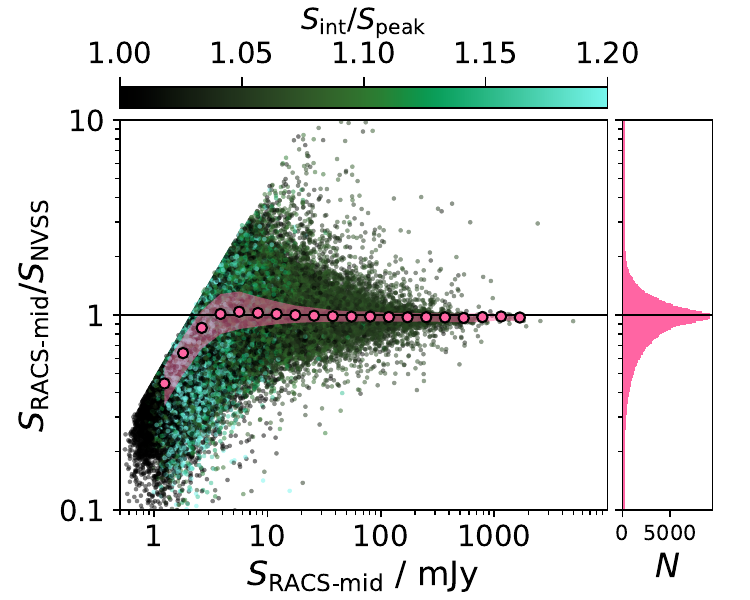}
    \caption{\label{fig:snr:td:nvss} Time-domain catalogue and NVSS.}
    \end{subfigure}\\%
    \begin{subfigure}[b]{0.32\linewidth}
    \includegraphics[width=1\linewidth]{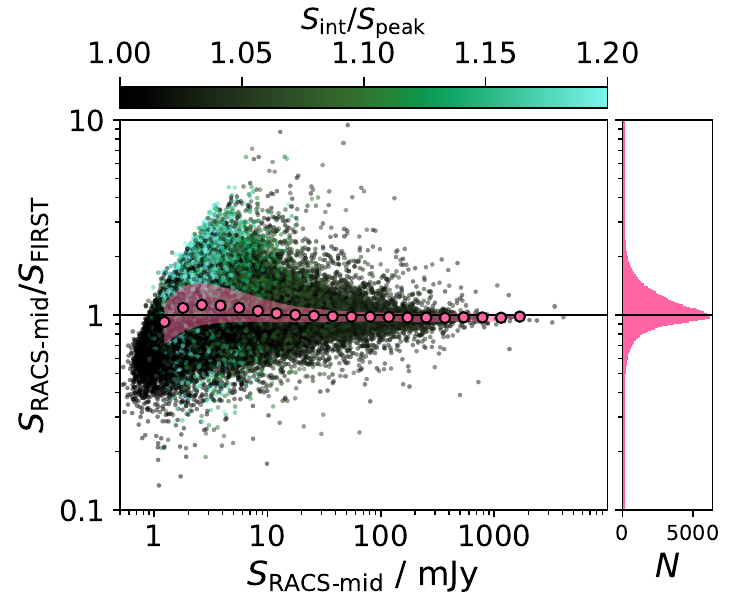}
    \caption{\label{fig:snr:fr:first} Primary catalogue and FIRST.}
    \end{subfigure}%
    \begin{subfigure}[b]{0.32\linewidth}
    \includegraphics[width=1\linewidth]{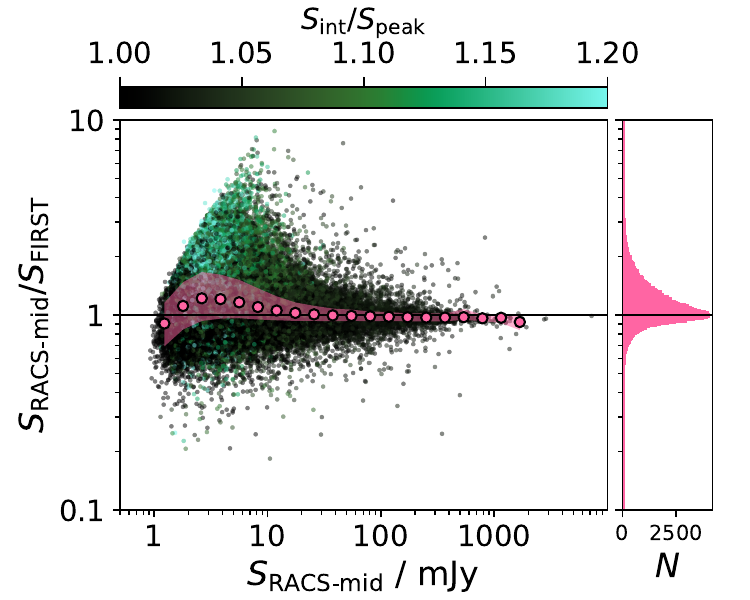}
    \caption{\label{fig:snr:25:first} 25-arcsec catalogue and FIRST.}
    \end{subfigure}%
    \begin{subfigure}[b]{0.32\linewidth}
    \includegraphics[width=1\linewidth]{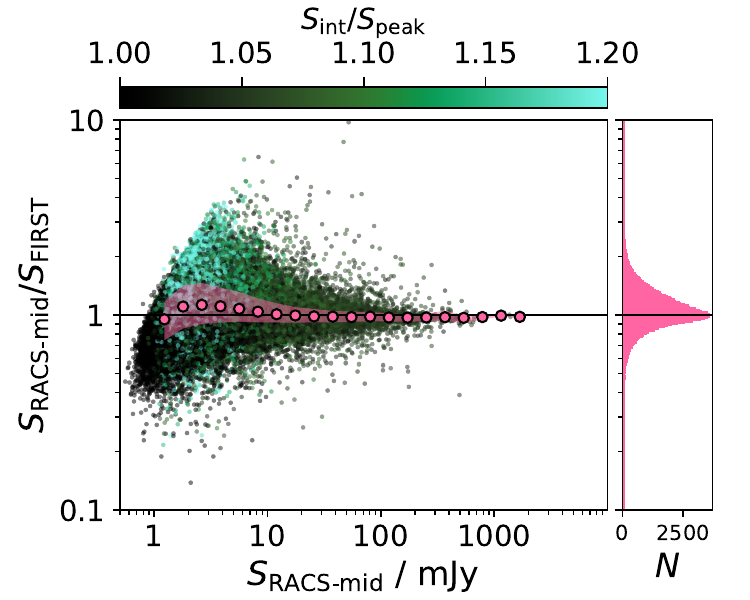}
    \caption{\label{fig:snr:td:first} Time-domain catalogue and FIRST.}
    \end{subfigure}\\%
    \begin{subfigure}[b]{0.32\linewidth}
    \includegraphics[width=1\linewidth]{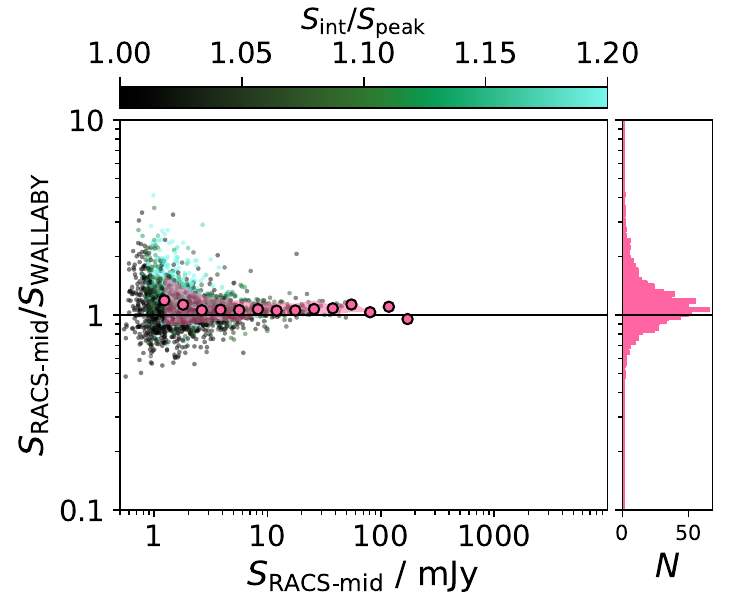}
    \caption{\label{fig:snr:fr:wallaby} Primary catalogue and WALLABY.}
    \end{subfigure}%
    \begin{subfigure}[b]{0.32\linewidth}
    \includegraphics[width=1\linewidth]{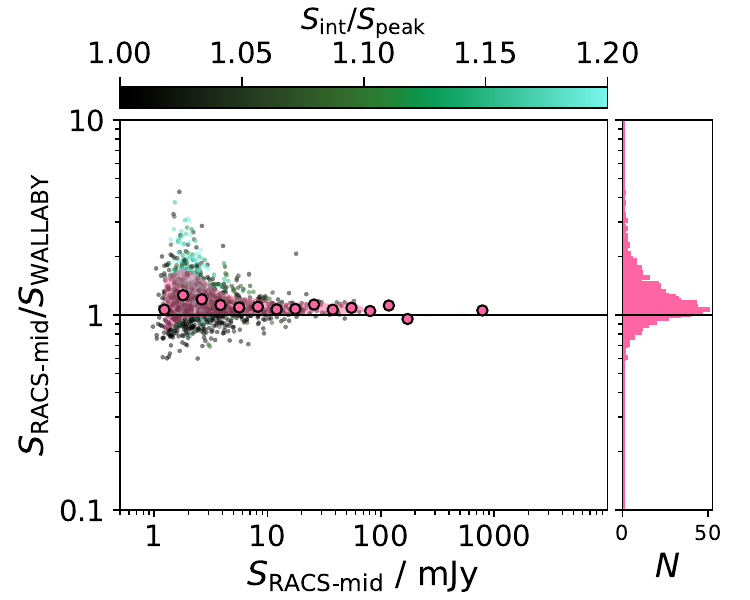}
    \caption{\label{fig:snr:25:wallaby} 25-arcsec catalogue and WALLABY.}
    \end{subfigure}%
    \begin{subfigure}[b]{0.32\linewidth}
    \includegraphics[width=1\linewidth]{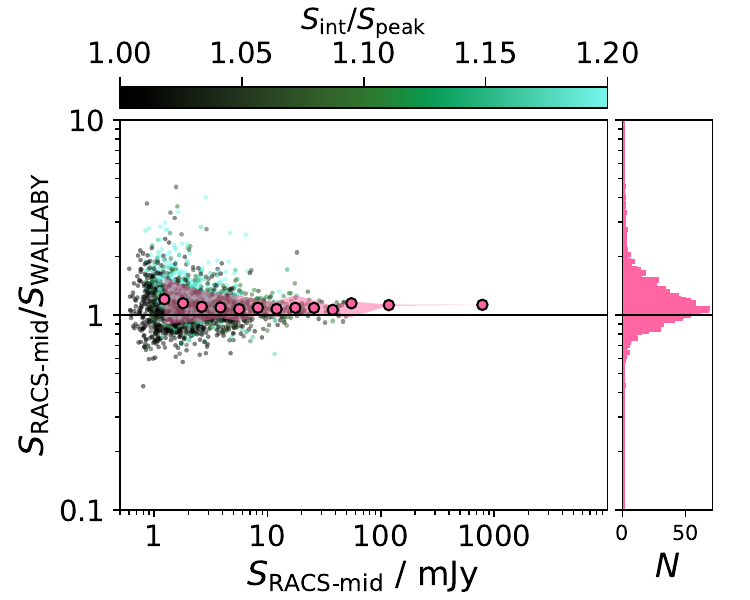}
    \caption{\label{fig:snr:td:wallaby} Time-domain catalogue and WALLABY.}
    \end{subfigure}\\%
    \caption{\label{fig:snr} Flux density ratio of unresolved sources cross-matched in the RACS-mid catalogues and the NVSS (\textit{top panels}), FIRST (\textit{centre panels}), and the WALLABY pre-pilot catalogue (\textit{bottom panels}) as a function of the RACS-mid flux density measurements. Background data points are coloured by their SNR in the RACS-mid data. The solid, black line in each panel indicates a ratio of 1. Pink points indicate median flux density ratios in bins, with the shaded pink region covering the $16^\text{th}$ and $84^\text{th}$ percentiles in each bin.}
\end{figure*}

For a direct brightness scale comparison, Figure~\ref{fig:snr} shows the ratio of integrated flux densities of sources cross-matched between NVSS, FIRST, and the WALLABY pre-pilot catalogue for the three RACS-mid catalogues independently as a function of the RACS-mid flux density. Sources in Figure~\ref{fig:snr} are coloured by $S_\text{int}/S_\text{peak}$ as measured in the respective RACS-mid catalogues. Generally $S_\text{int}/S_\text{peak}$ has a restricted range as these are intended to be unresolved sources as defined in Section~\ref{sec:resolved}. Figure~\ref{fig:snr} also displays a histogram of the distribution of the flux density ratios.

The results for the three RACS-mid catalogues are generally similar, and we report in Table~\ref{tab:photo} the median flux density ratios for each cross-match alongside the median flux density ratio after applying a 10\,mJy limit to the RACS-mid catalogue. For all three RACS-mid catalogues, the 10-mJy limit provides median ratios of $\sim 0.99$--$1.00$ for the NVSS and FIRST cross-matches broadly consistent with \citetalias{racs-mid}. The RACS-mid data shows marginally higher flux densities ($\sim 10$\%) than the WALLABY pre-pilot data, though the WALLABY pre-pilot data is already found to be underluminous with respect to the NVSS with additional positional-dependent variation \citep{For2021,Grundy2023}. The full sample shows more variation, with the low-brightness flux density ratios showing systematic deviation, different for the low-resolution NVSS and high-resolution FIRST and WALLABY pre-pilot data. In the case of the NVSS, the low-brightness cross-matches tend towards $<1$, which is a result of faint sources in the NVSS catalogue being pushed above the detection threshold due to scatter in the flux densities (i.e. Eddington bias). Conversely, the low-brightness end of the FIRST and WALLABY pre-pilot cross-matches shows a bump $>1$ in the ratios before also dropping $<1$. The increase in flux density as measured in RACS-mid can be explained by an increase in truly extended sources being cross-matched, which are not as well-detected in FIRST. This indicates our criteria for `unresolved' sources defined in Section~\ref{sec:resolved} has overestimated the true unresolved fraction for sources $<10$\,mJy.    

\subsubsection{Spectral indices}\label{sec:spectra}

\begin{figure*}[p]
    \centering
    \begin{subfigure}[b]{0.485\linewidth}
    \includegraphics[width=1\linewidth]{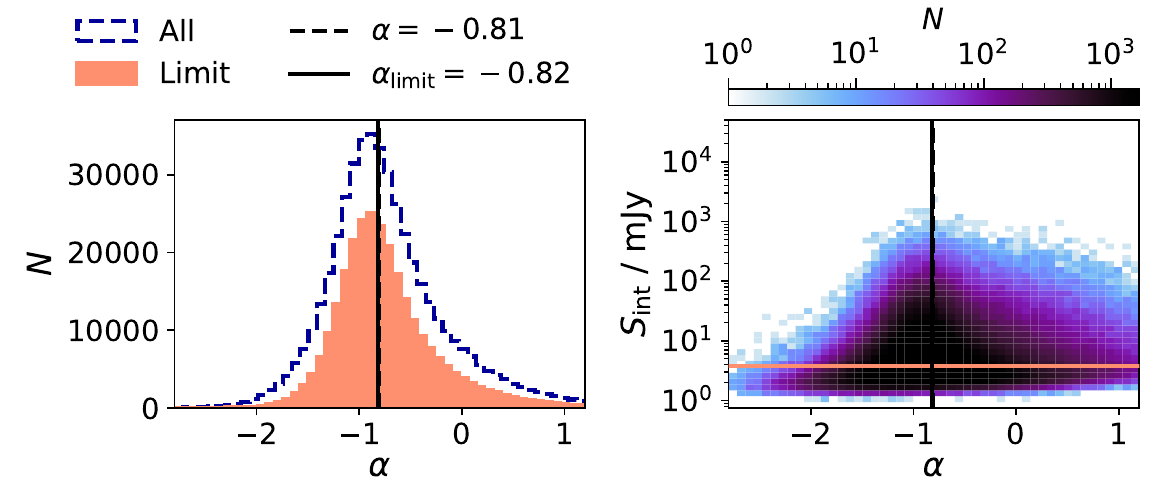}
    \caption{\label{fig:spindex:fr:racs} Primary catalogue and RACS-low.}
    \end{subfigure}\hfill%
    \begin{subfigure}[b]{0.485\linewidth}
    \includegraphics[width=1\linewidth]{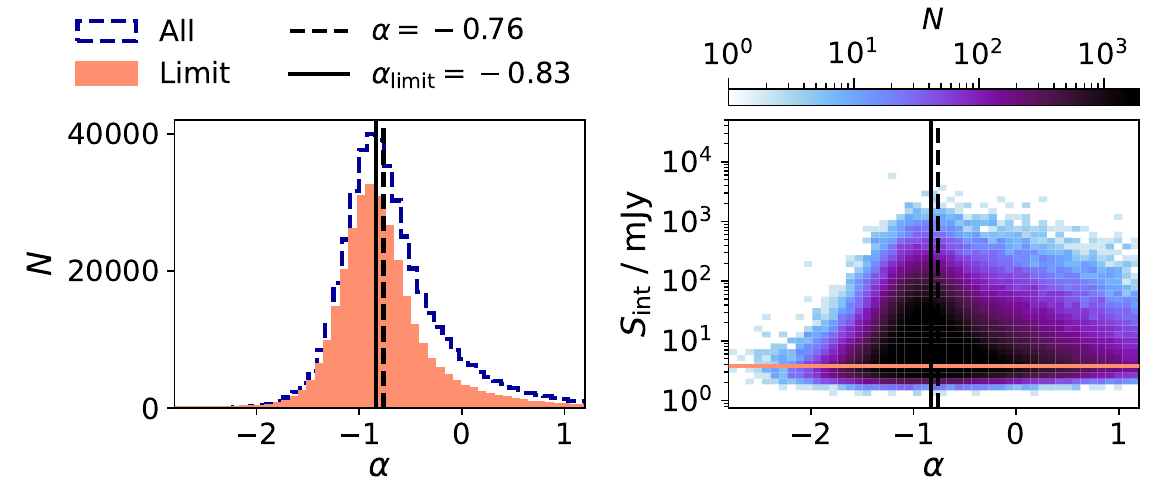}
    \caption{\label{fig:spindex:25:racs} 25-arcsec catalogue and RACS-low.}
    \end{subfigure}\\[0.5em]%
    \begin{subfigure}[b]{0.485\linewidth}
    \includegraphics[width=1\linewidth]{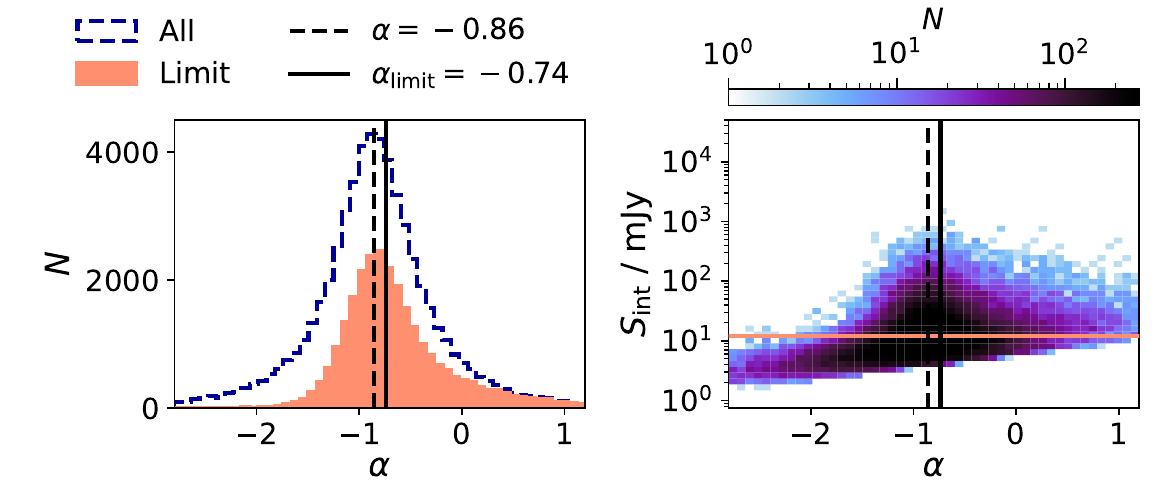}
    \caption{\label{fig:spindex:fr:sumss} Primary catalogue and SUMSS.}
    \end{subfigure}\hfill%
    \begin{subfigure}[b]{0.485\linewidth}
    \includegraphics[width=1\linewidth]{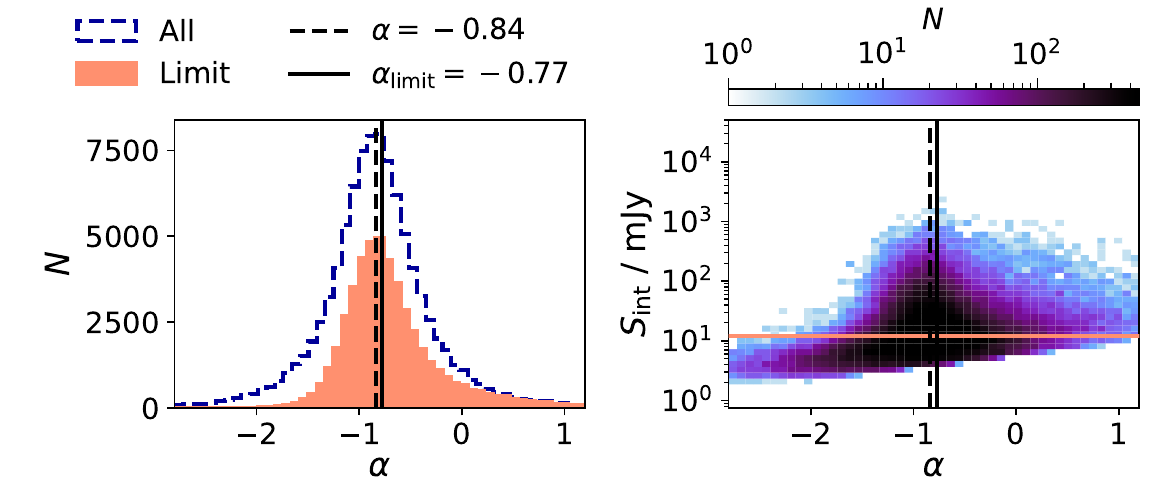}
    \caption{\label{fig:spindex:25:sumss} 25-arcsec catalogue and SUMSS.}
    \end{subfigure}\\[0.5em]%
    \begin{subfigure}[b]{0.485\linewidth}
    \includegraphics[width=1\linewidth]{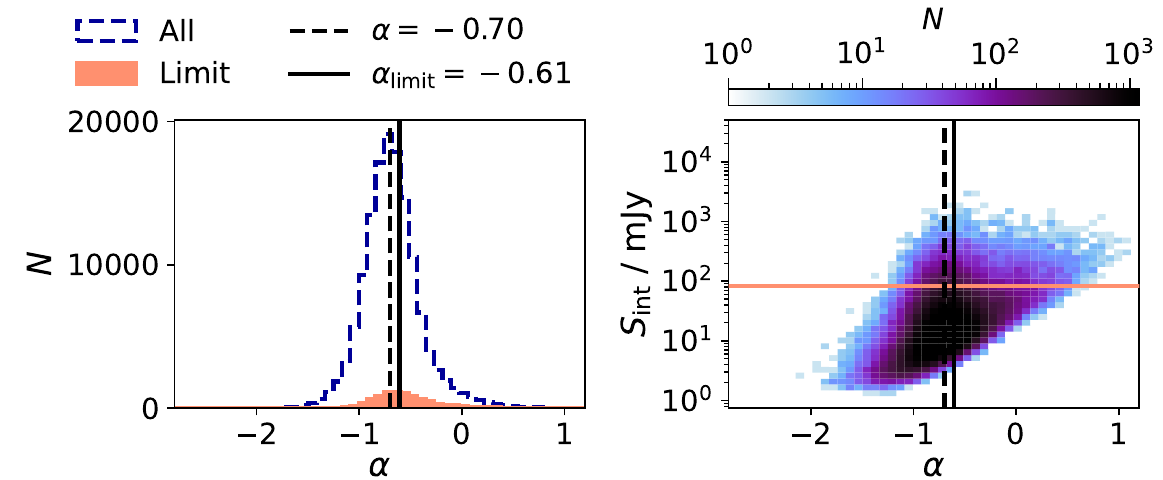}
    \caption{\label{fig:spindex:fr:tgss} Primary catalogue and TGSS.}
    \end{subfigure}\hfill%
    \begin{subfigure}[b]{0.485\linewidth}
    \includegraphics[width=1\linewidth]{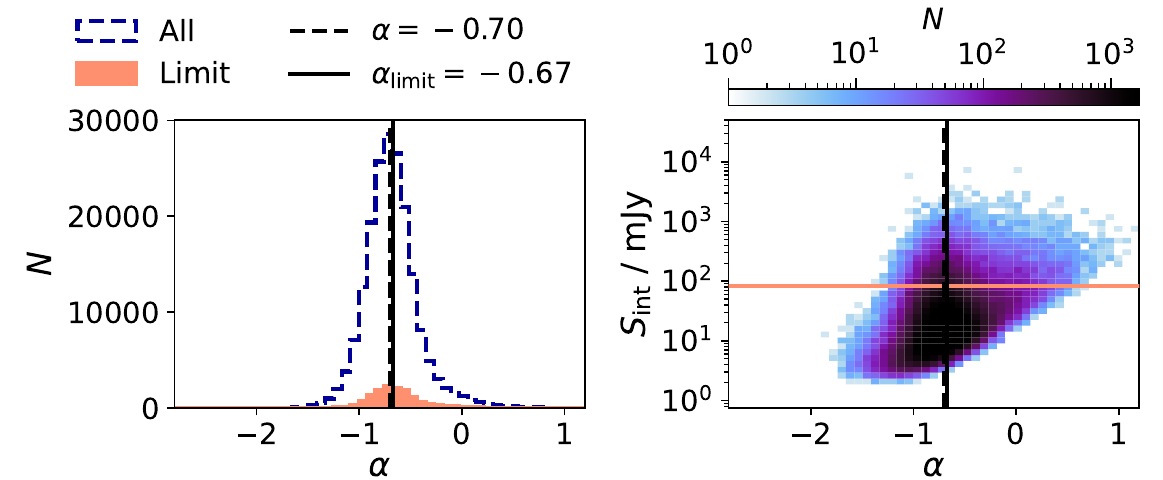}
    \caption{\label{fig:spindex:25:tgss} 25-arcsec catalogue and TGSS.}
    \end{subfigure}\\[0.5em]%
    \begin{subfigure}[b]{0.485\linewidth}
    \includegraphics[width=1\linewidth]{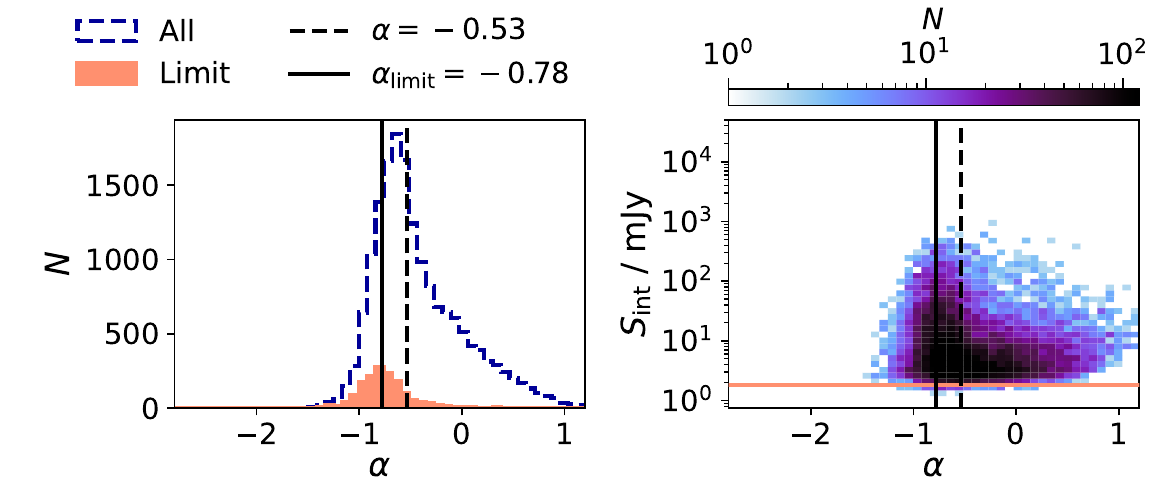}
    \caption{\label{fig:spindex:fr:lotss} Primary catalogue and LoTSS.}
    \end{subfigure}\hfill%
    \begin{subfigure}[b]{0.485\linewidth}
    \includegraphics[width=1\linewidth]{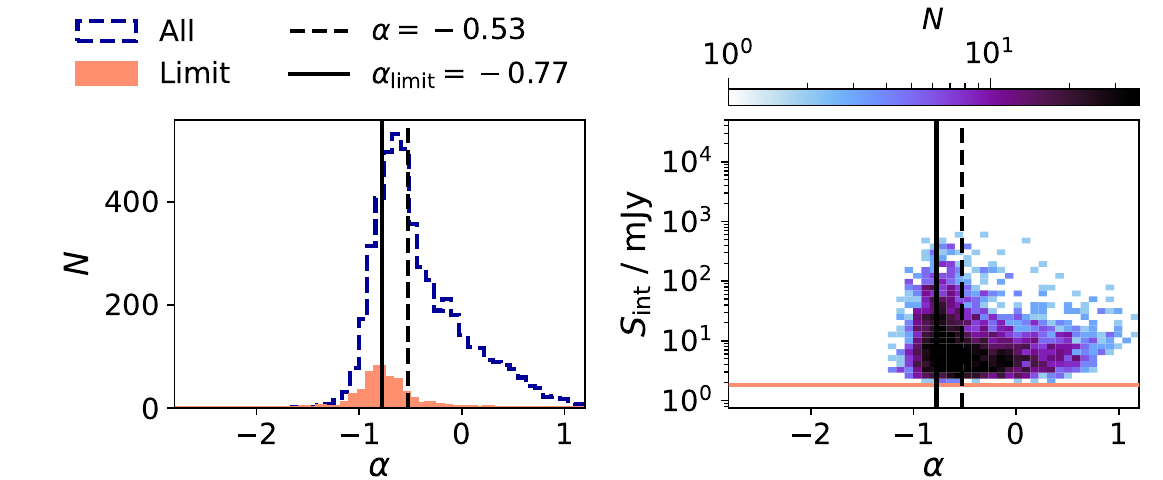}
    \caption{\label{fig:spindex:25:lotss} 25-arcsec catalogue and LoTSS.}
    \end{subfigure}\\[0.5em]%
    \caption{\label{fig:spindex} Spectral indices of sources cross-matched between the RACS-mid and RACS-low \subref{fig:spindex:fr:racs}--\subref{fig:spindex:25:racs}, SUMSS \subref{fig:spindex:fr:sumss}--\subref{fig:spindex:25:sumss}, TGSS-ADR1 \subref{fig:spindex:fr:tgss}--\subref{fig:spindex:25:tgss}, and LoTSS-DR2 \subref{fig:spindex:fr:lotss}--\subref{fig:spindex:25:lotss} for both the primary catalogue (\emph{left}) and the 25-arcsec catalogue (\emph{right}). \emph{Left panels.} Histogram of $\alpha$ with all sources (blue, dashed) and with flux density limits (orange, filled). Median $\alpha$ values are also shown. \emph{Right panels.} 2-D histogram of RACS-mid flux density measurements against spectral index. The  red line shows the RACS-mid flux density limit for the given catalogue comparison.}
\end{figure*}

For comparisons with surveys at significantly different frequencies, we follow \citetalias{racs2} and opt to assess the spectral indices of sources cross-matched between the RACS-mid primary catalogue and 25-arcsec catalogue and other surveys. For this purpose, we use {the} RACS-low catalogue \citepalias{racs2}, the SUMSS catalogue \citep{mmb+03}, the TGSS ADR1 \citep{ijmf16}, and the LoTSS-DR2 catalogue \citep{lotss:dr2}. A two-point spectral index, $\alpha$, is calculated for cross-matched sources assuming a power law, following \begin{equation}
\alpha = \frac{\log_\text{10}\left(S_\text{RACS-mid} / S_\text{survey}\right)}{\log_\text{10}\left(1367.5 / \nu_\text{survey}\right)} \, ,
\end{equation}
where $S_\text{RACS-mid}$ and $S_\text{survey}$ are the integrated flux densities of the given source in the RACS-mid catalogue and comparison survey, respectively, and $\nu_\text{survey}$ is the survey frequency. 

Following \citetalias{racs2} we also create a subsample by applying a flux density limit to each survey to remove some sensitivity bias---this helps reduce artificial flattening or steepening of the median spectra. We assume limits in the range $\alpha = -0.8\pm1.2$ and set a flux density limit for both the RACS-mid catalogues and comparison catalogues to be either side of that range, depending on which survey is higher in frequency. The resultant limits are reported in Table~\ref{tab:photo}. Table~\ref{tab:photo} also reports the median spectral indices for the full sample and flux-limited samples for each RACS-mid catalogue and comparison survey. We also plot the histogram of spectral indices (with and without the flux density limit) in Figure~\ref{fig:spindex} alongside a 2-D histogram showing RACS-mid flux density as a function of spectral index. The flux-limited subsample is included in the histograms to highlight the reduced number of sources, and the RACS-mid flux density limit is also shown in 2-D histogram for reference (though note the comparison survey limit is \textit{not} shown). {Where relevant the spectral indices of sources matched between catalogues are generally consistent with results from \citetalias{racs-mid}.}

\subsection{Astrometry}

\begin{figure}[t!]
    \centering
    \begin{subfigure}[b]{1\linewidth}
    \includegraphics[width=1\linewidth]{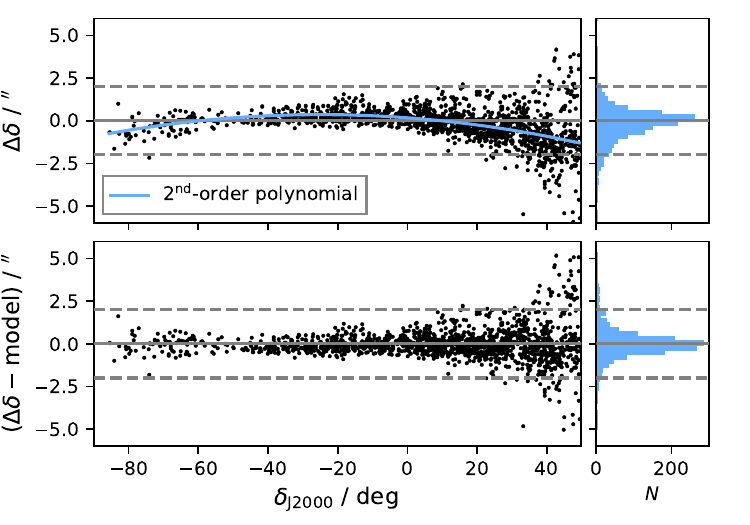}
    \caption{\label{fig:astrometry:dec:fr} Primary catalogue $-$ ICRF.}
    \end{subfigure}\\[2em]%
     \begin{subfigure}[b]{1\linewidth}
    \includegraphics[width=1\linewidth]{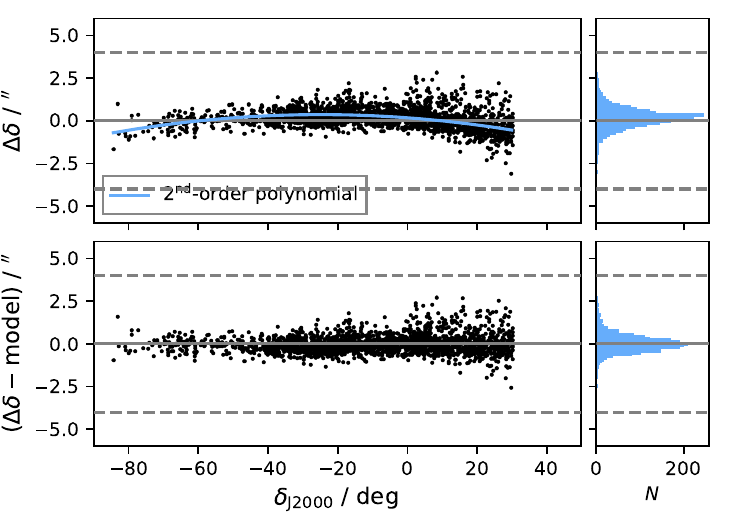}
    \caption{\label{fig:astrometry:dec:25} 25-arcsec catalogue $-$ ICRF.}
    \end{subfigure}\\[2em]%
     \begin{subfigure}[b]{1\linewidth}
    \includegraphics[width=1\linewidth]{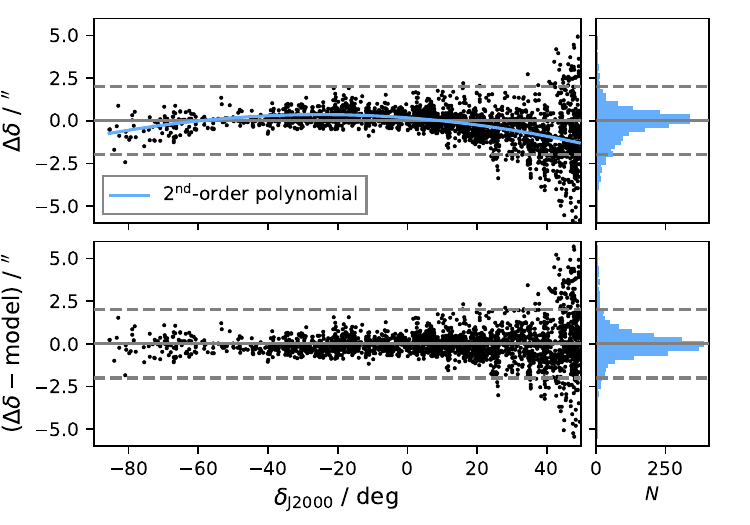}
    \caption{\label{fig:astrometry:dec:td} Time-domain catalogue $-$ ICRF.}
    \end{subfigure}\\[1em]%
    \caption{\label{fig:astrometry:dec} Declination ($\Delta\delta$) offsets from the ICRF3 as a function of declination for the primary catalogue \subref{fig:astrometry:dec:fr}, the auxiliary 25-arcsec catalogue \subref{fig:astrometry:dec:25}, and the auxiliary time-domain catalogue \subref{fig:astrometry:dec:td}. The fitted declination-dependent polynomial, derived from the auxiliary time-domain catalogue for all catalogues, is shown on the upper panels, and the lower panels show the residual offsets after subtraction of the derived model.}
\end{figure}

\begin{figure*}[t!]
    \centering
    \begin{subfigure}[b]{0.25\linewidth}
    \includegraphics[width=1\linewidth]{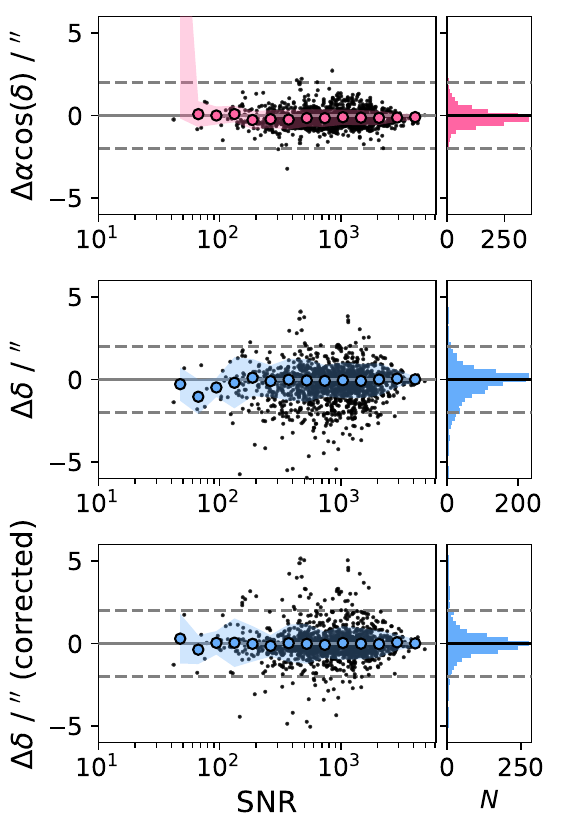}
    \caption{\label{fig:astrometry:snr:fr:icrf} Primary catalogue $-$ ICRF .}
    \end{subfigure}%
    \begin{subfigure}[b]{0.25\linewidth}
    \includegraphics[width=1\linewidth]{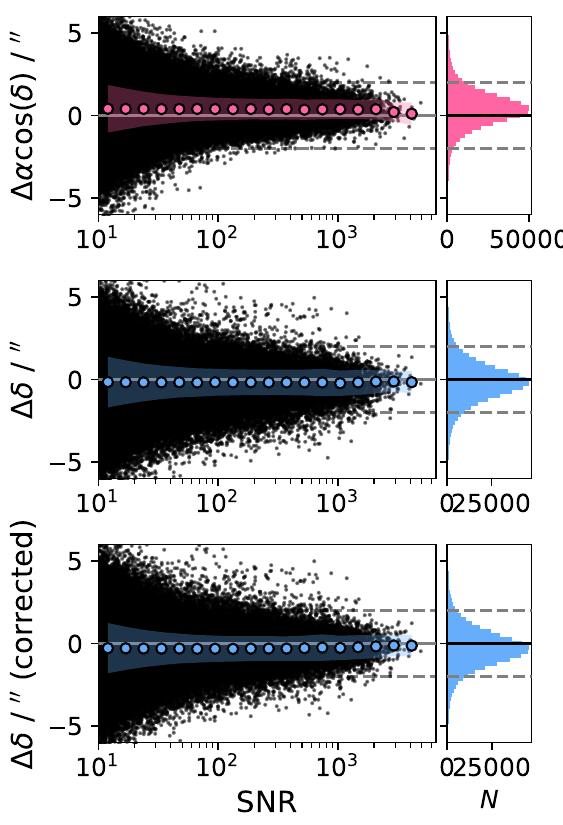}
    \caption{\label{fig:astrometry:snr:fr:racs} Primary catalogue $-$   RACS-low.}
    \end{subfigure}%
    \begin{subfigure}[b]{0.25\linewidth}
    \includegraphics[width=1\linewidth]{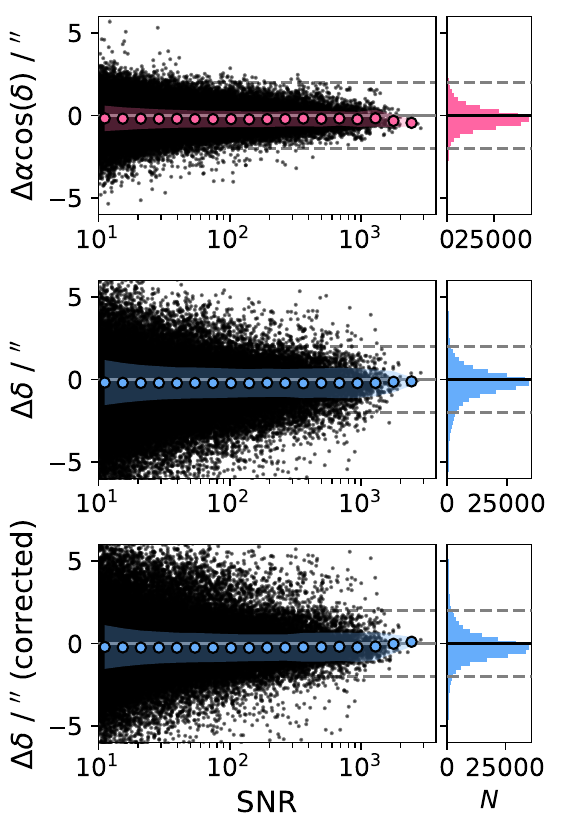}
    \caption{\label{fig:astrometry:snr:fr:vlass} Primary catalogue $-$ VLASS-QL.}
    \end{subfigure}%
    \begin{subfigure}[b]{0.25\linewidth}
    \includegraphics[width=1\linewidth]{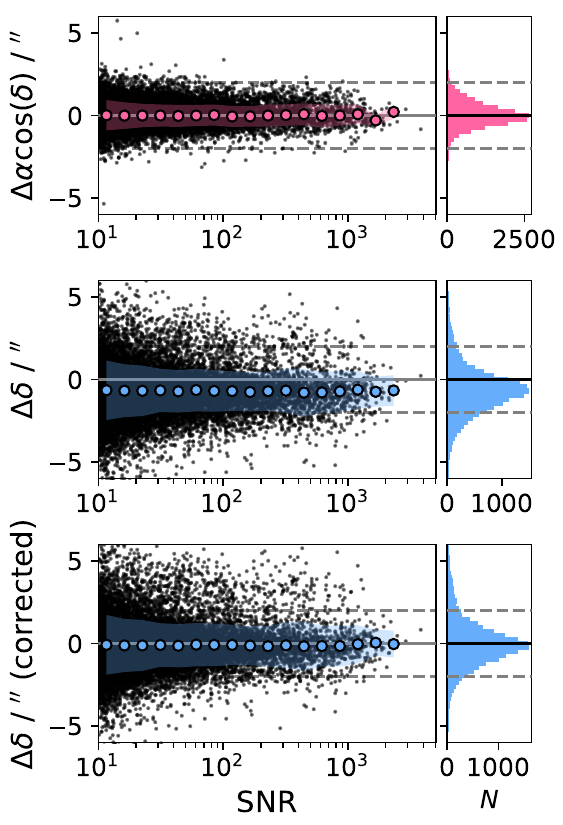}
    \caption{\label{fig:astrometry:snr:fr:lotss} Primary catalogue $-$  LoTSS-DR2.}
    \end{subfigure}\\[1em]%
    
    \begin{subfigure}[b]{0.25\linewidth}
    \includegraphics[width=1\linewidth]{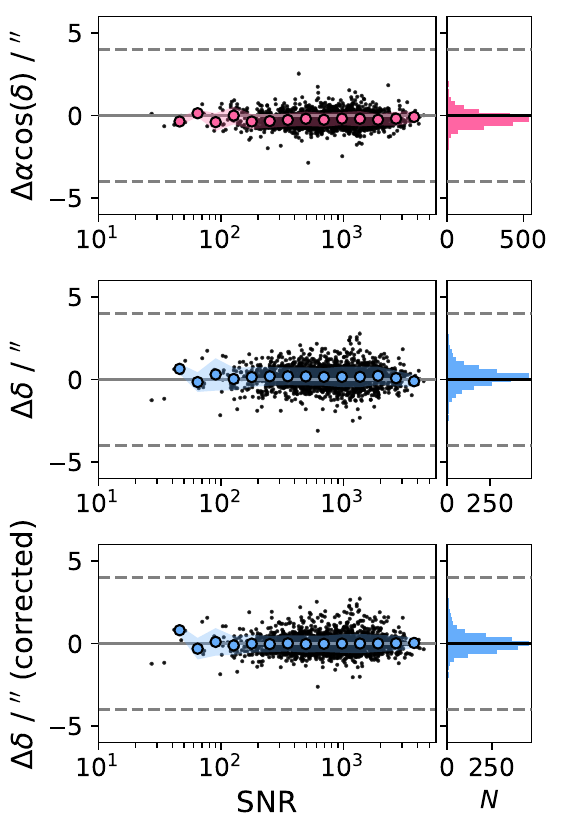}
    \caption{\label{fig:astrometry:snr:25:icrf} 25-arcsec $-$ ICRF .}
    \end{subfigure}%
    \begin{subfigure}[b]{0.25\linewidth}
    \includegraphics[width=1\linewidth]{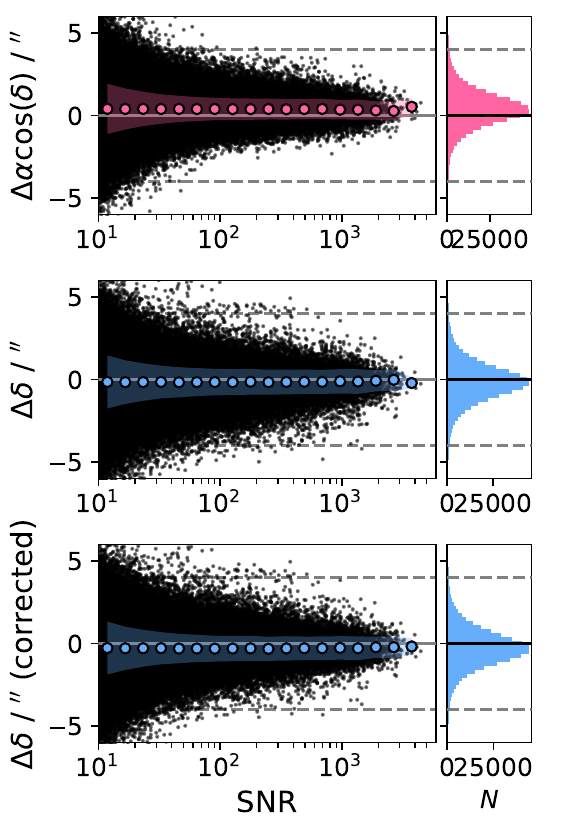}
    \caption{\label{fig:astrometry:snr:25:racs} 25-arcsec $-$   RACS-low.}
    \end{subfigure}%
    \begin{subfigure}[b]{0.25\linewidth}
    \includegraphics[width=1\linewidth]{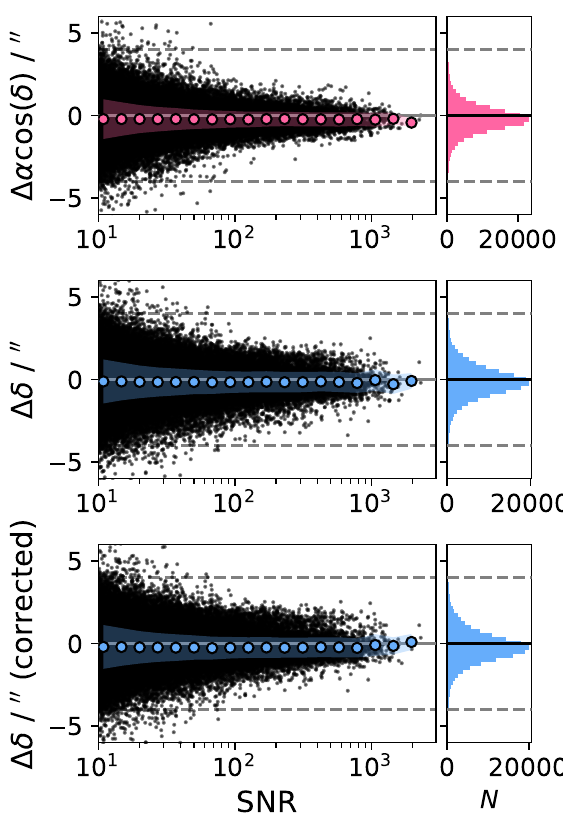}
    \caption{\label{fig:astrometry:snr:25:vlass} 25-arcsec $-$ VLASS-QL.}
    \end{subfigure}%
    \begin{subfigure}[b]{0.25\linewidth}
    \includegraphics[width=1\linewidth]{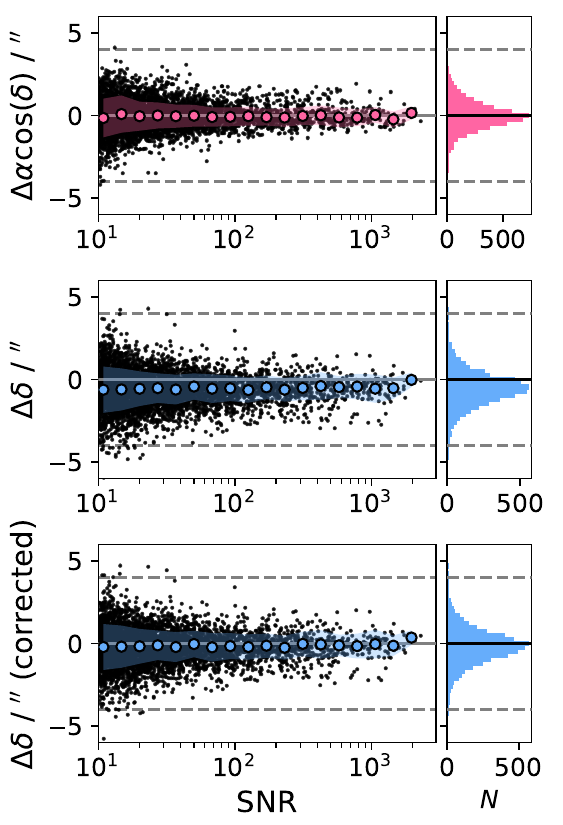}
    \caption{\label{fig:astrometry:snr:25:lotss} 25-arcsec $-$  LoTSS-DR2.}
    \end{subfigure}\\[1em]%

    \caption{\label{fig:astrometry:snr} R.~A. ($\Delta\alpha\cos\left(\delta\right)$) and declination ($\Delta\delta$) offsets as a function of SNR for the primary catalogue [\subref{fig:astrometry:snr:fr:icrf}--\subref{fig:astrometry:snr:fr:lotss}] and the 25-arcsec catalogue  [\subref{fig:astrometry:snr:25:icrf}--\subref{fig:astrometry:snr:25:lotss}]. \emph{Left.} ICRF cross-match. \emph{Centre-left.} RACS-low cross-match. \emph{Centre-right.} VLASS-QL cross-match. \emph{Right.} LoTSS-DR2 cross-match. In each panel, a dashed, grey line is drawn at $\pm2$\,arcsec corresponding to the pixel size of the \textit{original} RACS-mid images. Median values in SNR bins are also shown as coloured markers and the shaded regions indicate $\pm2\sigma$ for the SNR bins.}
\end{figure*}

\begin{table}[t!]
    \centering
    \caption{\label{tab:astrometry} Astrometric offsets of sources cross-matched between RACS-mid and external catalogues, with offsets defined as $\Delta\delta = \delta_\text{RACS-mid}-\delta_\text{survey}$ and $\Delta\alpha\cos\delta = (\alpha_\text{RACS-mid} - \alpha_\text{survey})\cos\delta$ for declination and R.~A. offsets, respectively.}
    \begin{tabular}{cc c c c}\toprule
         Survey & $N_\text{sources}$ & $\Delta\alpha\cos\delta$ & $\Delta\delta$ & $\Delta\delta_\text{corr}$ \\
          & & (arcsec) & (arcsec) & (arcsec) \\[0.5em]\midrule
          \multicolumn{5}{c}{Primary catalogue} \\[0.5em]\midrule
         ICRF3 & 1\,456 & $-0.14\pm0.65$ & $-0.04\pm1.08$ & $-0.03\pm0.97$ \\
         VLASS-QL & 237\,119 & $-0.20\pm0.61$ & $-0.21\pm1.07$ & $-0.25\pm1.02$ \\
         NVSS & 455\,474 & $-0.19\pm3.82$ & $-0.18\pm4.30$ & $-0.10\pm4.27$ \\
         FIRST & 131\,289 & $-0.09\pm0.61$ & $-0.18\pm1.15$ & $-0.10\pm1.07$ \\
         RACS-low & 464\,286 & $+0.39\pm1.07$ & $-0.17\pm1.17$ & $-0.30\pm1.15$ \\
         SUMSS & 57\,052 & $+0.63\pm2.89$ & $+0.06\pm4.06$ & $+0.07\pm4.05$ \\
         LoTSS-DR2 & 16\,840 & $-0.00\pm0.75$ & $-0.68\pm1.49$ & $-0.10\pm1.50$ \\[0.5em]\midrule
         \multicolumn{5}{c}{25-arcsec catalogue} \\[0.5em]\midrule
         ICRF3 & 2\,074 & $-0.23\pm0.42$ & $+0.17\pm0.58$ & $-0.01\pm0.53$ \\
         VLASS-QL & 160\,696 & $-0.22\pm0.83$ & $-0.15\pm0.97$ & $-0.24\pm0.96$ \\
         NVSS & 444\,206 & $-0.26\pm3.18$ & $+0.07\pm3.67$ & $-0.06\pm3.66$ \\
         FIRST & 84\,267 & $-0.14\pm0.80$ & $-0.06\pm0.94$ & $-0.05\pm0.90$ \\
         RACS-low & 451\,954 & $+0.39\pm1.09$ & $-0.15\pm1.18$ & $-0.29\pm1.17$ \\
         SUMSS & 94\,497 & $+0.64\pm2.72$ & $+0.14\pm3.95$ & $+0.08\pm3.94$ \\
         LoTSS-DR2 & 4\,973 & $-0.05\pm0.87$ & $-0.55\pm1.04$ & $-0.16\pm1.05$ 
         \\[0.5em]\midrule
         \multicolumn{5}{c}{Time-domain catalogue} \\[0.5em]\midrule
         ICRF3 & 2\,276 & $-0.17\pm0.51$ & $-0.06\pm1.16$ & $-0.05\pm1.04$ \\
         VLASS-QL & 137\,147 & $-0.19\pm0.59$ & $-0.22\pm0.95$ & $-0.25\pm0.89$ \\
         NVSS & 249\,104 & $-0.18\pm4.21$ & $-0.15\pm4.65$ & $-0.10\pm4.63$ \\
         FIRST & 75\,121 & $-0.08\pm0.59$ & $-0.22\pm1.03$ & $-0.13\pm0.90$ \\
         RACS-low & 268\,310 & $+0.41\pm1.12$ & $-0.20\pm1.21$ & $-0.32\pm1.19$ \\
         SUMSS & 26\,534 & $+0.65\pm2.97$ & $+0.04\pm4.20$ & $+0.03\pm4.19$ \\
         LoTSS-DR2 & 10\,334 & $-0.01\pm0.69$ & $-0.68\pm1.29$ & $-0.11\pm1.27$ \\[0.5em]\bottomrule
    \end{tabular}
    
\end{table}

As noted in \citetalias{racs-mid}, there are two astrometric errors that exist in the RACS-mid data (see figures 36 and 38 in \citetalias{racs-mid}). Firstly, the self-calibration process, performed independently for each PAF beam image prior to mosaicking, introduces up to pixel-scale offsets in both right ascension (R.~A.) and declination. This effect is not dependent on the SNR of the sources in the field. The result of this effect is an uncertainty in the astrometric accuracy up to $\pm 1$--2~arcsec. The second effect is a systematic declination-dependent offset in declination which is currently undiagnosed. This effect results in up to $\sim 2$\,arcsec offset at the highest declination of the survey. While we cannot reasonably correct the beam-to-beam astrometric precision at this stage, we do attempt to remove the systematic declination-dependent offets.

For removal of the declination-dependent declination offset, we cross-match the unresolved time-domain RACS-mid catalogue to sources in the third realisation of the International Celestial Reference Frame \citep[ICRF3;][]{Charlot2020}. The time-domain catalogue is used for this purpose to take advantage of sources in the overlap regions. Since the ICRF3 catalogue is sparse compared to RACS-mid, we do not use the isolated source restriction. We define the declination offsets as \begin{equation}
\Delta\delta = (\delta_\text{RACS-mid} - \delta_\text{ICRF3}) \quad \text{arcsec} ,
\end{equation} and fit a range of generic polynomials covering $0^\text{th}$ to $4^\text{th}$ orders. We use the SNR of the RACS-mid flux density measurement to weight the fitting procedure, and use the Aikaike Information Criterion \citep[AIC;][]{Akaike1974} as a means to select the appropriate order polynomial. As initial polynomial fitting only considers measurement uncertainties in $\delta_\text{J2000}$ reported by \texttt{PyBDSF}, we estimate uncertainties on the polynomial fit by repeating the fitting procedure after adding Gaussian noise to the model offsets. The added noise is drawn from a Gaussian distribution, assuming the mean and standard deviation of $\Delta\delta$ ($-0.26$ and $1.16$\,arcsec, respectively). With this, we find a $2^\text{nd}$-order polynomial of the form \begin{equation}\label{eq:decoffset}
 \Delta\delta = 0.175(34) - 0.0150(9)\delta - 0.000301(18) \delta^2  \quad \text{arcsec} ,
\end{equation}
where $\delta={\delta_\text{J2000}}^\circ$ and is the declination of the source in the RACS-mid catalogue. 

Figures~\ref{fig:astrometry:dec:fr}--\ref{fig:astrometry:dec:td} shows the RACS-mid $--$ ICRF3 declination offsets as a function of declination in the top panels, with medians in declination bins shown as blue circles. The upper panels highlight both the lower astrometric precision at high declination (above $\delta_\text{J2000} \approx 20^\circ$) and the general trend towards negative offsets at high- and low-declination. We show on the upper panels the selected $2^\text{nd}$-order polynomial, derived from the time-domain catalogue, and the lower panels show the same data but with removal of the offset model (i.e. the `corrected' declination). We report the median $\Delta\alpha$ and $\Delta\delta$ against the ICRF3 in Table~\ref{tab:astrometry} along with $1\sigma$ uncertainties. These remain similar to results from \citetalias{racs-mid}. We also report the median declination offset after applying the model correction---this results in no substantial difference to the overall median and uncertainty. 

Table~\ref{tab:astrometry} also reports similar median offsets for comparisons with a range of other surveys, both with and without the declination correction. Notably, the declination correction does not make a large difference in most cases as the correction is close to the level of the uncertainty (largely driven by the aforementioned beam-to-beam variation). The exception to this is for the comparison to LoTSS-DR2. For LoTSS-DR2, sky coverage begins at $\delta_\text{J2000}>15^\circ$ so suffers from the most extreme declination offsets. A column containing the `corrected' declination is added to all RACS-mid catalogues, though we retain the original declination column as well. Users may find cross-matching to high-declination regions more straightforward with the corrected declination measurements.

In Figure~\ref{fig:astrometry:snr} we show coordinate offsets for the primary and 25-arcsec catalogues against the ICRF3, RACS-low, VLASS-QL, and LoTSS-DR2 as a function of source SNR in the RACS-mid catalogues. In Figure~\ref{fig:astrometry:snr} we also show binned median offsets as a function of SNR, with $1\sigma$ error ranges shown as shaded regions. Note that the uncertainty remains fairly constant as a function of SNR, as mentioned in \citetalias{racs-mid} as the beam-to-beam astrometric offset introduced by self-calibration are direction- and source-independent. The lower panels of Figure~\ref{fig:astrometry:snr} show the declination offsets after the correction described in Equation~\ref{eq:decoffset}.

The overall median offsets for each catalogue comparison are also reported in Table~\ref{tab:astrometry}. Generally absolute median offsets are $\lesssim 0.3$~arcsec with the except of the aforementioned declination offset in LoTSS-DR2 and right ascension offsets with RACS-low ($\sim +0.4$~arcsec) and SUMSS ($\sim +0.6$~arcsec). A right ascension offset for RACS-low was reported by \citetalias{racs2} and is intrinsic to that catalogue. It is likely this is an elevation-dependent offset (the true form of the declination-dependent trend we see here) and with RACS-low being observed off the meridian the offset manifests in both declination and right ascension.

\subsection{RACS-mid and other ASKAP surveys}\label{sec:diffs}

\begin{figure}[t!]
    \centering
    \includegraphics[width=1\linewidth]{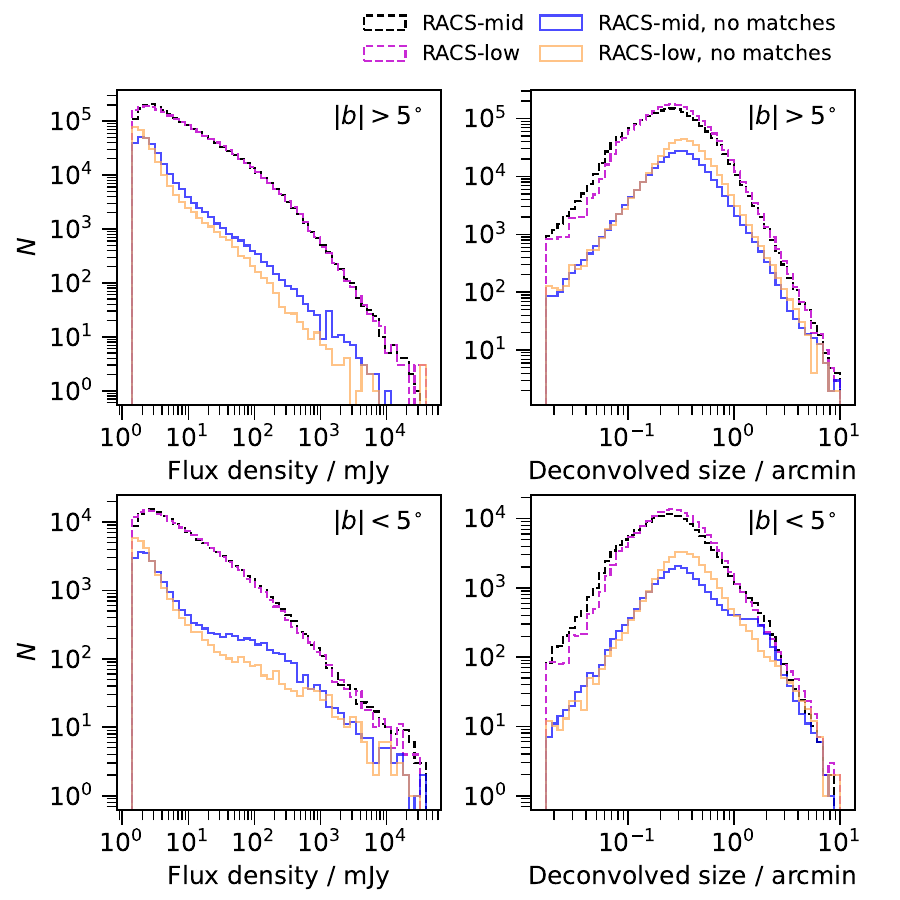}
    \caption{\label{fig:midlow} The distribution of sources as a function of flux density (\emph{top}) and fitted major axis (\emph{bottom}) for the RACS-mid 25-arcsec catalogue and the RACS-low catalogue. \emph{Left.} Sources outside of the Galactic Plane ($|b|>5^\circ$). \emph{Right.} Sources within the Galactic Plane ($|b|<5^\circ$). The distribution of sources that are not cross-matched between the two catalogues are also shown.}
\end{figure}

\begin{figure}[t!]
    \centering
    \includegraphics[width=1\linewidth]{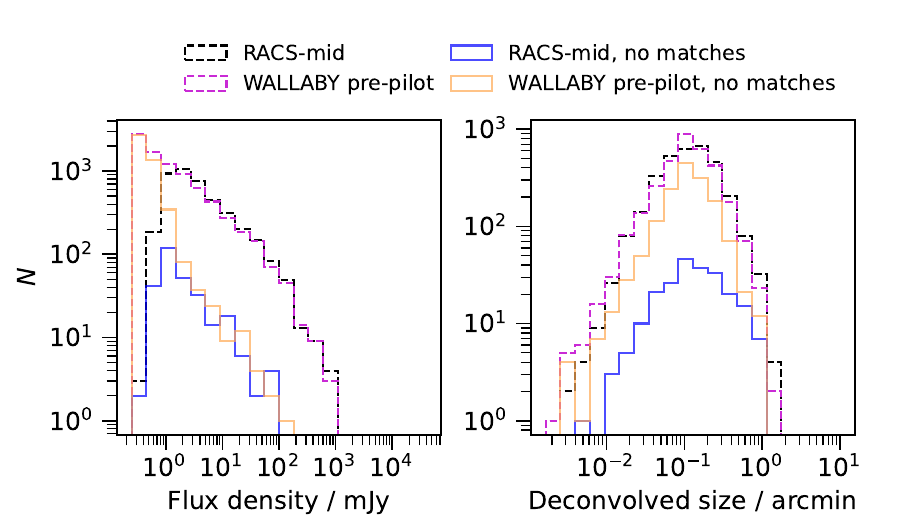}
    \caption{\label{fig:midwallaby} As in Figure~\ref{fig:midlow}, comaparing the RACS-mid primary catalogue to the WALLABY pre-pilot catalogue (only considering sources within the WALLABY pre-pilot image region). No Galactic Plane cut is applied as the region is not within the Galactic Plane.}
\end{figure}

\subsubsection{RACS-low and RACS-mid at 25 arcsec}

The goal of RACS is to create a global sky model for calibration and validation of the other ASKAP surveys and ASKAP observations in general. The global sky model is intended to be a combination of all three RACS bands from 887.5--1632.5\,MHz (low, mid, and high). While we are still awaiting RACS-high images and catalogues, we begin with an initial assessment of a sky model built between 887.5 and 1367.5\,MHz with the RACS-low and RACS-mid 25-arcsec catalogues. We perform simple cross-match between the source catalogues for each survey, taking sources that cross-match within $(\theta_\text{major}\theta_\text{minor})/4$, where $\theta_\text{major}$ and $\theta_\text{minor}$ are the fitted major and minor axes. This variable match criterion is to account for extended sources, which may have different centres in each catalogue. More robust catalogue matching processes may be implemented in future combinations of RACS \citep[e.g. the Positional Update and Matching Algorithm;][]{puma}, taking into account expected spectral information. We scale the flux densities of the RACS-low catalogue to 1367.5\,MHz assuming the flux-limit median $\alpha=-0.83$ (see Section~\ref{sec:spectra}). For this cross-match, we are interested in the region that completely overlaps with both the RACS-low and RACS-mid 25-arcsec catalogues and we only consider sources within the declination range $-80^\circ<\delta_\text{J2000}<+29^\circ$, including the Galactic Plane. Note that the RACS-low catalogue is missing a small region below $\delta_\text{J2000}=+30^\circ$ and has patchy coverage below $\delta_\text{J2000}=-80^\circ$.

Figure~\ref{fig:midlow} shows the distribution of sources of both the RACS-mid and RACS-low 25-arcsec catalogues, as a function of flux density (scaled to 1367.5\,MHz) and a function of the deconvolved source size. Also shown are the distributions of sources in each catalogue which do not match to sources in the other catalogue with the aforementioned cross-matching limit. The distributions of sources in the RACS-mid and RACS-low 25-arcsec catalogues is similar both in flux density and source size. RACS-low sources below $\sim 2$\,mJy (at 1367.5\,MHz) begin to fail to cross-match, which is due to the difference in sensitivity between the two catalogues. Conversely, as flux density increases there are more RACS-mid sources without matches. The Galactic Plane sources show a similar increase, particularly between $\sim 10$--1000\,mJy (and above 1\,arcmin in size), which is a consequence of the increase in artefacts as described in Section~\ref{sec:reliability}. 

\subsubsection{RACS-mid and WALLABY at the full resolution}

We also perform a similar variable-separation cross-match to the WALLABY pre-pilot catalogue \citep{Grundy2023} using the primary RACS-mid catalogue. For this cross-match, we first extract a cut-out of the RACS-mid primary catalogue that covers the same image region of the WALLABY pre-pilot mosaic. The WALLABY pre-pilot data is entirely outside of the Galactic Plane, so no distinction between Galactic latitudes is done for this comparison. Figure~\ref{fig:midwallaby} shows the distributions of sources in RACS-mid primary catalogue and the WALLABY pre-pilot catalogue along with non-matched sources as in Figure~\ref{fig:midlow}. Due to the increase in sensitivity of the WALLABY pre-pilot data ($\sigma_\text{rms} \approx 50$\,\textmu Jy\,PSF$^{-1}$), there are significantly more unmatched WALLABY pre-pilot sources at the low flux density end of the distribution, though no bias towards specific source sizes in the unmatched WALLABY pre-pilot sources. The RACS-mid primary catalogue, at least outside of the Galactic Plane and away from other particularly complex extended sources (e.g. Fornax~A and other large angular extent, nearby radio sources), provides a good interpretation of the 1367.5\,MHz sky.

\subsection{The brightness scale of the full-sensitivity images}\label{sec:tile:brightness}

While the main data products released with this work are the `all-sky' catalogues, we also provide the full-sensitivity, convolved maps generated prior to source-finding. With the exception of the time-domain catalogue (which uses the original images), these new images are mosaicked with nearby observations. This additional mosaicking provides a more uniform sensitivity pattern across the images, and some of the edge effects of each tile image discussed in \citetalias{racs-mid}, particularly for the brightness scale, can be reduced during this process.

\begin{figure*}[t!]
    \centering
    \begin{subfigure}[b]{0.5\linewidth}
    \includegraphics[width=1\linewidth]{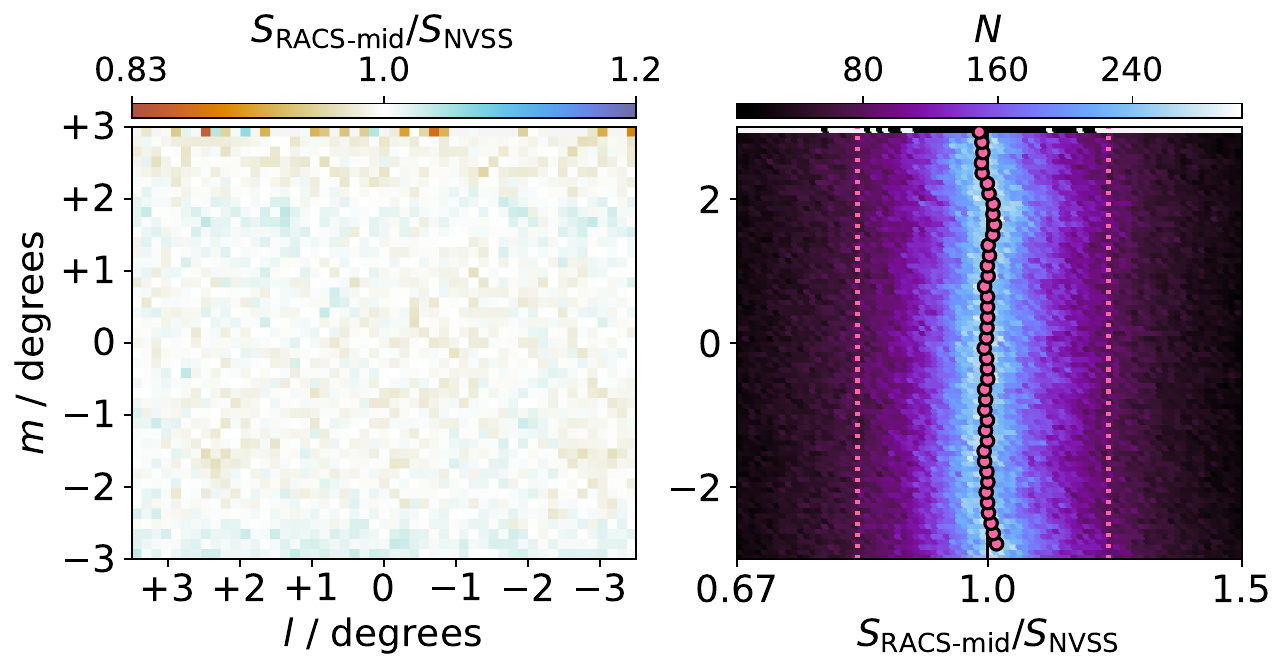}
    \caption{\label{fig:tileflux:nvss:fr} Primary / NVSS.}
    \end{subfigure}%
    \begin{subfigure}[b]{0.5\linewidth}
    \includegraphics[width=1\linewidth]{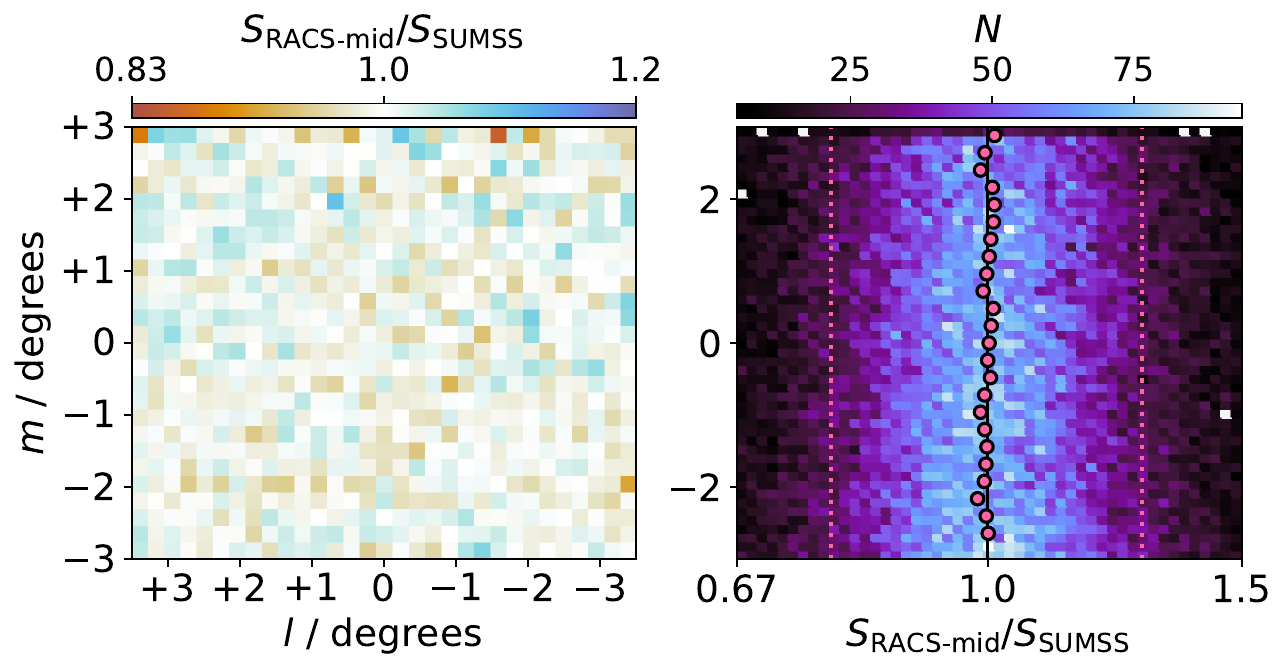}
    \caption{\label{fig:tileflux:sumss:fr} Primary / SUMSS.}
    \end{subfigure}\\%
    \begin{subfigure}[b]{0.5\linewidth}
    \includegraphics[width=1\linewidth]{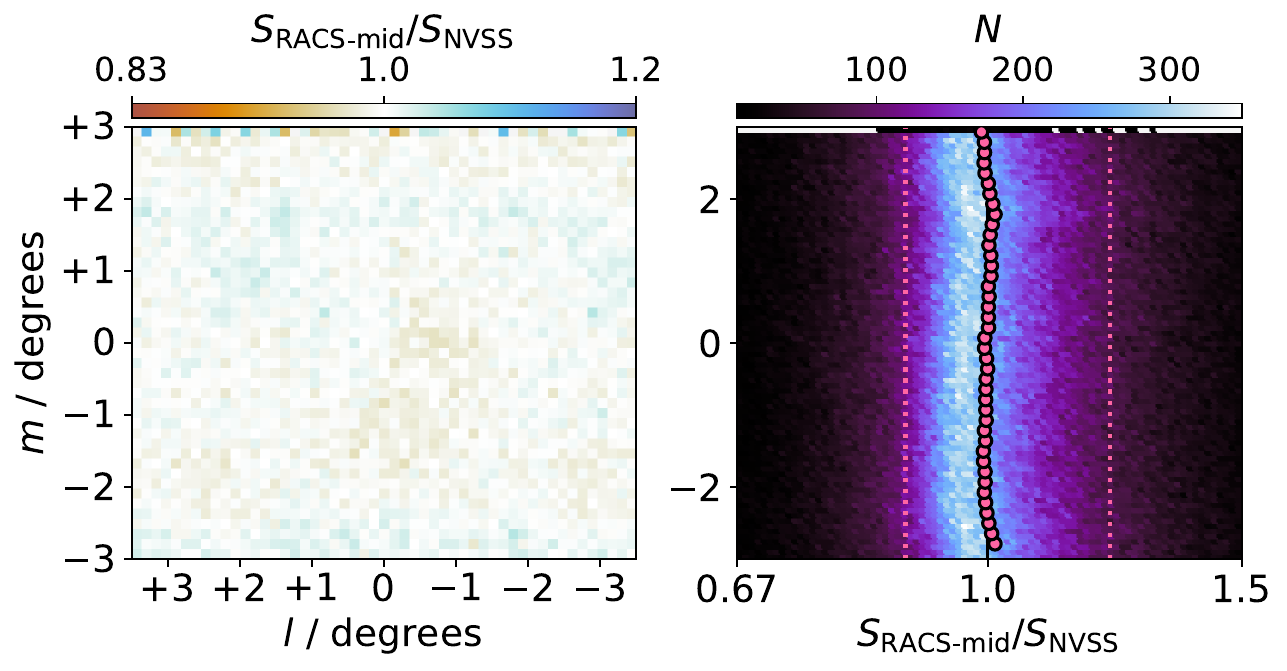}
    \caption{\label{fig:tileflux:nvss:25arcsec} 25-arcsec / NVSS.}
    \end{subfigure}%
    \begin{subfigure}[b]{0.5\linewidth}
    \includegraphics[width=1\linewidth]{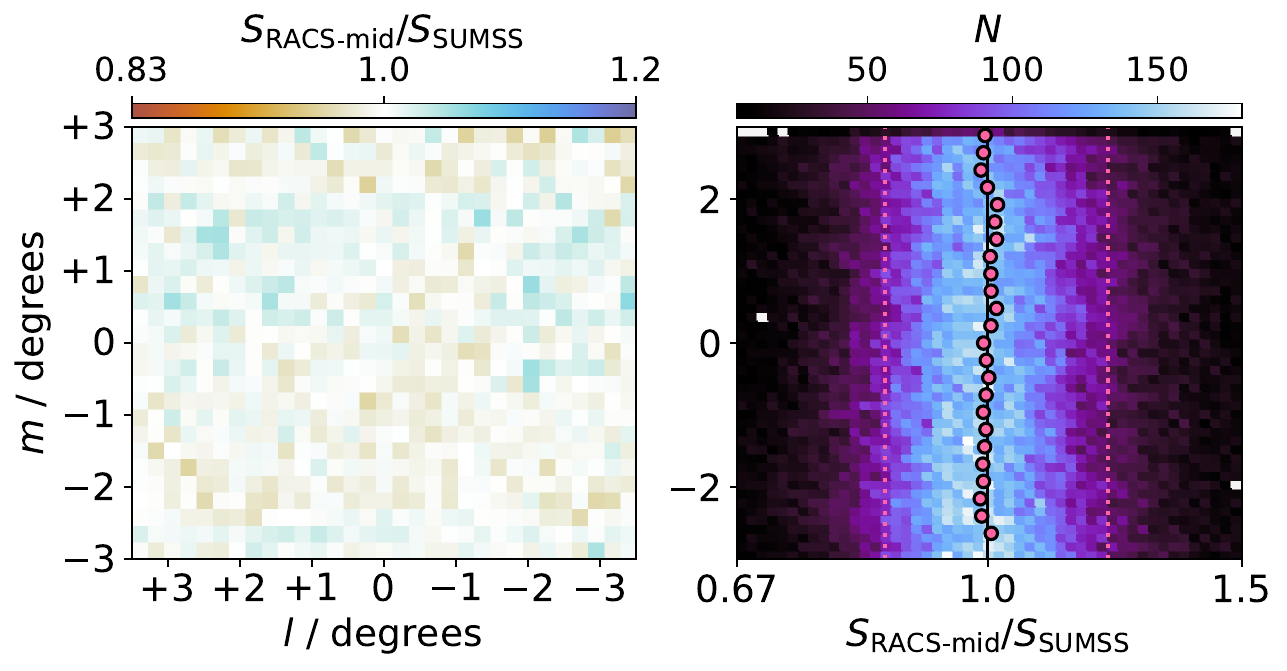}
    \caption{\label{fig:tileflux:sumss:25arcsec} 25-arcsec / SUMSS.}
    \end{subfigure}\\%
    \caption{\label{fig:tileflux} Brightness scale comparison as a function of tile reference coordinates for the primary catalogue and NVSS \subref{fig:tileflux:nvss:fr} and SUMSS \subref{fig:tileflux:sumss:fr}, and the 25-arcsec catalogue and NVSS \subref{fig:tileflux:nvss:25arcsec} and SUMSS \subref{fig:tileflux:sumss:25arcsec}. \emph{Left.} Median-binned flux density ratios. \emph{Right.} 2-D histogram of the flux density ratios and the tile coordinate $m$, with medians in $m$ show as pink circles. $16^\text{th}$ and $84^\text{th}$ percentiles are shown as pink, dashed lines.}
\end{figure*}

For assessing the brightness scale across the survey images, we apply the same unresolved flags described in Section~\ref{sec:resolved} to the individual source lists then cross-match the individual sources lists to the various external catalogues as described in Section~\ref{sec:matching}. Note that more sources will be cross-matched in this per--source-list assessment than when looking at the full catalogues, since we include sources in at the edge of images in overlap regions. For assessing the brightness scale, we look at the SUMSS and NVSS cross-matches to allow easy reference to the same comparisons made in \citetalias{racs-mid} for the original images. As part of the additional metadata added prior to source-list concatenation, we add tile-reference coordinates, $(l,m)$, for each source entry. In the case of the mosaicked images, these $(l,m)$ values are with reference to the mosaiced image, and \textit{not} the original image that was used in the mosaic. Following \citetalias{racs-mid} we then median-bin the flux density ratios ($S_\text{RACS-mid}/S_\text{survey}$, for NVSS and SUMSS) in bins of angular sizes $8.6 \times 8.6$\,arcmin$^{2}$ (NVSS) and $14.4 \times 14.4$\,arcmin$^{2}$ (SUMSS).

Figures~\ref{fig:tileflux:nvss:fr}--\ref{fig:tileflux:sumss:25arcsec} shows (\emph{left}) the flux density ratios of sources in the RACS-mid catalogues and comparison catalogues, median-binned in the image-based $(l, m)$ coordinates. Figure~\ref{fig:tileflux} also shows the 2-D histogram of flux density ratios in $m$, where the majority of the edge-effects were seen in \citetalias{racs-mid}. We normalise the ratios by the overall median flux density ratio to ensure bulk offsets introduced by a choice of spectral index do not detract from position-dependent effects. For both the primary and 25-arcsec catalogues, position-dependent brightness effects are almost completely removed due to the mosaicking process. Specifically, we refer to figures 30 and 31 in \citetalias{racs-mid} that highlight these residual brightness scales features. Residual variation in the mosaiced images is $\lesssim$ a few \%. 

With the reduction of edge effects, we determine that the brightness scale uncertainty for the primary and 25-arcsec catalogues retain the uncertainty reported for the central tile region in \citetalias{racs-mid}: 6\%. The time-domain catalogue retains both uncertainties (6\% and 14\% for the tile edges, equation 5 in \citetalias{racs-mid}), and the source entries will explicitly note the appropriate uncertainty based on its position in the image it is taken from.

\section{Data availability}
\citetalias{racs-mid} describes the raw survey data and how to access the original images and calibrated visibility datasets. This data release comprises both the full-sensitivity mosaicked images used for the primary catalogue and the fixed-resolution 25-arcsec catalogue, as well as the three catalogues described in Section~\ref{sec:catalogues}. For the catalogues and full-sensitivity images described in this paper, availability is similarly through the the CSIRO \footnote{Commonwealth Scientific and Industrial Research Organisation.} ASKAP Science Data Archive \citep[CASDA;][]{casda,Huynh2020}, and each product is summarised here for completeness:

\paragraph{Primary catalogue.} 
\begin{itemize}
    \item \emph{Source list.} The final merged source lists. This contains all sources detected during source-finding on the full-sensitivity images that are not removed due to the de-duplication process when constructing the catalogue. This catalogue features the full sky up to $\delta_\text{J2000} = +49\degree$ with sources detected in Stokes I images. This is the recommended catalogue for most use cases.
    \item \emph{Component list.} The Gaussian components associated with the source list. Generally only used for reconstruction of the source models.
    \item \emph{Full-sensitivity images.} Stokes I full-sensitivity images with associated weight maps. For most use-cases these are the preferred images to use over the original images.
    \item \emph{Link to images and catalogues.}\\\url{https://doi.org/10.25919/p524-xb81}. 
\end{itemize}

\paragraph{Auxiliary 25-arcsec catalogue.}
\begin{itemize}
    \item \emph{Source list.} The merged 25-arcsec source lists. This is similar to the primary catalogue, except with the fixed 25-arcsec resolution and with coverage up to $\delta_\text{J2000} = +30\degree$. 
    \item \emph{Component list.} The Gaussian components associated with the 25-arcsec source list.
    \item \emph{Full-sensitivity 25-arcsec images.} Stokes I full-sensitivity images with associated weight maps.
    \item \emph{Link for images and catalogues.}\\\url{https://doi.org/10.25919/p524-xb81}. 
\end{itemize}

\paragraph{Auxiliary time-domain catalogue.}
\begin{itemize}
    \item \emph{Source list.} The concatenated source list from the original RACS-mid images. This covers the entire survey region, and retains duplicate source entries.
    \item \emph{Component list.} The Gaussian components associated with the time-domain source list.
    \item \emph{Link to the catalogues.}\\\url{https://doi.org/10.25919/p8ns-da63}.  
\end{itemize}


\section{Summary}

The Australian SKA Pathfinder (ASKAP) has completed observing the sky in three bands centered at 887.5, 1367.5, and 1632.5\,MHz as part of the Rapid ASKAP Continuum Survey (RACS). RACS-low (at 887.5\,MHz) and RACS-mid (at 1367.5\,MHz) have been processed and released \citepalias[][]{racs1,racs-mid} with a catalogue for RACS-low also released \citepalias{racs2}. While the 1632.5-MHz data have been observed, these data are yet to be processed and will be released in the future as RACS-high. This work describes the release of an `all-sky' general-purpose catalogue for RACS-mid, along with two auxiliary catalogues for more specific use-cases. 

The general purpose, primary catalogue was created by mosaicking nearby images to improve sensitivity in overlapping regions, convolving to the lowest common resolution of neighbouring images, source-finding with \texttt{PyBDSF} on these full-sensitivity mosaics, then merging the output source-lists while removing duplicate entries. The primary catalogue covers $-90\degree < \delta_\text{J2000} \leq +49\degree$ ($\sim 36\,200$\,deg$^2$) and contains 3\,107\,143 sources (2\,863\,400 excluding the Galactic Plane) down to $5\sigma_\text{rms}$. We investigated the reliability and completeness of the catalogues, finding a significant decrease in reliability in the Galactic Plane due to higher incidence of artefacts from extended sources. The primary catalogue is found to be 95\% complete at 2\,mJy. We matched the catalogue to other external catalogues such as RACS-low, LoTSS-DR2, VLASS-QL, SUMSS, NVSS, and FIRST and find properties such as flux density and astrometry consistent with what is reported in \citetalias{racs-mid}, suggesting additional mosaicking done here is not introducing extra systematic issues into the images and catalogue. 

In addition to the primary catalogue, we also provide an auxiliary version of the catalogue with a fixed resolution of $25^{\prime\prime} \times 25^{\prime\prime}$. This catalogue is created in a similar way to the primary catalogue, except images are all convolved to $25^{\prime\prime} \times 25^{\prime\prime}$ prior to mosaicking and source-finding. Because of this, the coverage is reduced to $\delta_\text{2000} \leq +30\degree$. This matches closely the existing RACS-low catalogue described in \citetalias[][]{racs2}. This 25-arcsec catalogue features lessened sensitivity and consequently fewer sources---2\,156\,393 sources above $5\sigma_\text{rms}$ (1\,992\,406 excluding the Galactic Plane)---with a significant decrease in reliability in the Galactic Plane . Other properties and comparisons are similar to the primary catalogue.

A third auxiliary catalogue is provided for time-domain science. This version features no additional convolution or mosaicking of images prior to source-finding and retains all duplicate source entries in overlapping regions. The purpose of this auxiliary time-domain catalogue is to detect and characterise time-variable sources without averaging signal between epochs as will be the case for some sources in the primary catalogue. 

All data products, including catalogues and full-sensitivity images are being made available through the CSIRO ASKAP Science Data Archive. This data release focuses on total intensity with Stokes I and work is currently underway to catalogue the sources of circular polarization using the Stokes V images produced as part of RACS-mid \citep[e.g.][]{Rose2023}.

\begin{acknowledgement}
We would like to thank the anonymous referee for their feedback that has improved the quality of this work.
This scientific work uses data obtained from Inyarrimanha Ilgari Bundara / the Murchison Radio-astronomy Observatory. We acknowledge the Wajarri Yamaji People as the Traditional Owners and native title holders of the Observatory site. CSIRO’s ASKAP radio telescope is part of the Australia Telescope National Facility (\url{https://ror.org/05qajvd42}). Operation of ASKAP is funded by the Australian Government with support from the National Collaborative Research Infrastructure Strategy. ASKAP uses the resources of the Pawsey Supercomputing Research Centre. Establishment of ASKAP, Inyarrimanha Ilgari Bundara, the CSIRO Murchison Radio-astronomy Observatory and the Pawsey Supercomputing Research Centre are initiatives of the Australian Government, with support from the Government of Western Australia and the Science and Industry Endowment Fund.

We used a range of \texttt{python} software packages during this work and the production of this manuscript, including \texttt{aplpy} \citep{Robitaille2012}, \texttt{astropy} \citep{astropy:2018}, \texttt{matplotlib} \citep{Hunter2007}, \texttt{numpy} \citep{numpy}, \texttt{scipy} \citep{scipy}, and \texttt{cmasher} \citep{cmasher}.  Some of the results in this paper have been derived using the \texttt{healpy} \citep{healpy} and HEALPix package. We make use of \texttt{ds9} \citep{ds9} and \texttt{topcat} \citep{topcat} for visualisation, as well as the ``Aladin sky atlas'' developed at CDS, Strasbourg Observatory, France \citep{aladin1,aladin2} for obtaining catalogue data. For precision rounding in \LaTeX tables we used \texttt{to-precision}: \url{https://bitbucket.org/william_rusnack/to-precision/src/master/}.

This project used public archival data from the Dark Energy Survey (DES). Funding for the DES Projects has been provided by the U.S. Department of Energy, the U.S. National Science Foundation, the Ministry of Science and Education of Spain, the Science and Technology Facilities Council of the United Kingdom, the Higher Education Funding Council for England, the National Center for Supercomputing Applications at the University of Illinois at Urbana-Champaign, the Kavli Institute of Cosmological Physics at the University of Chicago, the Center for Cosmology and Astro-Particle Physics at the Ohio State University, the Mitchell Institute for Fundamental Physics and Astronomy at Texas A\&M University, Financiadora de Estudos e Projetos, Funda{\c c}{\~a}o Carlos Chagas Filho de Amparo {\`a} Pesquisa do Estado do Rio de Janeiro, Conselho Nacional de Desenvolvimento Cient{\'i}fico e Tecnol{\'o}gico and the Minist{\'e}rio da Ci{\^e}ncia, Tecnologia e Inova{\c c}{\~a}o, the Deutsche Forschungsgemeinschaft, and the Collaborating Institutions in the Dark Energy Survey.
The Collaborating Institutions are Argonne National Laboratory, the University of California at Santa Cruz, the University of Cambridge, Centro de Investigaciones Energ{\'e}ticas, Medioambientales y Tecnol{\'o}gicas-Madrid, the University of Chicago, University College London, the DES-Brazil Consortium, the University of Edinburgh, the Eidgen{\"o}ssische Technische Hochschule (ETH) Z{\"u}rich,  Fermi National Accelerator Laboratory, the University of Illinois at Urbana-Champaign, the Institut de Ci{\`e}ncies de l'Espai (IEEC/CSIC), the Institut de F{\'i}sica d'Altes Energies, Lawrence Berkeley National Laboratory, the Ludwig-Maximilians Universit{\"a}t M{\"u}nchen and the associated Excellence Cluster Universe, the University of Michigan, the National Optical Astronomy Observatory, the University of Nottingham, The Ohio State University, the OzDES Membership Consortium, the University of Pennsylvania, the University of Portsmouth, SLAC National Accelerator Laboratory, Stanford University, the University of Sussex, and Texas A\&M University.
Based in part on observations at Cerro Tololo Inter-American Observatory, National Optical Astronomy Observatory, which is operated by the Association of Universities for Research in Astronomy (AURA) under a cooperative agreement with the National Science Foundation.

The Digitized Sky Surveys were produced at the Space Telescope Science Institute under U.S. Government grant NAG W-2166. The images of these surveys are based on photographic data obtained using the Oschin Schmidt Telescope on Palomar Mountain and the UK Schmidt Telescope. The plates were processed into the present compressed digital form with the permission of these institutions.

\end{acknowledgement}


\bibliography{references}

\begin{thebibliography}{}
\makeatletter
\relax
\def\mn@urlcharsother{\let\do\@makeother \do\$\do\&\do\#\do\^\do\_\do\%\do\~}
\definecolor{darkblue}{RGB}{30, 144, 255}
\def\mndoi{\begingroup\mn@urlcharsother \@ifnextchar [ {\mndoi@} {\mndoi@[]}}
\def\mndoi@[#1]#2{\def\@tempa{#1}\ifx\@tempa\@empty \href
  {http://dx.doi.org/#2} {\textcolor{darkblue}{doi:#2}}\else \href
  {http://dx.doi.org/#2} {\textcolor{darkblue}{#1}}\fi \endgroup}
\def\mn@eprint#1#2{\mn@eprint@#1:#2::\@nil}
\def\mn@eprint@arXiv#1{\href {http://arxiv.org/abs/#1} {{\tt arXiv:#1}}}
\def\mn@eprint@dblp#1{\href {http://dblp.uni-trier.de/rec/bibtex/#1.xml}
  {dblp:#1}}
\def\mn@eprint@#1:#2:#3:#4\@nil{\def\@tempa {#1}\def\@tempb {#2}\def\@tempc
  {#3}\ifx \@tempc \@empty \let \@tempc \@tempb \let \@tempb \@tempa \fi \ifx
  \@tempb \@empty \def\@tempb {arXiv}\fi \@ifundefined
  {mn@eprint@\@tempb}{\@tempb:\@tempc}{\expandafter \expandafter \csname
  mn@eprint@\@tempb\endcsname \expandafter{\@tempc}}}

\bibitem[\protect\citeauthoryear{{Abazajian} et~al.,}{{Abazajian}
  et~al.}{2009}]{sdssdr7}
{Abazajian} K.~N.,  et~al., 2009, \mndoi [\apjs] {10.1088/0067-0049/182/2/543},
  \href {http://adsabs.harvard.edu/abs/2009ApJS..182..543A} {182, 543}

\bibitem[\protect\citeauthoryear{{Abbott} et~al.,}{{Abbott}
  et~al.}{2018}]{des1}
{Abbott} T.~M.~C.,  et~al., 2018, \mndoi [\apjs] {10.3847/1538-4365/aae9f0},
  \href {https://ui.adsabs.harvard.edu/abs/2018ApJS..239...18A} {239, 18}

\bibitem[\protect\citeauthoryear{{Abbott} et~al.,}{{Abbott}
  et~al.}{2021}]{des:dr2}
{Abbott} T.~M.~C.,  et~al., 2021, \mndoi [\apjs] {10.3847/1538-4365/ac00b3},
  \href {https://ui.adsabs.harvard.edu/abs/2021ApJS..255...20A} {255, 20}

\bibitem[\protect\citeauthoryear{Akaike}{Akaike}{1974}]{Akaike1974}
Akaike H.,  1974, \mndoi [IEEE Transactions on Automatic Control]
  {10.1109/TAC.1974.1100705}, 19, 716

\bibitem[\protect\citeauthoryear{{Astropy Collaboration} et~al.,}{{Astropy
  Collaboration} et~al.}{2018}]{astropy:2018}
{Astropy Collaboration} et~al., 2018, \mndoi [\aj] {10.3847/1538-3881/aabc4f},
  \href {https://ui.adsabs.harvard.edu/abs/2018AJ....156..123A} {156, 123}

\bibitem[\protect\citeauthoryear{{Becker}, {White}  \& {Helfand}}{{Becker}
  et~al.}{1995}]{Becker95}
{Becker} R.~H.,  {White} R.~L.,   {Helfand} D.~J.,  1995, \mndoi [\apj]
  {10.1086/176166}, \href
  {https://ui.adsabs.harvard.edu/abs/1995ApJ...450..559B} {450, 559}

\bibitem[\protect\citeauthoryear{{Bertin}, {Mellier}, {Radovich}, {Missonnier},
  {Didelon}  \& {Morin}}{{Bertin} et~al.}{2002}]{swarp}
{Bertin} E.,  {Mellier} Y.,  {Radovich} M.,  {Missonnier} G.,  {Didelon} P.,
  {Morin} B.,  2002, in {Bohlender} D.~A.,  {Durand} D.,   {Handley} T.~H.,
  eds,  Astronomical Society of the Pacific Conference Series Vol. 281,
  Astronomical Data Analysis Software and Systems XI. p.~228

\bibitem[\protect\citeauthoryear{{Boch} \& {Fernique}}{{Boch} \&
  {Fernique}}{2014}]{aladin2}
{Boch} T.,  {Fernique} P.,  2014, in {Manset} N.,  {Forshay} P.,  eds,
  Astronomical Society of the Pacific Conference Series Vol. 485, Astronomical
  Data Analysis Software and Systems XXIII. p.~277

\bibitem[\protect\citeauthoryear{{Bock}, {Large}  \& {Sadler}}{{Bock}
  et~al.}{1999}]{bls99}
{Bock} D.~C.-J.,  {Large} M.~I.,   {Sadler} E.~M.,  1999, \mndoi [\aj]
  {10.1086/300786}, \href {http://adsabs.harvard.edu/abs/1999AJ....117.1578B}
  {117, 1578}

\bibitem[\protect\citeauthoryear{{Bondi}, {Ciliegi}, {Schinnerer},
  {Smol{\v{c}}i{\'c}}, {Jahnke}, {Carilli}  \& {Zamorani}}{{Bondi}
  et~al.}{2008}]{Bondi2008}
{Bondi} M.,  {Ciliegi} P.,  {Schinnerer} E.,  {Smol{\v{c}}i{\'c}} V.,  {Jahnke}
  K.,  {Carilli} C.,   {Zamorani} G.,  2008, \mndoi [\apj] {10.1086/589324},
  \href {https://ui.adsabs.harvard.edu/abs/2008ApJ...681.1129B} {681, 1129}

\bibitem[\protect\citeauthoryear{{Bonnarel} et~al.,}{{Bonnarel}
  et~al.}{2000}]{aladin1}
{Bonnarel} F.,  et~al., 2000, \mndoi [\aaps] {10.1051/aas:2000331}, \href
  {https://ui.adsabs.harvard.edu/abs/2000A&AS..143...33B} {143, 33}

\bibitem[\protect\citeauthoryear{{Chapman}, {Dempsey}, {Miller}, {Heywood},
  {Pritchard}, {Sangster}, {Whiting}  \& {Dart}}{{Chapman}
  et~al.}{2017}]{casda}
{Chapman} J.~M.,  {Dempsey} J.,  {Miller} D.,  {Heywood} I.,  {Pritchard} J.,
  {Sangster} E.,  {Whiting} M.,   {Dart} M.,  2017, in {Lorente} N.~P.~F.,
  {Shortridge} K.,   {Wayth} R.,  eds,  Astronomical Society of the Pacific
  Conference Series Vol. 512, Astronomical Data Analysis Software and Systems
  XXV. p.~73

\bibitem[\protect\citeauthoryear{{Charlot} et~al.,}{{Charlot}
  et~al.}{2020}]{Charlot2020}
{Charlot} P.,  et~al., 2020, \mndoi [\aap] {10.1051/0004-6361/202038368}, \href
  {https://ui.adsabs.harvard.edu/abs/2020A&A...644A.159C} {644, A159}

\bibitem[\protect\citeauthoryear{{Cohen}}{{Cohen}}{2004}]{Cohen2004}
{Cohen} A.,  2004, Long Wavelength Array Memo~17, Estimates of the Classical
  Confusion Limit for the LWA.
{Naval Research Laboratory}, \url{https://www.faculty.ece.vt.edu/swe/lwa/}

\bibitem[\protect\citeauthoryear{{Condon}}{{Condon}}{1974}]{Condon1974}
{Condon} J.~J.,  1974, \mndoi [\apj] {10.1086/152714}, \href
  {https://ui.adsabs.harvard.edu/abs/1974ApJ...188..279C} {188, 279}

\bibitem[\protect\citeauthoryear{{Condon}, {Cotton}, {Greisen}, {Yin},
  {Perley}, {Taylor}  \& {Broderick}}{{Condon} et~al.}{1998}]{ccg+98}
{Condon} J.~J.,  {Cotton} W.~D.,  {Greisen} E.~W.,  {Yin} Q.~F.,  {Perley}
  R.~A.,  {Taylor} G.~B.,   {Broderick} J.~J.,  1998, \mndoi [\aj]
  {10.1086/300337}, \href {http://adsabs.harvard.edu/abs/1998AJ....115.1693C}
  {115, 1693}

\bibitem[\protect\citeauthoryear{{Condon} et~al.,}{{Condon}
  et~al.}{2012}]{Condon2012}
{Condon} J.~J.,  et~al., 2012, \mndoi [\apj] {10.1088/0004-637X/758/1/23},
  \href {https://ui.adsabs.harvard.edu/abs/2012ApJ...758...23C} {758, 23}

\bibitem[\protect\citeauthoryear{{Dabhade} et~al.,}{{Dabhade}
  et~al.}{2020}]{Dabhade2020}
{Dabhade} P.,  et~al., 2020, \mndoi [\aap] {10.1051/0004-6361/202038344}, \href
  {https://ui.adsabs.harvard.edu/abs/2020A&A...642A.153D} {642, A153}

\bibitem[\protect\citeauthoryear{{Driessen}, {Heald}, {Duchesne}, {Murphy},
  {Lenc}, {Leung}  \& {Moss}}{{Driessen} et~al.}{2023}]{Driessen2023}
{Driessen} L.~N.,  {Heald} G.,  {Duchesne} S.~W.,  {Murphy} T.,  {Lenc} E.,
  {Leung} J.~K.,   {Moss} V.~A.,  2023, \mndoi [\pasa] {10.1017/pasa.2023.26},
  \href {https://ui.adsabs.harvard.edu/abs/2023PASA...40...36D} {40, e036}

\bibitem[\protect\citeauthoryear{{Duchesne} et~al.,}{{Duchesne}
  et~al.}{2023}]{racs-mid}
{Duchesne} S.~W.,  et~al., 2023, \mndoi [\pasa] {10.1017/pasa.2023.31}, \href
  {https://ui.adsabs.harvard.edu/abs/2023PASA...40...34D} {40, e034}

\bibitem[\protect\citeauthoryear{{Flaugher} et~al.,}{{Flaugher}
  et~al.}{2015}]{decam}
{Flaugher} B.,  et~al., 2015, \mndoi [\aj] {10.1088/0004-6256/150/5/150}, \href
  {https://ui.adsabs.harvard.edu/abs/2015AJ....150..150F} {150, 150}

\bibitem[\protect\citeauthoryear{{For} et~al.,}{{For} et~al.}{2021}]{For2021}
{For} B.~Q.,  et~al., 2021, \mndoi [\mnras] {10.1093/mnras/stab2257}, \href
  {https://ui.adsabs.harvard.edu/abs/2021MNRAS.507.2300F} {507, 2300}

\bibitem[\protect\citeauthoryear{{Franzen}, {Hurley-Walker}, {White},
  {Hancock}, {Seymour}, {Kapi{\'n}ska}, {Staveley-Smith}  \& {Wayth}}{{Franzen}
  et~al.}{2021}]{Franzen2021}
{Franzen} T.~M.~O.,  {Hurley-Walker} N.,  {White} S.~V.,  {Hancock} P.~J.,
  {Seymour} N.,  {Kapi{\'n}ska} A.~D.,  {Staveley-Smith} L.,   {Wayth} R.~B.,
  2021, \mndoi [\pasa] {10.1017/pasa.2021.5}, \href
  {https://ui.adsabs.harvard.edu/abs/2021PASA...38...14F} {38, e014}

\bibitem[\protect\citeauthoryear{{Gordon} et~al.,}{{Gordon}
  et~al.}{2021}]{Gordon2021}
{Gordon} Y.~A.,  et~al., 2021, \mndoi [\apjs] {10.3847/1538-4365/ac05c0}, \href
  {https://ui.adsabs.harvard.edu/abs/2021ApJS..255...30G} {255, 30}

\bibitem[\protect\citeauthoryear{{G{\'o}rski}, {Hivon}, {Banday}, {Wandelt},
  {Hansen}, {Reinecke}  \& {Bartelmann}}{{G{\'o}rski}
  et~al.}{2005}]{Gorski2005}
{G{\'o}rski} K.~M.,  {Hivon} E.,  {Banday} A.~J.,  {Wandelt} B.~D.,  {Hansen}
  F.~K.,  {Reinecke} M.,   {Bartelmann} M.,  2005, \mndoi [\apj]
  {10.1086/427976}, \href
  {https://ui.adsabs.harvard.edu/abs/2005ApJ...622..759G} {622, 759}

\bibitem[\protect\citeauthoryear{{Grundy} et~al.,}{{Grundy}
  et~al.}{2023}]{Grundy2023}
{Grundy} J.~A.,  et~al., 2023, \mndoi [\pasa] {10.1017/pasa.2023.11}, \href
  {https://ui.adsabs.harvard.edu/abs/2023PASA...40...12G} {40, e012}

\bibitem[\protect\citeauthoryear{{Gulati} et~al.,}{{Gulati}
  et~al.}{2023}]{Gulati2023}
{Gulati} A.,  et~al., 2023, \mndoi [\pasa] {10.1017/pasa.2023.21}, \href
  {https://ui.adsabs.harvard.edu/abs/2023PASA...40...25G} {40, e025}

\bibitem[\protect\citeauthoryear{{Hale} et~al.,}{{Hale} et~al.}{2021}]{racs2}
{Hale} C.~L.,  et~al., 2021, \mndoi [\pasa] {10.1017/pasa.2021.47}, \href
  {https://ui.adsabs.harvard.edu/abs/2021PASA...38...58H} {38, e058}

\bibitem[\protect\citeauthoryear{{Hale} et~al.,}{{Hale}
  et~al.}{2023}]{Hale2023}
{Hale} C.~L.,  et~al., 2023, \mndoi [\mnras] {10.1093/mnras/stac3320}, \href
  {https://ui.adsabs.harvard.edu/abs/2023MNRAS.520.2668H} {520, 2668}

\bibitem[\protect\citeauthoryear{Harris et~al.,}{Harris et~al.}{2020}]{numpy}
Harris C.~R.,  et~al., 2020, \mndoi [Nature] {10.1038/s41586-020-2649-2}, 585,
  357

\bibitem[\protect\citeauthoryear{{Helfand}, {White}  \& {Becker}}{{Helfand}
  et~al.}{2015}]{hwb15}
{Helfand} D.~J.,  {White} R.~L.,   {Becker} R.~H.,  2015, \mndoi [\apj]
  {10.1088/0004-637X/801/1/26}, \href
  {http://cdsads.u-strasbg.fr/abs/2015ApJ...801...26H} {801, 26}

\bibitem[\protect\citeauthoryear{{Heywood} et~al.,}{{Heywood}
  et~al.}{2016}]{Heywood2016}
{Heywood} I.,  et~al., 2016, \mndoi [\mnras] {10.1093/mnras/stw186}, \href
  {https://ui.adsabs.harvard.edu/abs/2016MNRAS.457.4160H} {457, 4160}

\bibitem[\protect\citeauthoryear{{Hotan} et~al.,}{{Hotan}
  et~al.}{2014}]{Hotan2014}
{Hotan} A.~W.,  et~al., 2014, \mndoi [\pasa] {10.1017/pasa.2014.36}, \href
  {https://ui.adsabs.harvard.edu/abs/2014PASA...31...41H} {31, e041}

\bibitem[\protect\citeauthoryear{{Hotan} et~al.,}{{Hotan}
  et~al.}{2021}]{Hotan2021}
{Hotan} A.~W.,  et~al., 2021, \mndoi [\pasa] {10.1017/pasa.2021.1}, \href
  {https://ui.adsabs.harvard.edu/abs/2021PASA...38....9H} {38, e009}

\bibitem[\protect\citeauthoryear{{Hunter}}{{Hunter}}{2007}]{Hunter2007}
{Hunter} J.~D.,  2007, \mndoi [Computing in Science and Engineering]
  {10.1109/MCSE.2007.55}, \href
  {http://adsabs.harvard.edu/abs/2007CSE.....9...90H} {9, 90}

\bibitem[\protect\citeauthoryear{{Hurley-Walker} et~al.,}{{Hurley-Walker}
  et~al.}{2017}]{gleamegc}
{Hurley-Walker} N.,  et~al., 2017, \mndoi [\mnras] {10.1093/mnras/stw2337},
  \href {http://adsabs.harvard.edu/abs/2017MNRAS.464.1146H} {464, 1146}

\bibitem[\protect\citeauthoryear{{Huynh}, {Dempsey}, {Whiting}  \&
  {Ophel}}{{Huynh} et~al.}{2020}]{Huynh2020}
{Huynh} M.,  {Dempsey} J.,  {Whiting} M.~T.,   {Ophel} M.,  2020, in
  {Ballester} P.,  {Ibsen} J.,  {Solar} M.,   {Shortridge} K.,  eds,
  Astronomical Society of the Pacific Conference Series Vol. 522, Astronomical
  Data Analysis Software and Systems XXVII. p.~263

\bibitem[\protect\citeauthoryear{{Intema}, {Jagannathan}, {Mooley}  \&
  {Frail}}{{Intema} et~al.}{2017}]{ijmf16}
{Intema} H.~T.,  {Jagannathan} P.,  {Mooley} K.~P.,   {Frail} D.~A.,  2017,
  \mndoi [\aap] {10.1051/0004-6361/201628536}, \href
  {https://ui.adsabs.harvard.edu/abs/2017A&A...598A..78I} {598, A78}

\bibitem[\protect\citeauthoryear{{Joye} \& {Mandel}}{{Joye} \&
  {Mandel}}{2003}]{ds9}
{Joye} W.~A.,  {Mandel} E.,  2003, in {Payne} H.~E.,  {Jedrzejewski} R.~I.,
  {Hook} R.~N.,  eds,  Astronomical Society of the Pacific Conference Series
  Vol. 295, Astronomical Data Analysis Software and Systems XII. p.~489

\bibitem[\protect\citeauthoryear{{Koribalski} et~al.,}{{Koribalski}
  et~al.}{2020}]{wallaby1}
{Koribalski} B.~S.,  et~al., 2020, \mndoi [\apss] {10.1007/s10509-020-03831-4},
  \href {https://ui.adsabs.harvard.edu/abs/2020Ap&SS.365..118K} {365, 118}

\bibitem[\protect\citeauthoryear{{Lacy} et~al.,}{{Lacy} et~al.}{2020}]{vlass}
{Lacy} M.,  et~al., 2020, \mndoi [\pasp] {10.1088/1538-3873/ab63eb}, \href
  {https://ui.adsabs.harvard.edu/abs/2020PASP..132c5001L} {132, 035001}

\bibitem[\protect\citeauthoryear{{Leung} et~al.,}{{Leung}
  et~al.}{2021}]{Leung2021}
{Leung} J.~K.,  et~al., 2021, \mndoi [\mnras] {10.1093/mnras/stab326}, \href
  {https://ui.adsabs.harvard.edu/abs/2021MNRAS.503.1847L} {503, 1847}

\bibitem[\protect\citeauthoryear{{Levrier}, {Wilman}, {Obreschkow},
  {Kloeckner}, {Heywood}  \& {Rawlings}}{{Levrier} et~al.}{2009}]{Levrier2009}
{Levrier} F.,  {Wilman} R.~J.,  {Obreschkow} D.,  {Kloeckner} H.~R.,  {Heywood}
  I.~H.,   {Rawlings} S.,  2009, in Wide Field Astronomy \& Technology for the
  Square Kilometre Array. p.~5 (\mn@eprint {arXiv} {0911.4611}),
  \mndoi{10.22323/1.132.0005}

\bibitem[\protect\citeauthoryear{{Line}, {Webster}, {Pindor}, {Mitchell}  \&
  {Trott}}{{Line} et~al.}{2017}]{puma}
{Line} J.~L.~B.,  {Webster} R.~L.,  {Pindor} B.,  {Mitchell} D.~A.,   {Trott}
  C.~M.,  2017, \mndoi [\pasa] {10.1017/pasa.2016.58}, \href
  {https://ui.adsabs.harvard.edu/abs/2017PASA...34....3L} {34, e003}

\bibitem[\protect\citeauthoryear{{Mandal} et~al.,}{{Mandal}
  et~al.}{2021}]{Mandal2021}
{Mandal} S.,  et~al., 2021, \mndoi [\aap] {10.1051/0004-6361/202039998}, \href
  {https://ui.adsabs.harvard.edu/abs/2021A&A...648A...5M} {648, A5}

\bibitem[\protect\citeauthoryear{{Matthews}, {Condon}, {Cotton}  \&
  {Mauch}}{{Matthews} et~al.}{2021}]{Matthews2021}
{Matthews} A.~M.,  {Condon} J.~J.,  {Cotton} W.~D.,   {Mauch} T.,  2021, \mndoi
  [\apj] {10.3847/1538-4357/abdd37}, \href
  {https://ui.adsabs.harvard.edu/abs/2021ApJ...909..193M} {909, 193}

\bibitem[\protect\citeauthoryear{{Mauch}, {Murphy}, {Buttery}, {Curran},
  {Hunstead}, {Piestrzynski}, {Robertson}  \& {Sadler}}{{Mauch}
  et~al.}{2003}]{mmb+03}
{Mauch} T.,  {Murphy} T.,  {Buttery} H.~J.,  {Curran} J.,  {Hunstead} R.~W.,
  {Piestrzynski} B.,  {Robertson} J.~G.,   {Sadler} E.~M.,  2003, \mndoi
  [\mnras] {10.1046/j.1365-8711.2003.06605.x}, \href
  {http://adsabs.harvard.edu/abs/2003MNRAS.342.1117M} {342, 1117}

\bibitem[\protect\citeauthoryear{{McConnell} et~al.,}{{McConnell}
  et~al.}{2016}]{McConnell2016}
{McConnell} D.,  et~al., 2016, \mndoi [\pasa] {10.1017/pasa.2016.37}, \href
  {https://ui.adsabs.harvard.edu/abs/2016PASA...33...42M} {33, e042}

\bibitem[\protect\citeauthoryear{{McConnell} et~al.,}{{McConnell}
  et~al.}{2020}]{racs1}
{McConnell} D.,  et~al., 2020, \mndoi [\pasa] {10.1017/pasa.2020.41}, \href
  {https://ui.adsabs.harvard.edu/abs/2020PASA...37...48M} {37, e048}

\bibitem[\protect\citeauthoryear{{Mohan} \& {Rafferty}}{{Mohan} \&
  {Rafferty}}{2015}]{pybdsf}
{Mohan} N.,  {Rafferty} D.,  2015, {PyBDSF: Python Blob Detection and Source
  Finder} (\mn@eprint {ascl} {1502.007})

\bibitem[\protect\citeauthoryear{{Murphy} et~al.,}{{Murphy}
  et~al.}{2013}]{Murphy2013}
{Murphy} T.,  et~al., 2013, \mndoi [\pasa] {10.1017/pasa.2012.006}, \href
  {https://ui.adsabs.harvard.edu/abs/2013PASA...30....6M} {30, e006}

\bibitem[\protect\citeauthoryear{{Murphy} et~al.,}{{Murphy}
  et~al.}{2021}]{Murphy2021}
{Murphy} T.,  et~al., 2021, \mndoi [\pasa] {10.1017/pasa.2021.44}, \href
  {https://ui.adsabs.harvard.edu/abs/2021PASA...38...54M} {38, e054}

\bibitem[\protect\citeauthoryear{{Perley} \& {Butler}}{{Perley} \&
  {Butler}}{2017}]{Perley2017}
{Perley} R.~A.,  {Butler} B.~J.,  2017, \mndoi [\apjs]
  {10.3847/1538-4365/aa6df9}, \href
  {https://ui.adsabs.harvard.edu/abs/2017ApJS..230....7P} {230, 7}

\bibitem[\protect\citeauthoryear{{Pritchard} et~al.,}{{Pritchard}
  et~al.}{2021}]{Pritchard2021}
{Pritchard} J.,  et~al., 2021, \mndoi [\mnras] {10.1093/mnras/stab299}, \href
  {https://ui.adsabs.harvard.edu/abs/2021MNRAS.502.5438P} {502, 5438}

\bibitem[\protect\citeauthoryear{{Robitaille} \& {Bressert}}{{Robitaille} \&
  {Bressert}}{2012}]{Robitaille2012}
{Robitaille} T.,  {Bressert} E.,  2012, {APLpy: Astronomical Plotting Library
  in Python}, Astrophysics Source Code Library (\mn@eprint {ascl} {1208.017})

\bibitem[\protect\citeauthoryear{{Rose} et~al.,}{{Rose}
  et~al.}{2023}]{Rose2023}
{Rose} K.,  et~al., 2023, \mndoi [\apjl] {10.3847/2041-8213/ace188}, \href
  {https://ui.adsabs.harvard.edu/abs/2023ApJ...951L..43R} {951, L43}

\bibitem[\protect\citeauthoryear{{Shimwell} et~al.,}{{Shimwell}
  et~al.}{2019}]{lotss-dr1}
{Shimwell} T.~W.,  et~al., 2019, \mndoi [\aap] {10.1051/0004-6361/201833559},
  \href {https://ui.adsabs.harvard.edu/abs/2019A&A...622A...1S} {622, A1}

\bibitem[\protect\citeauthoryear{{Shimwell} et~al.,}{{Shimwell}
  et~al.}{2022}]{lotss:dr2}
{Shimwell} T.~W.,  et~al., 2022, \mndoi [\aap] {10.1051/0004-6361/202142484},
  \href {https://ui.adsabs.harvard.edu/abs/2022A&A...659A...1S} {659, A1}

\bibitem[\protect\citeauthoryear{{Smol{\v{c}}i{\'c}}
  et~al.,}{{Smol{\v{c}}i{\'c}} et~al.}{2017a}]{Smolcic2017a}
{Smol{\v{c}}i{\'c}} V.,  et~al., 2017a, \mndoi [\aap]
  {10.1051/0004-6361/201628704}, \href
  {https://ui.adsabs.harvard.edu/abs/2017A&A...602A...1S} {602, A1}

\bibitem[\protect\citeauthoryear{{Smol{\v{c}}i{\'c}}
  et~al.,}{{Smol{\v{c}}i{\'c}} et~al.}{2017b}]{Smolcic2017b}
{Smol{\v{c}}i{\'c}} V.,  et~al., 2017b, \mndoi [\aap]
  {10.1051/0004-6361/201630223}, \href
  {https://ui.adsabs.harvard.edu/abs/2017A&A...602A...2S} {602, A2}

\bibitem[\protect\citeauthoryear{{Taylor}}{{Taylor}}{2005}]{topcat}
{Taylor} M.~B.,  2005, in {Shopbell} P.,  {Britton} M.,   {Ebert} R.,  eds,
  Astronomical Society of the Pacific Conference Series Vol. 347, Astronomical
  Data Analysis Software and Systems XIV. p.~29

\bibitem[\protect\citeauthoryear{{Thomson} et~al.,}{{Thomson}
  et~al.}{2023}]{Thomson2023}
{Thomson} A. J.~M.,  et~al., 2023, \mndoi [\pasa] {10.1017/pasa.2023.38}, \href
  {https://ui.adsabs.harvard.edu/abs/2023PASA...40...40T} {40, e040}

\bibitem[\protect\citeauthoryear{{Vernstrom}, {Scott}, {Wall}, {Condon},
  {Cotton}, {Kellermann}  \& {Perley}}{{Vernstrom}
  et~al.}{2016}]{Vernstrom2016b}
{Vernstrom} T.,  {Scott} D.,  {Wall} J.~V.,  {Condon} J.~J.,  {Cotton} W.~D.,
  {Kellermann} K.~I.,   {Perley} R.~A.,  2016, \mndoi [\mnras]
  {10.1093/mnras/stw1836}, \href
  {https://ui.adsabs.harvard.edu/abs/2016MNRAS.462.2934V} {462, 2934}

\bibitem[\protect\citeauthoryear{Virtanen et~al.,}{Virtanen
  et~al.}{2020}]{scipy}
Virtanen P.,  et~al., 2020, \mndoi [Nature Methods]
  {10.1038/s41592-019-0686-2}, \href {https://rdcu.be/b08Wh} {17, 261}

\bibitem[\protect\citeauthoryear{{Wayth} et~al.,}{{Wayth}
  et~al.}{2015}]{wlb+15}
{Wayth} R.~B.,  et~al., 2015, \mndoi [\pasa] {10.1017/pasa.2015.26}, \href
  {http://adsabs.harvard.edu/abs/2015PASA...32...25W} {32, 25}

\bibitem[\protect\citeauthoryear{{White}, {Becker}, {Helfand}  \&
  {Gregg}}{{White} et~al.}{1997}]{White1997}
{White} R.~L.,  {Becker} R.~H.,  {Helfand} D.~J.,   {Gregg} M.~D.,  1997,
  \mndoi [\apj] {10.1086/303564}, \href
  {https://ui.adsabs.harvard.edu/abs/1997ApJ...475..479W} {475, 479}

\bibitem[\protect\citeauthoryear{{Wilman} et~al.,}{{Wilman}
  et~al.}{2008}]{Wilman2008}
{Wilman} R.~J.,  et~al., 2008, \mndoi [\mnras]
  {10.1111/j.1365-2966.2008.13486.x}, \href
  {https://ui.adsabs.harvard.edu/abs/2008MNRAS.388.1335W} {388, 1335}

\bibitem[\protect\citeauthoryear{Zonca, Singer, Lenz, Reinecke, Rosset, Hivon
  \& Gorski}{Zonca et~al.}{2019}]{healpy}
Zonca A.,  Singer L.,  Lenz D.,  Reinecke M.,  Rosset C.,  Hivon E.,   Gorski
  K.,  2019, \mndoi [Journal of Open Source Software] {10.21105/joss.01298}, 4,
  1298

\bibitem[\protect\citeauthoryear{{de Zotti}, {Massardi}, {Negrello}  \&
  {Wall}}{{de Zotti} et~al.}{2010}]{deZotti2010}
{de Zotti} G.,  {Massardi} M.,  {Negrello} M.,   {Wall} J.,  2010, \mndoi
  [\aapr] {10.1007/s00159-009-0026-0}, \href
  {https://ui.adsabs.harvard.edu/abs/2010A&ARv..18....1D} {18, 1}

\bibitem[\protect\citeauthoryear{{van der Velden}}{{van der
  Velden}}{2020}]{cmasher}
{van der Velden} E.,  2020, \mndoi [The Journal of Open Source Software]
  {10.21105/joss.02004}, \href
  {https://ui.adsabs.harvard.edu/abs/2020JOSS....5.2004V} {5, 2004}

\makeatother
\end{thebibliography}

\appendix

\renewcommand{\thesubfigure}{(\roman{subfigure})}
\renewcommand\thefigure{A\arabic{figure}}   
\renewcommand\thetable{A\arabic{table}}
\setcounter{figure}{0}
\setcounter{table}{0}

\section{Catalogue columns}\label{app:columns}
While the columns (both names and data) are similar to those used for the RACS-low catalogue (see section~4.3 in \citetalias{racs2}). Tables~\ref{tab:columns:source} and \ref{tab:columns:component} outline the catalogues columns used for the three RACS-mid source and component catalogues, respectively. Note that the source and component lists feature almost identical columns with the exception of \texttt{Source\_ID} (in the source list) and \texttt{Gaussian\_ID} (in the component list) for unique identifiers for sources and components, respectively. The consequence of this is that quantity columns such as \texttt{Total\_flux} correspond to the source in source list or the component in the component list. 

Some columns have information that refers to the original images (\texttt{SBID}, \texttt{Scan\_start\_MJD}, and \texttt{Scan\_length}). In the case of the main reference catalogue and auxiliary 25-arcsec catalogue these refer to the central image that was used in mosaicking but does not include information about neighbouring images included in the mosaic. Therefore these columns will likely be accurate for sources at the centre (i.e. with low \texttt{Tile\_sep}) but will not be accurate towards image edges. In the case of the auxiliary time-domain catalogue, sources are taken from only non-mosaicked images so these columns will be accurate for all sources.  

Tables~\ref{tab:app:exsource} and \ref{tab:app:excomp} show three example rows from the primary source catalogue and primary component catalogue, respectively. The three example rows are 0, 500\,000, and 1\,500\,000, chosen arbitrarily.

\begin{table*}[t]
    \centering
    \caption{\label{tab:columns:source} Columns in the RACS-mid source catalogues.}
    \begin{adjustbox}{max width=\textwidth}
    \begin{tabular}{l l c l}\toprule
        Index & Name & Unit & Description \\\midrule
        0 & \texttt{Name} &  & Source name following IAU source name convention: \texttt{RACS-MID1 JHHMMSS.S$\pm$DDMMSS} \\
        1 & \texttt{Source\_ID} &  & Unique identifier for the source based on the \texttt{Field\_ID}. \\
        2 & \texttt{Field\_ID} & & Field name: \texttt{RACS\_HHMM$\pm$DD}. \\
        3 & \texttt{RA} & $\degree$ & J2000 right ascension of the source. \\
        4 & \texttt{Dec} & $\degree$ & J2000 declination of the source. \\
        5 & \texttt{Dec\_corr} & $ \degree $ & J2000 declination, with declination-dependent offset correction (Section~\ref{fig:astrometry:dec}). \\
        6 & \texttt{E\_RA} & $\degree$ & Uncertainty on \texttt{RA} from fitting the source position. \\
        7 & \texttt{E\_Dec} & $\degree$ & Uncertainty on \texttt{Dec} from fitting the source position. \\
        8 & \texttt{E\_Dec\_corr} & $\degree$ & Uncertainty on \texttt{Dec\_corr} from fitting the source position and declination offset model. \\
        9 & \texttt{Total\_flux} & mJy & Total flux density of the source. \\
        10 & \texttt{E\_Total\_flux\_PyBDSF} & mJy & Uncertainty in the total flux density from PyBDSF fitting. \\
        11 & \texttt{E\_Total\_flux} & mJy & Quadrature sum of the brightness scale and PyBDSF uncertainties for total flux density. \\
        12 & \texttt{Peak\_flux} & mJy\,beam$^{-1}$ & Peak flux density of the source. \\
        13 & \texttt{E\_Peak\_flux\_PyBDSF} & mJy\,beam$^{-1}$ & Uncertainty in the peak flux density from PyBDSF fitting. \\
        14 & \texttt{E\_Peak\_flux} & mJy\,beam$^{-1}$ &  Quadrature sum of the brightness scale and PyBDSF uncertainties for peak flux density. \\
        15 & \texttt{Maj\_axis} & arcsec & Size of the major axis of the source. \\
        16 & \texttt{Min\_axis} & arcsec & Size of the minor axis of the source. \\
        17 & \texttt{PA} & deg & Position angle of the source, east of north. \\
        18 & \texttt{E\_Maj\_axis} & arcsec & Uncertainty in the source major axis. \\
        19 & \texttt{E\_Min\_axis} & arcsec & Uncertainty in the source minor axis. \\
        20 & \texttt{E\_PA} & deg & Uncertainty in the source PA. \\
        21 & \texttt{DC\_Maj\_axis} & arcsec & Deconvolved size of the major axis of the source. \\
        22 & \texttt{DC\_Min\_axis} & arcsec & Deconvolved size of the minor axis of the source. \\
        23 & \texttt{DC\_PA} & deg & Deconvolved position angle of the source, east of north. \\
        24 & \texttt{E\_DC\_Maj\_axis} & arcsec & Uncertainty in the deconvolved source major axis. \\
        25 & \texttt{E\_DC\_Min\_axis} & arcsec & Uncertainty in the deconvolved source minor axis. \\
        26 & \texttt{E\_DC\_PA} & deg & Uncertainty in the deconvolved source PA. \\
        27 & \texttt{Noise} & mJy\,beam$^{-1}$ & Local estimate of the rms noise. \\
        28 & \texttt{Tile\_l} & deg & Direction cosine $l$ of the source with respect to the field centre. \\
        29 & \texttt{Tile\_m} & deg & Direction cosine $m$ of the source with respect to the field centre. \\
        30 & \texttt{Tile\_sep} & deg & Angular distance from the field centre. \\
        31 & \texttt{Gal\_lon} & deg & Galactic longitude of the source. \\
        32 & \texttt{Gal\_lat} & deg & Galactic latitude of the source. \\
        33 & \texttt{PSF\_Maj} & arcsec & FWHM of the PSF major axis of the field. \\
        34 & \texttt{PSF\_Min} & arcsec & FWHM of the PSF minor axis of the field. \\
        35 & \texttt{PSF\_PA} & deg & PA of the PSF of the field. \\
        36 & \texttt{S\_Code} & & Source structure classification provided by PyBDSF. \\
        37 & \texttt{N\_Gaussians} & & Number of Gaussian components comprising the source. \\
        38 & \texttt{Flag} & &  Source type flag. See Section~\ref{sec:resolved}. \\
        39 & \texttt{Scan\_start\_MJD} & yr & MJD of the observation start time for the field. \\
        40 & \texttt{Scan\_length} & s & Observation length of the field. \\
        41 & \texttt{SBID} & & Scheduling block ID of the field. \\
        42 & \texttt{E\_Flux\_scale} & & Fractional brightness scale uncertainty. \\
         \bottomrule
    \end{tabular}
    \end{adjustbox}
\end{table*}

\begin{table*}[t]
    \centering
    \begin{adjustbox}{max width=\textwidth}
    \caption{\label{tab:columns:component} Columns in the RACS-mid component catalogues.}
    \begin{tabular}{l l c l}\toprule
        Index & Name & Unit & Description \\\midrule
        0 & \texttt{Gaussian\_ID} &  & Unique identifier for the component based on the \texttt{Field\_ID}. \\
        1 & \texttt{Source\_ID} &  & Unique identifier for the source the component belongs to (see Table~\ref{tab:columns:source}). \\
        2 & \texttt{Field\_ID} & & Field name: \texttt{RACS\_HHMM$\pm$DD}. \\
        3 & \texttt{RA} & $\degree$ & J2000 right ascension of the component. \\
        4 & \texttt{Dec} & $\degree$ & J2000 declination of the component. \\
        5 & \texttt{Dec\_corr} & $ \degree $ & J2000 declination, with declination-dependent offset correction (Section~\ref{fig:astrometry:dec}). \\
        6 & \texttt{E\_RA} & $\degree$ & Uncertainty on \texttt{RA} from fitting the component position. \\
        7 & \texttt{E\_Dec} & $\degree$ & Uncertainty on \texttt{DEC} from fitting the component position. \\
        8 & \texttt{E\_Dec\_corr} & $\degree$ & Uncertainty on \texttt{Dec\_corr} from fitting the component position and declination offset model. \\
        9 & \texttt{Total\_flux} & mJy & Total flux density of the component. \\
        10 & \texttt{E\_Total\_flux\_PyBDSF} & mJy & Uncertainty in the total flux density from PyBDSF fitting. \\
        11 & \texttt{E\_Total\_flux} & mJy & Quadrature sum of the brightness scale and PyBDSF uncertainties for total flux density. \\
        12 & \texttt{Peak\_flux} & mJy\,beam$^{-1}$ & Peak flux density of the component. \\
        13 & \texttt{E\_Peak\_flux\_PyBDSF} & mJy\,beam$^{-1}$ & Uncertainty in the peak flux density from PyBDSF fitting. \\
        14 & \texttt{E\_Peak\_flux} & mJy\,beam$^{-1}$ &  Quadrature sum of the brightness scale and PyBDSF uncertainties for peak flux density. \\
        15 & \texttt{Maj\_axis} & arcsec & Size of the major axis of the component. \\
        16 & \texttt{Min\_axis} & arcsec & Size of the minor axis of the component. \\
        17 & \texttt{PA} & deg & Position angle of the component, east of north. \\
        18 & \texttt{E\_Maj\_axis} & arcsec & Uncertainty in the component major axis. \\
        19 & \texttt{E\_Min\_axis} & arcsec & Uncertainty in the component minor axis. \\
        20 & \texttt{E\_PA} & deg & Uncertainty in the component PA. \\
        21 & \texttt{DC\_Maj\_axis} & arcsec & Deconvolved size of the major axis of the component. \\
        22 & \texttt{DC\_Min\_axis} & arcsec & Deconvolved size of the minor axis of the component. \\
        23 & \texttt{DC\_PA} & deg & Deconvolved position angle of the component, east of north. \\
        24 & \texttt{E\_DC\_Maj\_axis} & arcsec & Uncertainty in the deconvolved component major axis. \\
        25 & \texttt{E\_DC\_Min\_axis} & arcsec & Uncertainty in the deconvolved component minor axis. \\
        26 & \texttt{E\_DC\_PA} & deg & Uncertainty in the deconvolved component PA. \\
        27 & \texttt{Noise} & mJy\,beam$^{-1}$ & Local estimate of the rms noise. \\
        28 & \texttt{Tile\_l} & deg & Direction cosine $l$ of the component with respect to the field centre. \\
        29 & \texttt{Tile\_m} & deg & Direction cosine $m$ of the component with respect to the field centre. \\
        30 & \texttt{Tile\_sep} & deg & Angular distance from the field centre. \\
        31 & \texttt{Gal\_lon} & deg & Galactic longitude of the component. \\
        32 & \texttt{Gal\_lat} & deg & Galactic latitude of the component. \\
        33 & \texttt{PSF\_Maj} & arcsec & FWHM of the PSF major axis of the field. \\
        34 & \texttt{PSF\_Min} & arcsec & FWHM of the PSF minor axis of the field. \\
        35 & \texttt{PSF\_PA} & deg & PA of the PSF of the field. \\
        36 & \texttt{S\_Code} & & Host source structure classification provided by PyBDSF. \\
        37 & \texttt{Flag} & &  Host source type flag. See Section~\ref{sec:resolved}. \\
        38 & \texttt{Scan\_start\_MJD} & yr & MJD of the observation start time for the field. \\
        39 & \texttt{Scan\_length} & s & Observation length of the field. \\
        40 & \texttt{SBID} & & Scheduling block ID of the field. \\
        41 & \texttt{E\_Flux\_scale} & & Fractional brightness scale uncertainty. \\
         \bottomrule
    \end{tabular}
    \end{adjustbox}
\end{table*}

\begin{table*}[t!]
    \centering
    \caption{\label{tab:app:exsource} Example three rows from the primary catalogue source list. Rows with associated uncertainties are rounded to match the uncertainty.}
    \begin{tabular}{l l c c c}\toprule
    Index & Name & Row 0 & Row 500\,000 & Row 1\,500\,000 \\\midrule
0 & \texttt{Name}  & RACS-MID1 J001712.0+370623  & RACS-MID1 J035732.6-665321  & RACS-MID1 J113753.8-045512  \\
1 & \texttt{Source\_ID}  & RACS\_0000+37\_83  & RACS\_0352-64\_1676  & RACS\_1127-04\_616  \\
2 & \texttt{Field\_ID}  & RACS\_0000+37  & RACS\_0352-64  & RACS\_1127-04  \\
3 & \texttt{RA}  & $4.29993$  & $59.38583$  & $174.474010$  \\
4 & \texttt{Dec}  & $37.1065$  & $-66.88918$  & $-4.920058$  \\
5 & \texttt{Dec\_corr}  & $37.1067$  & $-66.8891$  & $-4.92013$  \\
6 & \texttt{E\_RA}  & $0.00018$  & $0.00006$  & $0.000008$  \\
7 & \texttt{E\_Dec}  & $0.0021$  & $0.00004$  & $0.000005$  \\
8 & \texttt{E\_Dec\_corr}  & $0.0021$  & $0.0005$  & $0.00005$  \\
9 & \texttt{Total\_flux}  & $2.2$  & $16.9$  & $35.5$  \\
10 & \texttt{E\_Total\_flux\_PyBDSF}  & $0.561215228166031$  & $0.8797639604032615$  & $0.3384799415922147$  \\
11 & \texttt{E\_Total\_flux}  & $0.6$  & $1.3$  & $2.2$  \\
12 & \texttt{Peak\_flux}  & $1.33$  & $8.7$  & $34.9$  \\
13 & \texttt{E\_Peak\_flux\_PyBDSF}  & $0.19189768395374718$  & $0.17519237007945776$  & $0.1920388364685295$  \\
14 & \texttt{E\_Peak\_flux}  & $0.21$  & $0.5$  & $2.1$  \\
15 & \texttt{Maj\_axis}  & $66$  & $20.5$  & $10.67$  \\
16 & \texttt{Min\_axis}  & $14.6$  & $12.14$  & $8.80$  \\
17 & \texttt{PA}  & $1$  & $148.1$  & $76.6$  \\
18 & \texttt{E\_Maj\_axis}  & $18$  & $0.6$  & $0.06$  \\
19 & \texttt{E\_Min\_axis}  & $1.2$  & $0.23$  & $0.04$  \\
20 & \texttt{E\_PA}  & $9$  & $2.4$  & $1.1$  \\
21 & \texttt{DC\_Maj\_axis}  & $46$  & $16.4$  & $1.42$  \\
22 & \texttt{DC\_Min\_axis}  & $7.4$  & $6.36$  & $1.02$  \\
23 & \texttt{DC\_PA}  & $1$  & $138.7$  & $130.4$  \\
24 & \texttt{E\_DC\_Maj\_axis}  & $18$  & $0.6$  & $0.06$  \\
25 & \texttt{E\_DC\_Min\_axis}  & $1.2$  & $0.23$  & $0.04$  \\
26 & \texttt{E\_DC\_PA}  & $9$  & $2.4$  & $1.1$  \\
27 & \texttt{Noise}  & $0.25890651158988476$  & $0.17519237007945776$  & $0.19256076484452933$  \\
28 & \texttt{Tile\_l}  & $3.428092135103292$  & $0.5186964762309064$  & $2.524380128180197$  \\
29 & \texttt{Tile\_m}  & $-0.06163316029025561$  & $-1.9909730968373767$  & $-0.24834681422013694$  \\
30 & \texttt{Tile\_sep}  & $3.4286461370640944$  & $2.0574304135947807$  & $2.536566808046622$  \\
31 & \texttt{Gal\_lon}  & $115.38988737461287$  & $280.6581922381863$  & $271.16315947584053$  \\
32 & \texttt{Gal\_lat}  & $-25.26333975258068$  & $-41.27975708193742$  & $53.354126131693235$  \\
33 & \texttt{PSF\_Maj}  & $47.0$  & $13.0$  & $11.0$  \\
34 & \texttt{PSF\_Min}  & $13.0$  & $9.0$  & $9.0$  \\
35 & \texttt{PSF\_PA}  & $-2.0$  & $-5.0$  & $76.0$  \\
36 & \texttt{S\_Code}  & S  & M  & S  \\
37 & \texttt{N\_Gaussians}  & 1  & 2  & 1  \\
38 & \texttt{Flag}  & 1  & 1  & 0  \\
39 & \texttt{Scan\_start\_MJD}  & $59209.424927151034$  & $59239.44535211112$  & $59218.89534635104$  \\
40 & \texttt{Scan\_length}  & $905.748479394226$  & $915.701760796814$  & $905.7484803479$  \\
41 & \texttt{SBID}  & 20376  & 21946  & 20857  \\
42 & \texttt{E\_Flux\_scale}  & 0.06  & 0.06  & 0.06  \\\bottomrule
    \end{tabular}
\end{table*}

\begin{table*}[t!]
    \centering
    \caption{\label{tab:app:excomp} Example three rows from the primary catalogue component list. Rows with associated uncertainties are rounded to match the uncertainty.}
    \begin{tabular}{l l c c c}\toprule
    Index & Name & Row 0 & Row 500\,000 & Row 1\,500\,000 \\\midrule
0 & \texttt{Gaussian\_ID}  & RACS\_0000+37\_85  & RACS\_0248-18\_2709  & RACS\_0830-64\_2823  \\
1 & \texttt{Source\_ID}  & RACS\_0000+37\_83  & RACS\_0248-18\_1824  & RACS\_0830-64\_1911  \\
2 & \texttt{Field\_ID}  & RACS\_0000+37  & RACS\_0248-18  & RACS\_0830-64  \\
3 & \texttt{RA}  & $4.29993$  & $42.86770$  & $129.12277$  \\
4 & \texttt{Dec}  & $37.1065$  & $-20.15107$  & $-66.17611$  \\
5 & \texttt{Dec\_corr}  & $37.1067$  & $-20.15117$  & $-66.1761$  \\
6 & \texttt{E\_RA}  & $0.00018$  & $0.00029$  & $0.00019$  \\
7 & \texttt{E\_Dec}  & $0.0021$  & $0.00017$  & $0.00026$  \\
8 & \texttt{E\_Dec\_corr}  & $0.0021$  & $0.00020$  & $0.0005$  \\
9 & \texttt{Total\_flux}  & $2.2$  & $0.77$  & $0.76$  \\
10 & \texttt{E\_Total\_flux\_PyBDSF}  & $0.561215228166031$  & $0.25748102921394295$  & $0.25233369043145587$  \\
11 & \texttt{E\_Total\_flux}  & $0.6$  & $0.26$  & $0.26$  \\
12 & \texttt{Peak\_flux}  & $1.33$  & $0.79$  & $0.89$  \\
13 & \texttt{E\_Peak\_flux\_PyBDSF}  & $0.19189768395374718$  & $0.1491429321029494$  & $0.15968835603753778$  \\
14 & \texttt{E\_Peak\_flux}  & $0.21$  & $0.16$  & $0.17$  \\
15 & \texttt{Maj\_axis}  & $66$  & $11.2$  & $11.4$  \\
16 & \texttt{Min\_axis}  & $14.6$  & $8.1$  & $9.2$  \\
17 & \texttt{PA}  & $1$  & $75$  & $161$  \\
18 & \texttt{E\_Maj\_axis}  & $18$  & $2.5$  & $2.3$  \\
19 & \texttt{E\_Min\_axis}  & $1.2$  & $1.3$  & $1.5$  \\
20 & \texttt{E\_PA}  & $9$  & $27$  & $36$  \\
21 & \texttt{DC\_Maj\_axis}  & $46$  & $0.0$  & $0.0$  \\
22 & \texttt{DC\_Min\_axis}  & $7.4$  & $0.0$  & $0.0$  \\
23 & \texttt{DC\_PA}  & $1$  & $0$  & $0$  \\
24 & \texttt{E\_DC\_Maj\_axis}  & $18$  & $2.5$  & $2.3$  \\
25 & \texttt{E\_DC\_Min\_axis}  & $1.2$  & $1.3$  & $1.5$  \\
26 & \texttt{E\_DC\_PA}  & $9$  & $27$  & $36$  \\
27 & \texttt{Noise}  & $0.25890651158988476$  & $0.1532413443783298$  & $0.16807088104542345$  \\
28 & \texttt{Tile\_l}  & $3.428092135103292$  & $0.6386338358213877$  & $0.5577563287990496$  \\
29 & \texttt{Tile\_m}  & $-0.06163316029025561$  & $-1.484598076230764$  & $-1.2785573714374245$  \\
30 & \texttt{Tile\_sep}  & $3.4286461370640944$  & $1.6161326134338543$  & $1.394919737609507$  \\
31 & \texttt{Gal\_lon}  & $115.38988737461287$  & $205.15857015376932$  & $280.9120339736348$  \\
32 & \texttt{Gal\_lat}  & $-25.26333975258068$  & $-61.71453287237826$  & $-14.954794335737184$  \\
33 & \texttt{PSF\_Maj}  & $47.0$  & $10.0$  & $13.0$  \\
34 & \texttt{PSF\_Min}  & $13.0$  & $9.0$  & $10.0$  \\
35 & \texttt{PSF\_PA}  & $-2.0$  & $93.0$  & $-11.0$  \\
36 & \texttt{S\_Code}  & S  & S  & S  \\
37 & \texttt{Flag}  & 1  & 0  & 0  \\
38 & \texttt{Scan\_start\_MJD}  & $59209.424927151034$  & $59231.43342123114$  & $59250.64567851111$  \\
39 & \texttt{Scan\_length}  & $905.748479394226$  & $905.7484803479$  & $905.7484803479$  \\
40 & \texttt{SBID}  & 20376  & 21459  & 22458  \\
41 & \texttt{E\_Flux\_scale}  & 0.06  & 0.06  & 0.06  \\\bottomrule
    \end{tabular}
\end{table*}

\end{document}